\documentclass[amsmath, amssymb, aps, prb, reprint, superscriptaddress]{revtex4-2}

\usepackage{graphicx}
\usepackage{dcolumn}

\usepackage{bm}
\usepackage{physics}
\usepackage[svgnames]{xcolor}
\usepackage[colorlinks=true, linkcolor=blue, citecolor=blue, urlcolor=blue]{hyperref}

\begin{document}

\title{Superbunching in cathodoluminescence: a master equation approach}

\author{Tatsuro Yuge}
\email[]{yuge.tatsuro@shizuoka.ac.jp}
\affiliation{Department of Physics, Shizuoka University, Shizuoka 422-8529, Japan}

\author{Naoki Yamamoto}
\affiliation{Department of Materials Science and Engineering,
  School of Materials and Chemical Technology, Tokyo Institute of Technology,
  4259 Nagatsuta, Midoriku, Yokohama 226-8503, Japan}

\author{Takumi Sannomiya}
\affiliation{Department of Materials Science and Engineering,
  School of Materials and Chemical Technology, Tokyo Institute of Technology,
  4259 Nagatsuta, Midoriku, Yokohama 226-8503, Japan}

\author{Keiichirou Akiba}
\email[]{akiba.keiichiro@qst.go.jp}
\affiliation{Takasaki Advanced Radiation Research Institute,
  National Institutes for Quantum Science and Technology,
  1233 Watanuki, Takasaki, Gunma 370-1292, Japan}

\begin{abstract}
  We propose a theoretical model of a master equation for cathodoluminescence (CL).
  The master equation describes simultaneous excitation of multiple emitters
  by an incoming electron and radiative decay of individual emitters.
  We investigate the normalized second-order correlation function, $g^{(2)}(\tau)$, of this model.
  We derive the exact formula for the zero-time delay correlation, $g^{(2)}(0)$,
  and show that the model successfully describes giant bunching (superbunching) in the CL.
  We also derive an approximate form of $g^{(2)}(\tau)$, which is valid for small excitation rate.
  Furthermore, we discuss the state of the radiation field of the CL.
  We reveal that the superbunching results from a mixture of an excited photon state and the vacuum state
  and that this type of state is realized in the CL.
\end{abstract}

\maketitle

\section{Introduction}

In electron microscopy, cathodoluminescence (CL) visualizes optical properties beyond diffraction limit of light.
A wide range of materials can be investigated by this approach, for instance,
defect or luminescence centers in semiconductors \cite{Kalceff_etal_1995,Mitsui_etal_1996,Tararan_etal_2018,Bidaud_etal_2021},
quantum-confined structures \cite{Gustafsson_Samuelson_1994,Akiba_etal_2004,Merano_etal_2005},
surface plasmon polaritons \cite{Kuttge_etal_2009,Yamamoto_etal_2015,Sannomiya_etal_2020},
and fluorescent proteins \cite{Fisher_etal_2008,Nagayama_etal_2016,Akiba_etal_2020}.
Thus, the electron microscopy-based CL measurement is a powerful tool to analyze various materials on nanoscale.

The optical state of CL itself has not been in the spot light for a long time
though CL had been used in displays with cathode ray tubes for more than a century.
By the recent introduction of Hanbury Brown-Twiss (HBT) interferometry to CL,
the quantum character ``antibunching'' of the emitted states of CL
has been revealed with a deep subwavelength spatial resolution
in the measurement of a single nitrogen-vacancy (NV) center in a nanodiamond \cite{Tizei_Kociak_2013}.
Since antibunching is a result of the particle nature of a photon,
this HBT-CL technique opens a way to measure quantum optical phenomena on the nanoscale.

However, the HBT measurement of CL from multiple defect centers
has presented strong bunching \cite{Meuret_etal_2015},
which has not been observed in photoluminescence (PL) experiments for the same kind of sample.
Although there are differences between optical and electron-beam excitations such as absence of the NV$^-$ spectrum in CL \cite{Garcia_etal_2020},
this bunching observation raises a question on the origin of the bunching.
In addition, the observed bunching in CL is often huge,
where the normalized second-order correlation function, $g^{(2)}(\tau)$,
at time delay $\tau=0$ is larger than 2, i.e., superthermal values.
This is known as superbunching and a peculiar state of light.
A representative example of the superbunching is spontaneous parametric down-converted light (a squeezed vacuum),
which is the quantum light widely used as heralded single photons and entangled photon pairs \cite{Loudon_2000}.
Other examples are superradiant coupling of the emitters \cite{Auffeves_etal_2011,Leymann_etal_2015,Jahnke_etal_2016},
quantum dot--metal nanoparticles \cite{Ridolfo_etal_2010,Zhao_etal_2015},
and bimodal lasers \cite{Leymann_etal_2013a,Redlich_etal_2016,Marconi_etal_2018,Leymann_etal_2017,Schmidt_etal_2021}.

The bunching in CL has already enabled us to measure the luminescent lifetime well below the optical diffraction limit without a pulsed electron beam \cite{Meuret_etal_2016}.
This time-resolved measurement not only demonstrated the Purcell effect on the nanoscale \cite{Lourenco-Martins_etal_2018,Yanagimoto_etal_2021}
but also quantified excitation and emission efficiencies of optical nanostructures \cite{Meuret_etal_2017,Meuret_etal_2018}.
These practical applications prove that the CL photon correlation has great potential to access intrinsic nanophotonic properties in a direct manner and offer important insights into nanophotonic devices. Therefore, it is important to clarify how the strong bunching emerges in CL from both basic and applied aspects. The deeper understanding of CL photon correlation should progress nanoscale optical imaging to the next stage.

There were studies on CL photon statistics about half a century ago \cite{van_Rijswijk_1976a,van_Rijswijk_1976b}.
These pioneering investigations presented a theoretical description of the photon statistics and an experimental observation of the strong intensity correlation.
However, the feature of $g^{(2)}$ in CL was not focused.
In the first report of the superbunching in CL \cite{Meuret_etal_2015},
Meuret \textit{et al.} assumed that plasmons induce a synchronized excitation of multiple emitters
and proposed a stochastic model to perform a Monte Carlo simulation on $g^{(2)}(\tau)$.
On the basis of similar assumptions, CL excitation efficiency was estimated \cite{Meuret_etal_2017},
and an analytical model was constructed \cite{Garcia_etal_2021}.
Feldman \textit{et al.} claimed that the bunching in the nanodiamond CL is mediated by the phonon sidebands
and explained $g^{(2)}(\tau)$ using another Monte Carlo model \cite{Feldman_etal_2018}.
Besides these models, Yanagimoto \textit{et al.} derived an expression of $g^{(2)}(\tau)$ using a rate equation for multiple two-level systems \cite{Yanagimoto_etal_2021}.
However, the models in all the previous studies are essentially classical,
and no quantum model has been proposed that explains the superbunching in CL.
Even if photon bunching can be described by classical electromagnetic waves, the lack of photon picture would obscure the understanding of the essence of CL photon correlation.

In this study, we introduce a model of quantum master equation (QME)
to describe the dynamics of multiple emitters in CL.
The excitation by incident electrons is incorporated phenomenologically in the QME.
We find that the QME is reduced to a semi-classical master equation for the distribution of number of excited emitters.
From this master equation, we exactly obtain the stationary distribution
and the formula for zero-time delay correlation $g^{(2)}(0)$.
We also derive an approximate equation for delay time-dependent correlation $g^{(2)}(\tau)$.
We show that these results successfully reproduce several features of CL,
in particular, the superbunching and the decaying behavior of $g^{(2)}(\tau)$.
Moreover, we extend the model so that it is applicable to the case of large electron-beam current.
We also deduce the state of the radiation field from a possible sequence of pulses of the field in the CL.
From the model calculation and the deduced argument for the radiation field,
we shed light on a universal aspect of the superbunching.

\section{Model}
\label{sec:Model}

\subsection{Excitation process in cathodoluminescence}

Before describing the quantum master equation of our model,
we briefly explain the excitation process in a material by fast incident electrons in CL.

In CL, it is considered that an incident electron excites multiple emitters (defect centers)
not directly but via several steps of elementary processes mainly due to bulk plasmons and/or secondary electrons \cite{Egerton_2011,Yamamoto_2010,Rothwarf_1973,Yacobi_Holt_1986,Meuret_etal_2015,Meuret_etal_2017,Varkentina_etal_2022}.
The timescale required to excite emitters ranges from femtoseconds (bulk plasmon) to picoseconds (secondary electrons).
Therefore, when the radiative lifetime $\tau_\mathrm{rad}$ of the emitters is on the order of nanoseconds we consider this case in the present study),
the excitation timescale is sufficiently smaller than the emitter lifetime.

The region excited by the electron beam extends from the beam path.
Its size depends on
the beam diameter, the generation range of secondary (quasi-)particles,
and the mean free path and/or diffusion length of secondary (or higher-order) carriers.
The typical length scale of the excited region is several tens of nanometers when using a thin sample.

\subsection{Quantum master equation}

Considering the above excitation process by electron beam, we propose the following quantum model of CL.

The system of our interest is composed of $N$ emitters, which are located in the excited region.
Each emitter is modeled by a two-level system (TLS) with transition energy $\hbar \omega_\mathrm{e}$.
We assume that the density of emitters is low and thus the interaction among them is absent.
Therefore the system Hamiltonian is given by
\begin{align}
  \hat{H} = \sum_{j=1}^N \frac{\hbar \omega_\mathrm{e}}{2} \hat{\sigma}_j^z.
  \label{Hamiltonian}
\end{align}
Here $\hat{\sigma}_j^z$ is the $z$ component of the Pauli matrix for the $j$th TLS.
This is expressed as $\hat{\sigma}_j^z = \ket{1}_j\bra{1} - \ket{0}_j\bra{0}$
with the lower level state $\ket{0}_j$ and the upper one $\ket{1}_j$ of the $j$th TLS.

In this study, we consider CL emission with a continuous electron beam.
The emitters are continuously excited by the incoming electrons and decay with photon emission.
The Lindblad-type quantum master equation (QME)
\cite{Gorini_etal_1976,Lindblad_1976,Breuer_Petruccione_2002,Carmichael_1999}
is a suitable method for describing dynamics in this situation.
The QME has the following form:
\begin{align}
  \frac{d}{dt} \hat{\rho}(t) & = \mathcal{L} \hat{\rho}(t),
  \label{QME}
\end{align}
where $\hat{\rho}(t)$ is the state (density matrix) of the system at time $t$ and the Liouvillian $\mathcal{L}$ is given by
\begin{align}
  \mathcal{L} \hat{\rho} & = \frac{1}{i\hbar} [\hat{H}, \hat{\rho}]
  + \mathcal{D}_{\mathrm{rad}} \hat{\rho}
  + \mathcal{D}_{\mathrm{ex}} \hat{\rho}.
  \label{Liouvillian}
\end{align}
The first term represents the unitary part of the time evolution with the system Hamiltonian \eqref{Hamiltonian}.
The second and third terms represent the non-unitary parts due to the decay and excitation, respectively, as explained below.

The second term in the Liouvillian~\eqref{Liouvillian} describes the radiative decay of the emitters.
We here assume that the dipole moments of the emitters are randomly oriented.
In this case, they are independently damped
even though the emitters excited by the electron beam are located
within the excited region, which is smaller than the wavelength $2 \pi c / \omega_e$.
Therefore, as in the quantum optical master equation \cite{Breuer_Petruccione_2002, Carmichael_1999,Scully_Zubairy_1997}
under the assumption that the reservoir temperature is sufficiently smaller than $\hbar \omega_e$,
the second term is given in the following Lindblad form:
\begin{align}
  \mathcal{D}_{\mathrm{rad}} \hat{\rho}
   & = \frac{1}{\tau_\mathrm{rad}} \sum_{j=1}^N
  \qty( \hat{\sigma}_j^- \hat{\rho} \hat{\sigma}_j^+
  - \frac{1}{2} \qty{ \hat{\sigma}_j^+ \hat{\sigma}_j^-, \hat{\rho} } ).
  \label{QME_2nd_term}
\end{align}
Here $\hat{\sigma}_j^+ = \ket{1}_j\bra{0}$ and $\hat{\sigma}_j^- = \ket{0}_j\bra{1}$ are the raising and lowering operators of the $j$th TLS, respectively.
And $\tau_\mathrm{rad}$ is the radiative lifetime of each emitter.
We note that we can incorporate the non-radiative decay in the same Lindblad form,
in which case we should replace the prefactor $1 / \tau_\mathrm{rad}$ with
$1 / \tau_\mathrm{tot} =  1 / \tau_\mathrm{rad} + 1 / \tau_\text{non-rad}$
to include the non-radiative lifetime $\tau_\text{non-rad}$.

The third term in the Liouvillian~\eqref{Liouvillian} describes the excitation of the emitters by an electron beam.
As explained in Sec.\ref{sec:Model}A, the excitation timescale for each electron is sufficiently smaller than the radiative lifetime $\tau_\mathrm{rad}$.
Therefore, we can consider that the multiple emitters are excited \textit{simultaneously} by a single incident electron.
In this study, for simplicity, we assume that the number of emitters simultaneously excited by an electron is constant and is equal to $N$ that is introduced in Eq.~\eqref{Hamiltonian}.
On the other hand, the excitation rate $\gamma$ is connected to the electron-beam current $I$.
The unit-time number of electrons incident on the sample is $(I/e)$
($e$ is the elementary charge).
And the excitation occurs $(I/e) p_{\mathrm{ex}}$ times per unit time,
where $p_{\mathrm{ex}}$ is the probability for the excitation by an electron.
Therefore, $\gamma = (I/e) p_{\mathrm{ex}}$.
To incorporate this simultaneous excitation of the $N$ emitters at the rate of $\gamma$,
we introduce the following third term:
\begin{align}
  \mathcal{D}_{\mathrm{ex}} \hat{\rho}
   & = \gamma \qty( \hat{\Pi}^+ \hat{\rho} \hat{\Pi}^-
  - \frac{1}{2} \qty{ \hat{\Pi}^- \hat{\Pi}^+, \hat{\rho} } ),
  \label{QME_3rd_term}
  \\
  \hat{\Pi}^{\pm}
   & = \bigotimes_{j=1}^{N} \hat{\sigma}_j^{\pm}.
  \label{Lindblad_op_3rd_term}
\end{align}
The Lindblad operator $\hat{\Pi}^+$ of Eq.~\eqref{Lindblad_op_3rd_term} raises all the TLSs to the upper levels
if all of them are in the lower levels.
This expresses the situation that the emitters are simultaneously excited by an incident electron.
We note that this term induces no excitation when some of the TLSs are already in the upper levels.
However, such a no-excitation event does not occur
if the excitation rate $\gamma$ is much smaller than the radiative damping rate $1 / \tau_{\mathrm{rad}}$ since in such cases all the TLSs are in the lower levels for most of the time.
Therefore, this model is valid for $\tau_{\mathrm{rad}} \gamma \ll 1$.
If $\tau_{\mathrm{rad}}$ is around 10~ns,
this condition is fulfilled for the beam current $I$ less than 10~pA.

Here, we make three remarks on this model.
First, the number $N$ of emitters excited by a single incident electron is independent of the current $I$
while the excitation rate $\gamma$ is proportional to $I$ (as explained above).
Instead, $N$ depends on the energy of the electron (acceleration voltage) and on the sample parameters (film thickness, density of emitters, and so on).

Second, as explained for Eq.~\eqref{QME_3rd_term}, we assume that the incident electrons always excite the same $N$ emitters.
In actual experiments of CL,
the number of emitters excited by each electron varies around the average.
We give an extension of the model to incorporate this effect in Sec.~\ref{sec:generalization} and the Supplemental Material \cite{supplement} (Refs. \cite{Meuret_etal_2015, Breuer_Petruccione_2002, Nakajima_1958, Zwanzig_1960, Shibata_Arimitsu_1980, Gardiner_2009} are included therein).
We note that
this effect does not essentially alter the results of the present model
if its validity condition $\tau_{\mathrm{rad}} \gamma \ll 1$ is satisfied.
Moreover, the extended model can be well approximated by the present model for $\tau_{\mathrm{rad}} \gamma \ll 1$,
where $N$ is regarded as the average number of emitters excited by an electron.

Finally, small excitation volume
(i.e., high spatial resolution of the electron beam),
one of the characteristics of CL, is taken into consideration in the model:
the same emitters are excited every time.
Combining this with the second remark,
the situation we assume in this model is as follows:
we consider the emitters located within the excited region ($N_{\mathrm{tot}}$ emitters in total),
each incident electron in the beam excites a part of them (say, $N_{\mathrm{ex}}$ emitters, where $N_{\mathrm{ex}}$ varies for each electron),
and its average number $\overline{N_{\mathrm{ex}}}$ is $N$.

\subsection{Semi-classical master equation for the number of excited emitters}

As seen in the next section, the statistics of the number of excited emitters
$\hat{n} = \sum_{j=1}^N \hat{\sigma}_j^+ \hat{\sigma}_j^-$
is useful to investigate the second-order correlation function $g^{(2)}$.
The statistics is governed by the probability $P(n,t)$ that the number of excited emitters is $n$ at time $t$.
As derived in \ref{sec:derivation_of_ME},
QME \eqref{QME} is exactly reduced to the following semi-classical master equation for $P(n,t)$:
\begin{align}
  \frac{d}{dt} P(0,t) & = \frac{1}{\tau_{\mathrm{rad}}} P(1,t) - \gamma P(0,t),
  \label{ME1}
  \\
  \frac{d}{dt} P(n,t) & = \frac{n+1}{\tau_{\mathrm{rad}}} P(n+1,t) - \frac{n}{\tau_{\mathrm{rad}}} P(n,t),
  \label{ME2}
  \\
  \frac{d}{dt} P(N,t) & = \gamma P(0,t) - \frac{N}{\tau_{\mathrm{rad}}} P(N,t),
  \label{ME3}
\end{align}
where Eq.~\eqref{ME2} is for $1 \le n \le N - 1$.

\section{Second-order correlation function}

\subsection{Steady state}

In CL with a continuous beam, the system is in the steady state $\hat{\rho}_{\mathrm{ss}}$,
which is determined by $\mathcal{L} \hat{\rho}_{\mathrm{ss}} = 0$.
In the following, we write the steady-state average $\Trace [ \hat{\rho}_{\mathrm{ss}} \cdots ]$ as $\expval*{\cdots}_{\mathrm{ss}}$.

In the steady state, $P(n,t)$ also becomes the stationary distribution $P_{\mathrm{ss}}(n)$.
We obtain the equations that determine $P_{\mathrm{ss}}(n)$
by setting the left hand sides of Eqs.~\eqref{ME1}--\eqref{ME3} to zero.
We can exactly solve these equations with the normalization condition $\sum_{n=0}^N P_{\mathrm{ss}}(n) = 1$ to obtain
\begin{align}
  P_{\mathrm{ss}}(0) & = \frac{1}{1 + z_N \tau_{\mathrm{rad}} \gamma},
  \label{Pss_0}
  \\
  P_{\mathrm{ss}}(n) & = \frac{\tau_{\mathrm{rad}} \gamma}{n(1 + z_N \tau_{\mathrm{rad}} \gamma)}
  \quad (1 \le n \le N),
  \label{Pss_n}
\end{align}
where $z_N = \sum_{m=1}^N (1/m)$.
From $P_{\mathrm{ss}}(n)$, we can calculate the steady-state moments of $\hat{n}$.
The first two are:
\begin{align}
  \expval{\hat{n}}_{\mathrm{ss}}
   & = \sum_{n=0}^N n P_{\mathrm{ss}}(n)
  = \frac{N \tau_{\mathrm{rad}} \gamma}{1 + z_N \tau_{\mathrm{rad}} \gamma},
  \label{n_avg}
  \\
  \expval{\hat{n}^2}_{\mathrm{ss}}
   & = \sum_{n=0}^N n^2 P_{\mathrm{ss}}(n)
  = \frac{N (N+1) \tau_{\mathrm{rad}} \gamma}{2 (1 + z_N \tau_{\mathrm{rad}} \gamma)}.
  \label{n2_avg}
\end{align}
We use these moments in calculating $g^{(2)}$.

We now investigate the normalized second-order correlation function $g^{(2)}(\tau)$ in the steady state.
This is defined by
$g^{(2)}(\tau) = \expval*{\mathcal{T} :\hat{I}_{\mathrm{rad}} \hat{I}_{\mathrm{rad}}(\tau):}_{\mathrm{ss}} / \expval*{\hat{I}_{\mathrm{rad}}}_{\mathrm{ss}}^2$,
where $\hat{I}_{\mathrm{rad}}$ is the intensity operator of the radiation field,
$\mathcal{T}$ is the time ordering, and $: :$ is the normal ordering \cite{Mandel_Wolf_1995}.
Thanks to the normalization factor $\expval*{\hat{I}_{\mathrm{rad}}}_{\mathrm{ss}}^2$,
collection and detection efficiencies of light and the linear loss of an optical system do not affect the value of $g^{(2)}$,
and thus defined $g^{(2)}(\tau)$ describes the second-order correlation that is obtained in the HBT experiments.
To proceed further, we again use the assumption that the dipole moments of the emitters are randomly oriented.
In this case, the intensity operator reads $\hat{I}_{\mathrm{rad}} \propto \sum_{j=1}^N \hat{\sigma}_j^+ \hat{\sigma}_j^- = \hat{n}$.
Therefore $g^{(2)}(\tau)$ is given by
\begin{align}
  g^{(2)}(\tau)
   & = \frac{\sum_{j_1,j_2} \expval{\hat{\sigma}_{j_1}^+ \hat{\sigma}_{j_2}^+(\tau) \hat{\sigma}_{j_2}^-(\tau) \hat{\sigma}_{j_1}^-}_{\mathrm{ss}}}
  {\expval{\hat{n}}_{\mathrm{ss}}^2}.
  \label{g2_def}
\end{align}
Since the steady-state correlation function is symmetric at $\tau = 0$,
we analyze $g^{(2)}(\tau)$ for $\tau \ge 0$ in the following.

\subsection{Zero-time delay correlation: Superbunching}

First, we derive the exact formula for the zero-time delay correlation function $g^{(2)}(0)$.
At $\tau = 0$, we can rewrite the numerator of Eq.~\eqref{g2_def} as follows:
\begin{align}
  \sum_{j_1,j_2} \expval{\hat{\sigma}_{j_1}^+ \hat{\sigma}_{j_2}^+ \hat{\sigma}_{j_2}^- \hat{\sigma}_{j_1}^-}_{\mathrm{ss}}
   & = \sum_{j_1 \neq j_2}
  \expval{\hat{\sigma}_{j_1}^+ \hat{\sigma}_{j_2}^+ \hat{\sigma}_{j_2}^- \hat{\sigma}_{j_1}^-}_{\mathrm{ss}}
  \notag
  \\
   & = \expval{\hat{n}^2}_{\mathrm{ss}} - \expval{\hat{n}}_{\mathrm{ss}},
  \label{n2-n}
\end{align}
noting that $\hat{\sigma}_{j_1}^\pm$ and $\hat{\sigma}_{j_2}^\pm$ are commutative only if $j_1 \neq j_2$.
Applying Eqs.~\eqref{n_avg} and \eqref{n2_avg}, we obtain the exact formula for $g^{(2)}(0)$:
\begin{align}
  g^{(2)}(0)
   & = \frac{1}{2} \qty(z_N + \frac{1}{\tau_{\mathrm{rad}} \gamma} ) \qty( 1 - \frac{1}{N} ).
  \label{g2zero_analytic}
\end{align}

\begin{figure}[t]
  \includegraphics[width=\linewidth]{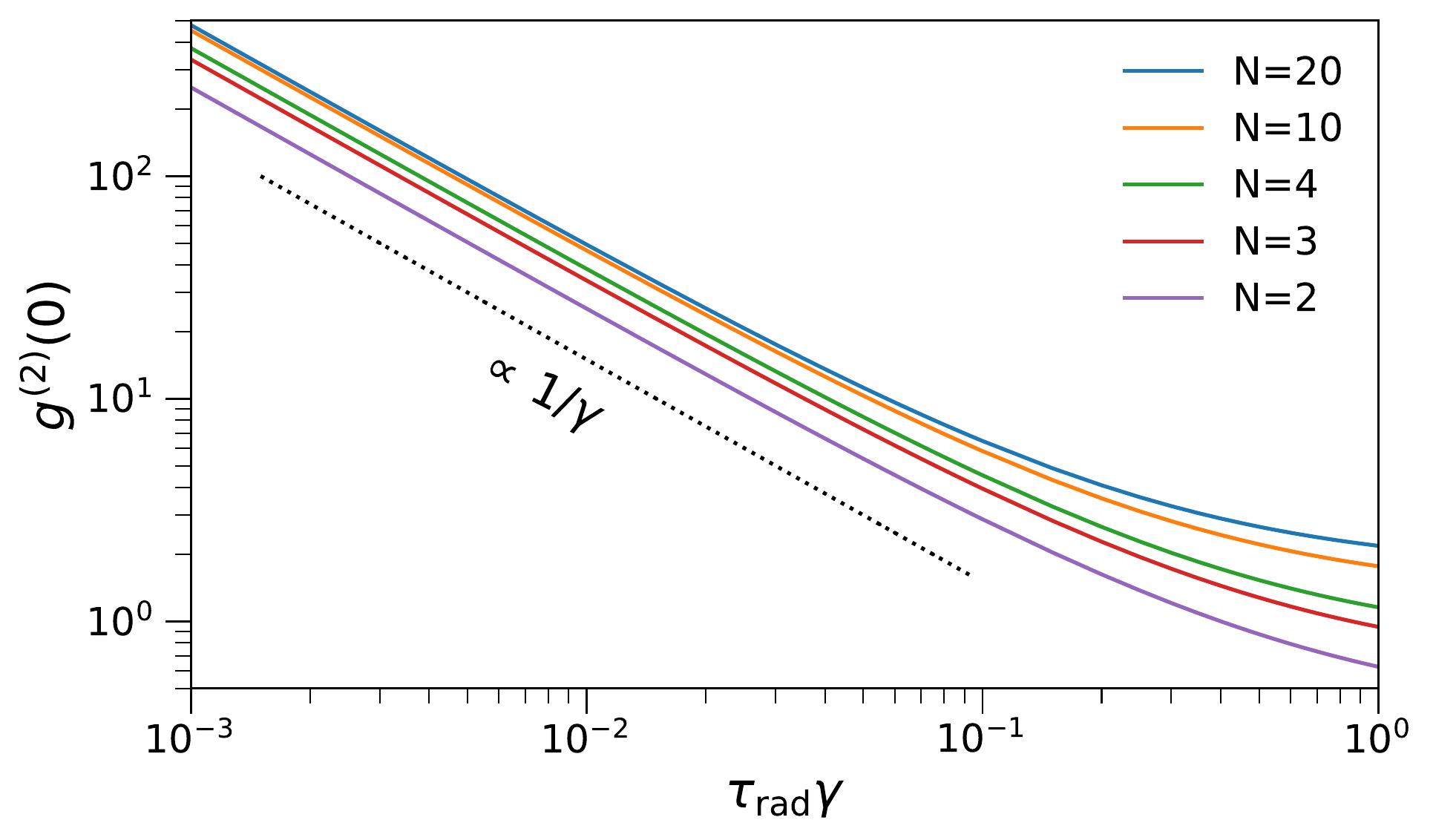}
  \caption{
    Zero-time delay correlation $g^{(2)}(0)$ [Eq.~\eqref{g2zero_analytic}] as a function of the excitation rate $\gamma$ (normalized by the radiative lifetime $\tau_{\mathrm{rad}}$).
    The curves from bottom to top correspond to $N =$ 2, 3, 4, 10, and 20, respectively.
    The dotted line is a visual guide proportional to $1 / \gamma$.
  }
  \label{fig:g2zero}
\end{figure}

From this formula, we can easily show that the superbunching, $g^{(2)}(0) \gg 2$,
is observed for $\tau_{\mathrm{rad}} \gamma \ll 1$.
Figure~\ref{fig:g2zero}, which shows the $\gamma$ dependence of $g^{(2)}(0)$ for $N \ge 2$,
illustrates this feature clearly.
On the other hand, when $N=1$, we have $g^{(2)}(0) = 0$ indicating the antibunching.
This result implies that excitation of multiple emitters ($N \ge 2$) by a single incoming electron are necessary for the superbunching.

We note that, in formula~\eqref{g2zero_analytic},
$g^{(2)}(0)$ is proportional to $1 / \gamma$ for $\tau_{\mathrm{rad}} \gamma \ll 1$.
In Fig.~\ref{fig:g2zero}, this seems valid for $\tau_{\mathrm{rad}} \gamma \lesssim 0.1$.
Since the excitation rate $\gamma$ is proportional to the electron current $I$ and the excitation efficiency $p_{\mathrm{ex}}$ as explained above Eq.~\eqref{QME_3rd_term},
this means that $g^{(2)}(0)$ is proportional to $1 / I$ and $1 / p_{\mathrm{ex}}$ for $\tau_{\mathrm{rad}} \gamma \ll 1$.
Therefore, formula~\eqref{g2zero_analytic} reproduces the properties of $g^{(2)}(0)$ discussed in Refs.~\cite{Meuret_etal_2017,Garcia_etal_2021,Yanagimoto_etal_2021}.

We also make a remark on the limitation of this formula.
In experiments, $g^{(2)}(0)$ approaches 1 for large electron current
\cite{Meuret_etal_2015,Meuret_etal_2017,Feldman_etal_2018,Garcia_etal_2021}.
In comparison, the theoretical formula~\eqref{g2zero_analytic} of $g^{(2)}(0)$ approaches $(1/2) z_N (1 - 1/N) \neq 1$ for large $\gamma$ (thus for large electron current).
This discrepancy is attributed to the limited validity range of  $\tau_{\mathrm{rad}} \gamma$ in the present model.
As explained below Eq.~\eqref{Lindblad_op_3rd_term},
the model is applicable for sufficiently small $\tau_{\mathrm{rad}} \gamma$
because the excitation term in Eq.~\eqref{QME_3rd_term} does not work for large $\tau_{\mathrm{rad}} \gamma$.
However, note that,
in Sec.~\ref{sec:generalization} and the Supplemental Material \cite{supplement} (Refs. \cite{Meuret_etal_2015, Breuer_Petruccione_2002, Nakajima_1958, Zwanzig_1960, Shibata_Arimitsu_1980, Gardiner_2009} are included therein),
we generalize the present model to apply it even to large $\tau_{\mathrm{rad}} \gamma$
and show that this discrepancy is resolved in the generalized model.

\subsection{Finite-time delay correlation}

Next, we derive an approximate form of $g^{(2)}(\tau)$ under the assumption of $N \tau_{\mathrm{rad}} \gamma \ll 1$.
To this end, we apply the quantum regression theorem (QRT) \cite{Breuer_Petruccione_2002, Carmichael_1999}
to the correlation function
$\sum_{j_1 = 1}^N \sum_{j_2 = 1}^N \expval{\hat{\sigma}_{j_1}^+ \hat{\sigma}_{j_2}^+(\tau) \hat{\sigma}_{j_2}^-(\tau) \hat{\sigma}_{j_1}^-}_{\mathrm{ss}}
  = \sum_{j = 1}^N \expval{\hat{\sigma}_j^+ \hat{n}(\tau) \hat{\sigma}_j^-}_{\mathrm{ss}}$
in the numerator of Eq.~\eqref{g2_def}.
We can derive that this correlation function shows a multiple exponential decay
and approaches $\expval{\hat{n}}_{\mathrm{ss}}^2$ [thus $g^{(2)}(\tau) \to 1$] for $\tau \to \infty$.
Furthermore, if we assume $N \tau_{\mathrm{rad}} \gamma \ll 1$,
we can show that the lowest decay rate is approximately equal to $\lambda_1 \simeq (1 / \tau_{\mathrm{rad}}) (1 + N \tau_{\mathrm{rad}} \gamma)$
and the second lowest is $\lambda_2 \simeq (2 / \tau_{\mathrm{rad}}) [1 - N(N-1) \tau_{\mathrm{rad}} \gamma / 4]$ (see \ref{sec:derivation_of_decay_rates} for derivation).
Therefore, the decaying behavior of $\sum_{j = 1}^N \expval{\hat{\sigma}_j^+ \hat{n}(\tau) \hat{\sigma}_j^-}_{\mathrm{ss}}$ is dominated by $e^{-\lambda_1 \tau}$.

Combining this decaying behavior with the asymptotic value $\lim_{\tau \to \infty} g^{(2)}(\tau) = 1$,
we arrive at an approximate expression for $g^{(2)}(\tau)$:
\begin{align}
   & g^{(2)}(\tau) \simeq \bigl[ g^{(2)}(0) - 1 \bigr] e^{-\tau / \tau_{\mathrm{rad}}^{\mathrm{eff}}} + 1
  \notag
  \\[4pt]
   & = C \qty( 1 - \frac{1}{N} ) e^{-\tau / \tau_{\mathrm{rad}}^{\mathrm{eff}}}
  + \qty( 1 - \frac{1}{N} e^{-\tau / \tau_{\mathrm{rad}}^{\mathrm{eff}}} ),
  \label{g2tau_approximate}
\end{align}
In the second line, we used formula~\eqref{g2zero_analytic} for $g^{(2)}(0)$.
Here, the effective life time $\tau_{\mathrm{rad}}^{\mathrm{eff}}$ is given by
\begin{align}
  \frac{1}{\tau_{\mathrm{rad}}^{\mathrm{eff}}}
  = \frac{1}{\tau_{\mathrm{rad}}} + N \gamma,
\end{align}
and the prefactor $C$ is
\begin{align}
  C = \frac{1}{2} \qty(z_N + \frac{1}{\tau_{\mathrm{rad}} \gamma} ) - 1.
  \label{prefactor_g2tau_approximate}
\end{align}

We note that our approximate expression~\eqref{g2tau_approximate} has a form similar to that in Ref.~\cite{Meuret_etal_2015}, which is valid for the small $N$ region.
In their expression,
the decay time is the bare lifetime $\tau_{\mathrm{rad}}$ instead of $\tau_{\mathrm{rad}}^{\mathrm{eff}}$
and the prefactor $C'$ corresponding to $C$ in ours reads $C' = I_0 / (I \times P_{\mathrm{el}}^1)$.
In the notation of this paper,
$I_0 = e / \tau_{\mathrm{rad}}$ and the probability of creating electron--hole pairs by an incoming electron $P_{\mathrm{el}}^1$ should be proportional to $p_{\mathrm{ex}}$.
Since $\gamma = (I / e) p_{\mathrm{ex}}$ as explained above Eq.~\eqref{QME_3rd_term},
we have $C' \propto 1 / (\tau_{\mathrm{rad}} \gamma)$.
In the small $N$ region ($N \tau_{\mathrm{rad}} \gamma \ll 1$),
the effective lifetime becomes $\tau_{\mathrm{rad}}^{\mathrm{eff}} \simeq \tau_{\mathrm{rad}}$
and our prefactor $C$ of Eq.~\eqref{prefactor_g2tau_approximate} yields
$C \simeq 1 / (2 \tau_{\mathrm{rad}} \gamma) \propto 1 / (\tau_{\mathrm{rad}} \gamma)$.
Therefore, our result from the master equation perspective
validates the formula in Ref.~\cite{Meuret_etal_2015}.

\subsection{Numerical demonstration}

\begin{figure}[t]
  \includegraphics[width=\linewidth]{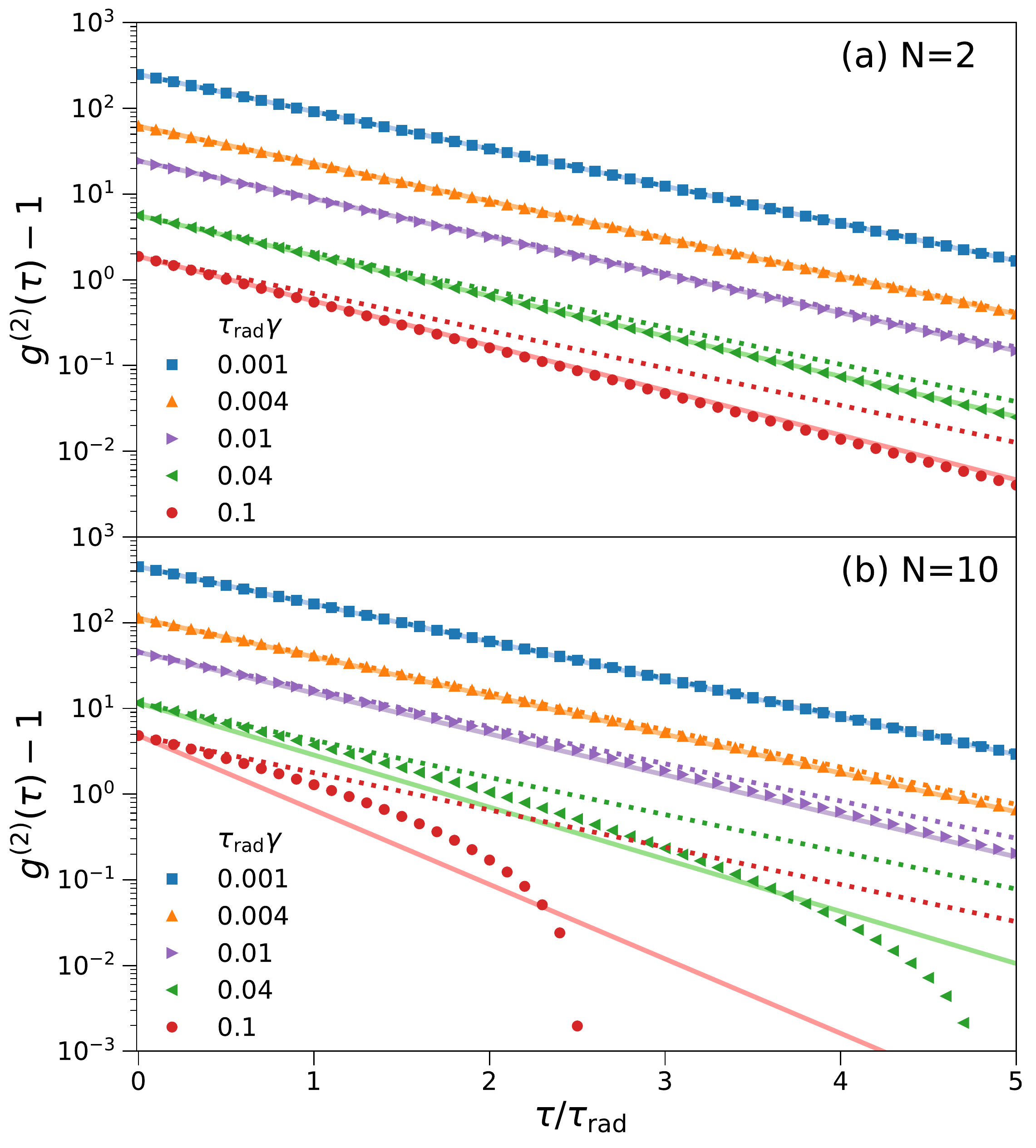}
  \caption{
    Normalized second-order correlation function $g^{(2)}(\tau)$ for (a) $N=2$ and (b) $N=10$.
    Semilogarithmic plots of $g^{(2)}(\tau) - 1$ are shown
    for excitation rates ranging from $\tau_{\mathrm{rad}} \gamma =$ 0.001 to 0.1.
    The symbols, solid lines, and dotted lines represent the numerical results, approximate formula \eqref{g2tau_approximate}, and decay curves proportional to $e^{-\tau / \tau_{\mathrm{rad}}}$ [Eq.~\eqref{g2tau_approximate_2}], respectively.
    For $\tau_{\mathrm{rad}} \gamma = 0.1$ and $N=10$ [filled circles in (b)],
    data points for $\tau / \tau_{\mathrm{rad}} \gtrsim 2.5$ are not shown
    because $g^{(2)}(\tau) - 1$ is extremely small [$g^{(2)}(\tau) \approx 1$] in this region.
  }
  \label{fig:g2tau}
\end{figure}

To demonstrate the applicability of the approximate expression~\eqref{g2tau_approximate},
we compare it with results of numerical simulation.
In the simulation, we numerically solve the eigenvalue problem of $\mathcal{L}$
to obtain the steady state $\hat{\rho}_{\mathrm{ss}}$ as the right eigenvector corresponding to its zero eigenvalue.
Then, we use the QRT to compute $g^{(2)}(\tau)$.
We plot the results for $N=2$ and $N=10$ in Fig.~\ref{fig:g2tau}.
We also plot Eq.~\eqref{g2tau_approximate} (solid lines)
and its variant where $\tau_{\mathrm{rad}}^{\mathrm{eff}}$ is replaced with $\tau_{\mathrm{rad}}$ (dotted lines):
\begin{align}
  g^{(2)}(\tau) \simeq
  C \qty( 1 - \frac{1}{N} ) e^{-\tau / \tau_{\mathrm{rad}}}
  + \qty( 1 - \frac{1}{N} e^{-\tau / \tau_{\mathrm{rad}}} ).
  \label{g2tau_approximate_2}
\end{align}

In the case of $N=2$, Fig.~\ref{fig:g2tau}(a) shows that
the approximate expression~\eqref{g2tau_approximate} describes the numerical data well.
In comparison, although Eq.~\eqref{g2tau_approximate_2} deviates from the numerical results
for $\tau_{\mathrm{rad}} \gamma \ge 0.04$ ($N \tau_{\mathrm{rad}} \gamma \ge 0.08$),
it also describes the numerical data well for smaller $\tau_{\mathrm{rad}} \gamma$
because $\tau_{\mathrm{rad}}^{\mathrm{eff}} \simeq \tau_{\mathrm{rad}}$ in this regime.

In the case of $N=10$, Fig.~\ref{fig:g2tau}(b) shows that
Eq.~\eqref{g2tau_approximate} well approximates the numerical data
for $\tau_{\mathrm{rad}} \gamma \le 0.004$ ($N\tau_{\mathrm{rad}} \gamma \le 0.04$).
As $\tau_{\mathrm{rad}} \gamma$ becomes larger,
we observe clearer deviations between the numerical results and Eq.~\eqref{g2tau_approximate} as well as Eq.~\eqref{g2tau_approximate_2}.

From the results of $N=2$ and $N=10$,
we conclude that the approximation by Eq.~\eqref{g2tau_approximate} is valid for small $N \tau_{\mathrm{rad}} \gamma$.
We also note that the cruder approximation by Eq.~\eqref{g2tau_approximate_2} is valid if $N \tau_{\mathrm{rad}} \gamma$ is sufficiently small, and the decay rate of $g^{(2)}(\tau)$ can be used for estimation of the lifetime $\tau_{\mathrm{rad}}$ of the emitters.
In experiments, one should carefully choose electron-beam current $I$ in order to obtain the lifetime $\tau_{\mathrm{rad}}$ from the HBT measurement.

\section{Generalization of model}
\label{sec:generalization}

In the model in Sec.~\ref{sec:Model}, the same $N$ emitters are excited by each incident electron.
In experiments, however, the emitters excited are different for each electron.
Here, we generalize the model to incorporate this effect.

Let $N_\mathrm{tot}$ be the number of emitters that the electron beam can excite,
so that these emitters are located within the excited region.
We assume that $N_\mathrm{tot}$ is a fixed number.
A part of these emitters are excited by an incoming electron.
Similarly to Eq.~\eqref{QME_3rd_term},
if $j_1$th, $j_2$th, ... and $j_{N_{\mathrm{ex}}}$th emitters are simultaneously excited by an electron,
the excitation term in the QME should read
\begin{align*}
  \Gamma_{N_{\mathrm{ex}}} \Bigl(
   & \hat{\Pi}_{j_1,j_2,...,j_{N_{\mathrm{ex}}}}^+ ~ \hat{\rho} ~ \hat{\Pi}_{j_1,j_2,...,j_{N_{\mathrm{ex}}}}^-
  \notag
  \\
   & - \frac{1}{2} \qty{ \hat{\Pi}_{j_1,j_2,...,j_{N_{\mathrm{ex}}}}^- \hat{\Pi}_{j_1,j_2,...,j_{N_{\mathrm{ex}}}}^+ ,\mspace{2mu}\hat{\rho}} \Bigr),
\end{align*}
where $\hat{\Pi}_{j_1,j_2,...,j_{N_{\mathrm{ex}}}}^\pm = \bigotimes_{i=1}^{N_{\mathrm{ex}}} \hat{\sigma}_{j_i}^{\pm}$.
$\Gamma_{N_{\mathrm{ex}}}$ is the rate of this excitation;
for simplicity, we assume $N_{\mathrm{ex}}$-emitter excitations have the same rate
(i.e., $\Gamma_{N_{\mathrm{ex}}}$ depends only on the number of emitters $N_{\mathrm{ex}}$ but not on the indices $j_1,j_2,...,j_{N_{\mathrm{ex}}}$).
If the emitters are independently excited by an incident electron,
the number $N_{\mathrm{ex}}$ of emitters excited by the electron follows the binomial distribution,
$\eta^{N_{\mathrm{ex}}} (1 - \eta)^{N_{\mathrm{tot}}-N_{\mathrm{ex}}}$,
where $\eta$ is the probability that a single emitter is excited by an electron ($ 0 < \eta < 1$).
Therefore, it is reasonable to assume
\begin{align*}
  \Gamma_{N_{\mathrm{ex}}} = \gamma_{2} \eta^{N_{\mathrm{ex}}} (1 - \eta)^{N_{\mathrm{tot}}-N_{\mathrm{ex}}},
\end{align*}
where $\gamma_{2}$ is a positive constant that is proportional to the rate of the incoming electron, $I / e$.

Summing up all the possible excitations,
we obtain a generalized excitation term in the QME
\begin{widetext}
  \begin{align}
    \mathcal{D}_{\mathrm{ex}}^{(2)} \mspace{2mu} \hat{\rho}
     & = \gamma_{2} \sum_{N_{\mathrm{ex}}=1}^{N_\mathrm{tot}}
    \eta^{N_{\mathrm{ex}}} (1 - \eta)^{N_{\mathrm{tot}}-N_{\mathrm{ex}}}
    \mspace{-6mu} \sum_{j_1 < j_2 < ... < j_{N_{\mathrm{ex}}}} \mspace{-3mu}
    \qty( \hat{\Pi}_{j_1,j_2,...,j_{N_{\mathrm{ex}}}}^+ ~ \hat{\rho} ~ \hat{\Pi}_{j_1,j_2,...,j_{N_{\mathrm{ex}}}}^-
    - \frac{1}{2} \qty{ \hat{\Pi}_{j_1,j_2,...,j_{N_{\mathrm{ex}}}}^- \hat{\Pi}_{j_1,j_2,...,j_{N_{\mathrm{ex}}}}^+ ,\mspace{2mu}\hat{\rho}} ),
    \label{model2_dissipator}
  \end{align}
\end{widetext}
where each index in the second sum on the right-hand side runs from $1$ to $N_\mathrm{tot}$
satisfying the constraint of $j_1 < j_2 < \cdots < j_{N_{\mathrm{ex}}}$.
We thus obtain a generalized model
by replacing $\mathcal{D}_{\mathrm{ex}} \hat{\rho}$ in Eq.~\eqref{Liouvillian} with $\mathcal{D}_{\mathrm{ex}}^{(2)} \hat{\rho}$
and $N$ with $N_\mathrm{tot}$ in Eqs.~\eqref{Hamiltonian} and \eqref{QME_2nd_term}.

Hereafter, we refer to the model in  Sec.~\ref{sec:Model} as Model 1
and the generalized model in this section as Model 2.

\subsection{Relation between Models}

We can interpret Model 1 as a simplified description of Model 2,
where $N$ in Model 1 corresponds to an average number $\overline{N_{\mathrm{ex}}}$ of excitations by an incoming electron in Model 2.
Since $\overline{N_{\mathrm{ex}}}
  = \sum_{N_{\mathrm{ex}}=0}^{N_{\mathrm{tot}}} \eta^{N_{\mathrm{ex}}} (1 - \eta)^{N_{\mathrm{tot}}-N_{\mathrm{ex}}} \binom{N_{\mathrm{tot}}}{N_{\mathrm{ex}}} N_{\mathrm{ex}}
  = \eta N_{\mathrm{tot}}$ in the binomial distribution,
$N$ in Model 1 is connected to Model 2 by
\begin{align}
  N = \eta N_{\mathrm{tot}}.
  \label{relation_1_2_N}
\end{align}
Furthermore, for Model 1 to be an effective description of Model 2,
the average number of emitters excited by the electron beam per unit time must be equal:
$\gamma N = \gamma_2 \overline{N_{\mathrm{ex}}}$.
Combining this equation with Eq.~\eqref{relation_1_2_N}, we have
\begin{align}
  \gamma = \gamma_2.
  \label{relation_1_2_gamma}
\end{align}

To investigate the condition that Model 1 well approximates Model 2,
we note the relative fluctuation of the excitation number in the binomial distribution:
\begin{align}
  \frac{\sqrt{\overline{N_{\mathrm{ex}}^2} - \overline{N_{\mathrm{ex}}}^2}}{\overline{N_{\mathrm{ex}}}}
  = \sqrt{\frac{1 - \eta}{\eta N_{\mathrm{tot}}}}.
  \label{relative_fluctuation_Nex}
\end{align}
This implies that the relative fluctuation of the number of emitters excited by an incident electron becomes smaller as $N_{\mathrm{tot}}$ or $\eta$ increases.
Therefore, if the total number $N_{\mathrm{tot}}$ of emitters in the excited region is sufficiently large
or if the single-emitter-excitation probability $\eta$ is near 1,
we can assume that the excitation number is approximately the same single value, $\overline{N_{\mathrm{ex}}}$ ($=N$), for each electron.
This is the condition for Model 1 to approximate Model 2.
In the Supplemental Material \cite{supplement} (Refs. \cite{Meuret_etal_2015, Breuer_Petruccione_2002, Nakajima_1958, Zwanzig_1960, Shibata_Arimitsu_1980, Gardiner_2009} are included therein),
we numerically demonstrate this approximate relation between Models 1 and 2.

We also note the difference between Models for large excitation rate.
Unlike the excitation term \eqref{QME_3rd_term} of Model 1,
Eq.~\eqref{model2_dissipator} of Model 2 can excite emitters even for large $\tau_{\mathrm{rad}} \gamma$.
This suggests that Model 2 is applicable even for $\tau_{\mathrm{rad}} \gamma > 1$.
In the Supplemental Material \cite{supplement}, we numerically demonstrate that this is the case.
There, we find that $g^{(2)}(0)$ in Model 2 approaches $1 - 1 / N_{\mathrm{tot}}$ for large $\tau_{\mathrm{rad}} \gamma$.
Therefore, if $N_{\mathrm{tot}}$ is sufficiently large,
$g^{(2)}(0)$ is nearly equal to 1, which is consistent with the experimental results
\cite{Meuret_etal_2015,Meuret_etal_2017,Feldman_etal_2018,Garcia_etal_2021}.

\section{State of radiation field}

In the previous section,
our analysis on $g^{(2)}$ of the radiation field is based on the steady state $\hat{\rho}_{\mathrm{ss}}$ of the emitters.
Understanding the state of the radiation field itself is also an interesting problem.
In this section, to qualitatively understand the radiation field state in CL,
we give a heuristic argument on this problem
under the assumption of $z_N \tau_{\mathrm{rad}} \gamma \ll 1$
[note that $z_N = \sum_{m=1}^N (1/m) \approx \log N + \delta$ with an irrelevant constant $\delta$ ($0 < \delta < 1$)].

\begin{figure}[tb]
  \includegraphics[width=\linewidth]{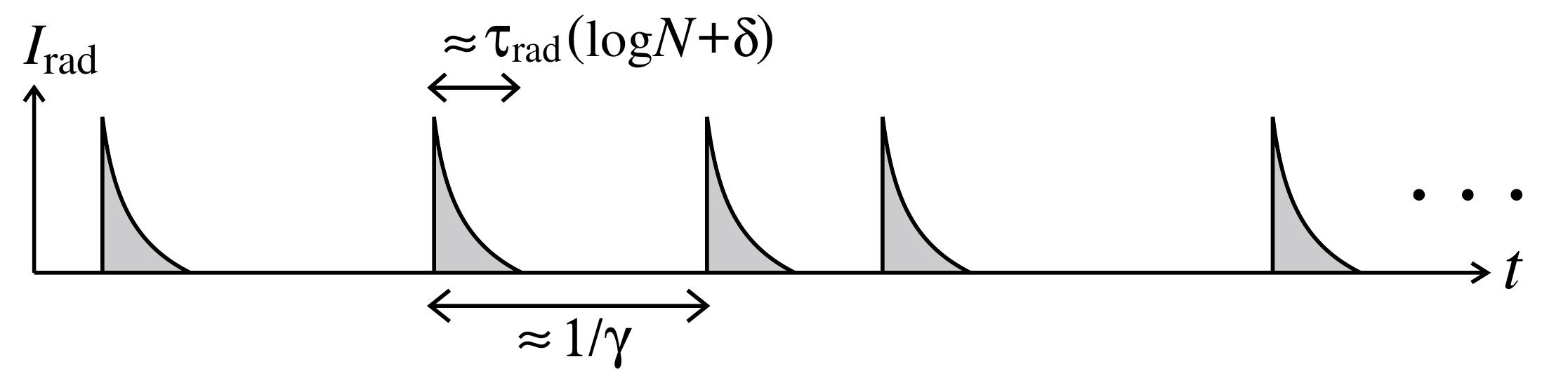}
  \caption{A sequence of random pulses of radiation intensity $I_{\mathrm{rad}}$.
    The intensity profile of each pulse is $I_{\mathrm{rad}}\approx N e^{-\tau / \tau_{\mathrm{rad}}}$ on average, so that the mean duration of the pulse is roughly equal to $\tau_{\mathrm{rad}} (\log N + \delta)$.}
  \label{fig:pulses}
\end{figure}

We first revisit the process of the radiation in CL.
Even though an electron beam irradiates a sample continuously, each electron in the beam exists discretely.
An incoming electron excites multiple (say, $N$) emitters and the emitters decay with radiating photons.
The radiation generated in this process is considered to have a time profile of the intensity $I_{\mathrm{rad}}(t)$
that is composed of a sequence of random pulses, as schematically depicted in Fig.~\ref{fig:pulses}.
In each pulse, the excited emitters radiate photons within the duration of the emission process.
The duration is random due to the spontaneous emission process of the emitters,
and the mean duration is roughly equal to $\tau_{\mathrm{rad}} (\log N + \delta)$
because $I_{\mathrm{rad}} \approx N e^{-\tau / \tau_{\mathrm{rad}}} > \epsilon$
(with some small positive constant $\epsilon$) should be satisfied for $\tau$ within the duration.
The instance at which a single pulse starts is also random reflecting the randomness of the incoming electrons,
and the mean time between successive pulses is equal to  $1 / \gamma$, where $\gamma$ is the excitation rate.

From the above argument, we can classify total time of photodetection into pulse-existing regions and zero-intensity regions,
and we estimate the ratio $q$ of the former regions to the total as
$q \approx \tau_{\mathrm{rad}} \gamma (\log N + \delta)$ $(\ll 1)$.
In the former regions, the radiation field is in a certain photonic state $\hat{\rho}^{\mathrm{rad}}_N$ whose average photon number is around $N$.
In the latter, it is in the vacuum state $\hat{\rho}^{\mathrm{rad}}_{\mathrm{vac}} = \ketbra*{\mathrm{vac}}{\mathrm{vac}}$.
Therefore, we can consider the steady state of the radiation field in CL as the average of $\hat{\rho}^{\mathrm{rad}}_N$ and $\hat{\rho}^{\mathrm{rad}}_{\mathrm{vac}}$:
\begin{align}
  \hat{\rho}^{\mathrm{rad}}_{\mathrm{avg}}
  = q \hat{\rho}^{\mathrm{rad}}_N + (1 - q) \hat{\rho}^{\mathrm{rad}}_{\mathrm{vac}}.
  \label{steady_state_radiation}
\end{align}
In this case, the zero-time delay correlation function of the radiation field $\hat{\rho}^{\mathrm{rad}}_{\mathrm{avg}}$
yields a $(1/ q)$ multiple of that of $\hat{\rho}^{\mathrm{rad}}_N$.
In fact, for single-mode radiation, we have
\begin{align}
  g^{(2)}(0)
  = \frac{\Trace [ \hat{\rho}^{\mathrm{rad}}_{\mathrm{avg}} \hat{a}^\dag \hat{a}^\dag \hat{a} \hat{a} ]}{\Trace [ \hat{\rho}^{\mathrm{rad}}_{\mathrm{avg}} \hat{a}^\dag \hat{a} ]^2}
  = \frac{1}{q} \times \frac{\Trace [ \hat{\rho}^{\mathrm{rad}}_N \hat{a}^\dag \hat{a}^\dag \hat{a} \hat{a} ]}{\Trace [ \hat{\rho}^{\mathrm{rad}}_N \hat{a}^\dag \hat{a} ]^2},
  \label{g2_zero_single_mode}
\end{align}
where $\hat{a}^\dag$ and $\hat{a}$ are the creation and destruction operators of the mode, respectively.
This result with $q \approx \tau_{\mathrm{rad}} \gamma (\log N + \delta)$
reproduces the approximate proportionality of formula~\eqref{g2zero_analytic} to $1 / \tau_{\mathrm{rad}} \gamma$
and gives rise to the superbunching, $g^{(2)}(0) \gg 2$.

To be more concrete, we assume that the emitters' population is directly transferred to the single-mode photonic population as
\begin{align}
  \hat{\rho}^{\mathrm{rad}}_{\mathrm{avg}}
   & = \sum_{n=0}^N P_{\mathrm{ss}}(n) \ketbra*{n}{n},
  \label{rho_N_avg}
\end{align}
where $\ket{n}$ is the $n$-photon state of the single mode (in particular, $\ket{0} = \ket{\mathrm{vac}}$)
and $P_{\mathrm{ss}}(n)$ is the steady-state probability of the number of excited emitters [Eqs.~\eqref{Pss_0} and \eqref{Pss_n}].
This state with Eq.~\eqref{g2_zero_single_mode} exactly reproduces formula~\eqref{g2zero_analytic} for $g^{(2)}(0)$.
Moreover, we can write this state in the form of Eq.~\eqref{steady_state_radiation},
where $q = 1 - P_{\mathrm{ss}}(0) = z_N \tau_{\mathrm{rad}} \gamma / (1 + z_N \tau_{\mathrm{rad}} \gamma)$
and
\begin{align}
  \hat{\rho}^{\mathrm{rad}}_N = \sum_{n=1}^N \frac{P_{\mathrm{ss}}(n)}{1 - P_{\mathrm{ss}}(0)} \ketbra*{n}{n}
  = \sum_{n=1}^N \frac{1}{z_N n} \ketbra*{n}{n}.
  \label{rho_N_rad}
\end{align}
For $z_N \tau_{\mathrm{rad}} \gamma \ll 1$, $q \approx z_N \tau_{\mathrm{rad}} \gamma \approx \tau_{\mathrm{rad}} \gamma (\log N + \delta)$ holds as expected.

To interpret the superbunching effect in different systems,
similar arguments are given in Refs.~\cite{Lettau_etal_2018,Grunwald_2019}
(see also Ref.~\cite{Loudon_1983}):
a classical (incoherent) mixture of high- and low-intensity states with a large weight on the lower
as in Eq.~\eqref{steady_state_radiation} leads to an enhancement of $g^{(2)}(0)$.
We note that a quantum superposition state
$\sqrt{q}\ket{\mathrm{rad}_N} + \sqrt{1-q}\ket{\mathrm{vac}}$
with some pure photonic state $\ket{\mathrm{rad}_N}$ such as $\ket{N}$
also leads to a similar enhancement \cite{Grunwald_2019}
and, only from $g^{(2)}(0)$, it is impossible to distinguish
whether the radiation state is a classical or quantum mixture.
In our model of CL, however, it should be a classical one because the steady state of the emitters is diagonal in the standard basis.

On the basis of this argument, we also discuss
the difference between CL and PL photon statistics:
the superbunching is observed in the CL whereas $g^{(2)}_{\mathrm{PL}}(0) \simeq 1$ for the PL of the same sample \cite{Meuret_etal_2015}.
While an electron instantly excites multiple emitters in CL, a continuous wave laser for PL has a typical excitation duration time, which reflects the coherence time. Thus, the interval of successive excitations in PL is much shorter than the radiation lifetime, such that the radiation in PL is continuous without pulsing (unlike Fig.~\ref{fig:pulses} of CL).
Therefore, the state of the PL radiation field can be considered simply as $\hat{\rho}^{\mathrm{rad}}_N$ and not mixed with the vacuum state.
Although the form of $\hat{\rho}^{\mathrm{rad}}_N$ given by Eq.~\eqref{rho_N_rad} cannot correctly express the state of PL
because the present model is not valid under continuous condition,
the form of $\hat{\rho}^{\mathrm{rad}}_{\mathrm{avg}}$ given by Eq.~\eqref{rho_N_avg} approaches Eq.~\eqref{rho_N_rad} for $\tau_{\mathrm{rad}} \gamma \gg 1$.
This supports the above statement of $g^{(2)}_{\mathrm{PL}}(0) \simeq 1$.
If there is almost no difference between CL and PL in the material response related to the upper state excitation,
once the state of PL is known, we can estimate the state of CL, providing the calculation of optical systems such as an interferometer with CL.
This opens up novel application of CL itself (e.g., superbunched light source) and nanoscale analysis using the CL photon state property.

\section{Discussion and summary}

In this study,
we have constructed a QME model that captures essential aspects of CL: simultaneous excitation and individual decay of emitters.
We have derived the exact formula for the zero-time delay correlation $g^{(2)}(0)$, which successfully describes the superbunching.
We have also derived an approximate form for the finite-time delay correlation $g^{(2)}(\tau)$,
which shows that the radiative life time $\tau_{\mathrm{rad}}$ can be extracted from its $\tau$ dependence for small $N \tau_{\mathrm{rad}} \gamma$.

In the present model, we have assumed that the properties (transition energy $\hbar \omega_{\mathrm{e}}$, radiative lifetime $\tau_{\mathrm{rad}}$, and excitation strength) of all the emitters are the same and that there is no interaction among them.
Also, they have randomly-oriented dipole moments.
It is expected that introducing inhomogeneity of emitters does not drastically alter the main conclusion of the model (for small $\tau_{\mathrm{rad}} \gamma$) on the superbunching in CL.
One interesting direction for future research is introducing interaction.
If there is interaction,
the QME is not simply reduced to the semi-classical master equation
and some genuine quantum effects may emerge.
The present model provides a simple basis for this direction.

We also note that, although we have not presented the spectrum of CL in this work, we can also investigate it in the model introducing the spectral resolution \cite{Eberly_Wodkiewicz_1977,Yamaguchi_etal_2021}.
In calculating it, we can apply the QRT \cite{Breuer_Petruccione_2002,Carmichael_1999} to steady-state correlation functions.

In the model, the simultaneous excitation of multiple emitters by an incoming electron is phenomenologically introduced in Eq.~\eqref{QME_3rd_term}.
In the Supplemental Material \cite{supplement} (Refs. \cite{Meuret_etal_2015, Breuer_Petruccione_2002, Nakajima_1958, Zwanzig_1960, Shibata_Arimitsu_1980, Gardiner_2009} are included therein),
we give a justification of this excitation term [Eq.~\eqref{QME_3rd_term}],
where we derive another QME from a microscopic model with a stochastic interaction term emulating the process that an
electron, randomly incident on the sample, randomly excites the emitters while traveling through the sample.
We numerically demonstrate that our model [Eqs.~\eqref{QME}--\eqref{Lindblad_op_3rd_term}] can approximately describe the features of the microscopically derived QME in the Supplemental Material.
It is an important future work to derive this excitation effect from a first-principle analysis on the complicated elementary processes in CL starting from non-stochastic interactions.

Another implication of this phenomenological incorporation
is that the superbunching is not specific to CL experiments;
we can observe superbunching if there could be simultaneous excitation and individual radiative decay of multiple emitters and if their timescales are largely different.
Indeed, giant photon bunching can be observed in other systems that have cooperative emission \cite{Temnov_Woggon_2009,Kamide_etal_2014,Jahnke_etal_2016}.
In these systems, we can interpret that excitations via dark states play the role of the simultaneous excitation of multiple emitters.
In a similar manner, we can explain another example of giant bunching in a system composed of a quantum dot and a metal nanoparticle \cite{Ridolfo_etal_2010}:
in this system, since the transition rate is suppressed by the Fano destructive interference,
multiple photons can be efficiently excited through the dark state.
This plays the role of (nearly) simultaneous multiple excitation.

We have also discussed a possible state of radiation field in the CL.
Through a heuristic argument, we have proposed the state in Eq.~\eqref{steady_state_radiation}
and confirmed that it is consistent with the QME analysis and well describes superbunching.
This result implies that we may observe superbunching with mechanisms
other than the simultaneous excitation of multiple emitters.
Indeed, in Ref.~\cite{Loudon_1983},
an enhancement of $g^{(2)}(0)$ is discussed in a train of pulses with a regular interval.
Our argument, schematically depicted in Fig.~\ref{fig:pulses}, shows that the same enhancement is observed in random pulses
and that the state of the radiation field in CL should be similar to that of the randomly modulated optical beam.
Also, Ref.~\cite{Lettau_etal_2018} discusses superbunching with states similar to Eq.~\eqref{steady_state_radiation}
and shows that a bimodal microcavity laser with an emitter achieves such a state as its steady state.
Indeed, superbunching is reported in bimodal microlaser systems
\cite{Leymann_etal_2013a,Redlich_etal_2016,Marconi_etal_2018,Leymann_etal_2017,Schmidt_etal_2021}.

In other words,
a potent way to observe superbunching is generating a mixture of photonic states as in Eq.~\eqref{steady_state_radiation},
and there are several systems and methods which can generate this type of state.
The present study by the master equation reveals that the simultaneous excitation of multiple emitters in CL is one of them
and gives quantum insight into CL photon statistics.
Since the time correlation measurement of CL has given new functionalities to nanoscale optical imaging, the obtained results imply a potential of CL
to reveal and even utilize the quantum nature of light-matter interaction on the nanoscale.

\begin{acknowledgments}
  We acknowledge Kenji Kamide, Sotatsu Yanagimoto, and Makoto Yamaguchi for helpful comments.
  One of the authors (K. A.) is grateful to Takeshi Ohshima for his kind support of the research.
  This work was supported by Research Foundation for Opto-Science and Technology
  and JSPS KAKENHI Grants No. JP18K03454, No. JP21K18195, No. JP22H01963, and No. JP22H05032.
\end{acknowledgments}

\setcounter{section}{0}
\renewcommand{\thesection}{Appendix~\Alph{section}}

\section{Master equation for $P(n,t)$}
\label{sec:derivation_of_ME}

In this appendix, we derive the master equation for the number of excited emitters [Eqs.~\eqref{ME1}--\eqref{ME3}].
For this purpose, we introduce some notation.
We denote each of the standard basis states as $\ket{s} =\otimes_{j=1}^N \ket*{s_j}_j$ with $s = \sum_{j=1}^N 2^{j-1} s_j$.
Then, the standard basis is represented as $\{ \ket{s} \mid s=0,1,2,...,2^N - 1 \}$.
We write the diagonal elements of the system state $\hat{\rho}(t)$ as $\rho(s,t) = \bra{s} \hat{\rho}(t) \ket{s}$.
In addition, we define $\ket{s^{j+}} = \hat{\sigma}_j^+ \ket{s}$, which is also one of the standard basis states if $s_j=0$.

From QME~\eqref{QME},
we can show $\bra{s} \{ d\hat{\rho}(t) / dt \} \ket{s} = \bra{s} \mathcal{L} \hat{\rho}(t) \ket{s}$ yields
\begin{align}
   & \frac{\partial}{\partial t} \rho(0, t)
  = \frac{1}{\tau_{\mathrm{rad}}} \sum_{j=1}^N \rho(0^{j+}, t) - \gamma \rho(0, t),
  \label{ME_diagonal_1}
  \\
   & \frac{\partial}{\partial t} \rho(s, t)
  = \frac{1}{\tau_{\mathrm{rad}}} \sum_{j=1}^N \Bigl[ \rho(s^{j+}, t) \delta_{s_j, 0} - \rho(s, t) \delta_{s_j, 1} \Bigr],
  \label{ME_diagonal_2}
  \\
   & \frac{\partial}{\partial t} \rho(2^N-1, t)
  = -\frac{N}{\tau_{\mathrm{rad}}} \rho(2^N-1, t) + \gamma \rho(0, t),
  \label{ME_diagonal_3}
\end{align}
where Eq.~\eqref{ME_diagonal_2} is for $1 \le s \le 2^N-2$.
We note that Eqs.~\eqref{ME_diagonal_1}--\eqref{ME_diagonal_3} form a closed set of equations for the diagonal elements
(Pauli master equation \cite{Breuer_Petruccione_2002,Vorberg_etal_2015}).

To transform Eqs.~\eqref{ME_diagonal_1}--\eqref{ME_diagonal_3} to the master equation for $P(n, t)$,
we note the connection between $P(n, t)$ and $\rho(s,t)$:
\begin{align}
  P(n,t) = \sum_{s=0}^{2^N - 1} \rho(s, t) \delta_{n_s, n},
  \label{P_n_t_and_rho_s_t}
\end{align}
where
\begin{align}
  n_s = \sum_{j=1}^N s_j = \sum_{j=1}^N \delta_{s_j, 1}
\end{align}
is the number of excited emitters in the state $\ket{s}$.
Differentiating Eq.~\eqref{P_n_t_and_rho_s_t} with respect to $t$ and using Eqs.~\eqref{ME_diagonal_1}--\eqref{ME_diagonal_3},
we obtain the master equation for $P(n,t)$ [Eqs.~\eqref{ME1}--\eqref{ME3}].

\section{A perturbative analysis of decay rates}
\label{sec:derivation_of_decay_rates}

In this appendix, we evaluate the decay rates of $\sum_{j=1}^N \expval{\hat{\sigma}_j^+ \hat{n}(\tau) \hat{\sigma}_j^-}_{\mathrm{ss}}$.
To this end, it is sufficient to investigate each term $\expval{\hat{\sigma}_j^+ \hat{n}(\tau) \hat{\sigma}_j^-}_{\mathrm{ss}}$ in the sum.
In the estimation, we use the QRT and a perturbative analysis under the assumption of $\tau_{\mathrm{rad}} \gamma \ll 1$.

We first note that $\hat{n} = \sum_{n=0}^N n \hat{P}_n$,
where $\hat{P}_n$ is the projection operator onto the subspace of states with $n$ excited emitters.
This leads to $\expval{\hat{\sigma}_j^+ \hat{n}(\tau) \hat{\sigma}_j^-}_{\mathrm{ss}} = \sum_n n \expval*{\hat{\sigma}_j^+ \hat{P}_n(\tau) \hat{\sigma}_j^-}_{\mathrm{ss}}$,
so that it is reasonable to investigate the decaying behavior of $\expval*{\hat{\sigma}_j^+ \hat{P}_n(\tau) \hat{\sigma}_j^-}_{\mathrm{ss}}$.

According to the QRT,
$\expval*{\hat{\sigma}_j^+ \hat{P}_n(\tau) \hat{\sigma}_j^-}$
obeys the differential equation whose form is the same as that of
$\expval*{\hat{P}_n(\tau)} = P(n,\tau)$.
The latter is the master equation for $P(n,t)$ [Eqs.~\eqref{ME1}--\eqref{ME3}],
which is expressed as $d\bm{P}(t) / dt = -\Gamma \bm{P}(t)$
with $\bm{P}(t) \equiv (P(0,t), P(1,t), ..., P(N,t))^\mathsf{T}$ ($\mathsf{T}$ stands for transpose)
and an $(N+1) \times (N+1)$ matrix $\Gamma$:
\begin{align}
   & \Gamma =
  \begin{pmatrix}
    \gamma  & -1 / \tau_{\mathrm{rad}} &                          &        &
    \\
            & 1 / \tau_{\mathrm{rad}}  & -2 / \tau_{\mathrm{rad}} &        &
    \\
            &                          & 2 / \tau_{\mathrm{rad}}  & \ddots &
    \\
            &                          &                          & \ddots & -N / \tau_{\mathrm{rad}}
    \\
    -\gamma &                          &                          &        & N / \tau_{\mathrm{rad}}
  \end{pmatrix}.
  \label{ME_matrix}
\end{align}
Therefore, by using the QRT, we obtain the differential equation
$d\bm{Q}_j(\tau) / d\tau = -\Gamma \bm{Q}_j(\tau)$
for $\bm{Q}_j(\tau) \equiv (\expval*{\hat{\sigma}_j^+ \hat{P}_0(\tau) \hat{\sigma}_j^-}_{\mathrm{ss}}, \expval*{\hat{\sigma}_j^+ \hat{P}_1(\tau) \hat{\sigma}_j^-}_{\mathrm{ss}}, ..., \expval*{\hat{\sigma}_j^+ \hat{P}_N(\tau) \hat{\sigma}_j^-}_{\mathrm{ss}})^{\mathsf{T}}$.
Moreover, using the eigenvalues $\lambda_n$ and its corresponding left and right eigenvectors, $\bm{\ell}_n$ and $\bm{r}_n$, of the non-Hermitian matrix $\Gamma$,
we can show
\begin{align}
  \bm{Q}_j(\tau)
  = \sum_{n=0}^N \bigl[ \bm{\ell}_n \cdot \bm{Q}_j(0) \bigr] \bm{r}_n e^{-\lambda_n \tau}.
\end{align}
This implies that $\expval{\hat{\sigma}_j^+ \hat{n}(\tau) \hat{\sigma}_j^-}_{\mathrm{ss}}$ exhibits a multiple exponential decay with the rates $\lambda_n$ ($n=1,2,...,N$).

Note that, as shown in Eqs.~\eqref{Pss_0} and \eqref{Pss_n},
$\Gamma$ has a zero eigenvalue $\lambda_0 = 0$
and the corresponding right eigenvector is
$\bm{r}_0 = \bm{P}_{\mathrm{ss}} \equiv (P_{\mathrm{ss}}(0), P_{\mathrm{ss}}(1), ..., P_{\mathrm{ss}}(N) )^\mathsf{T}$.
And it is straightforward to show that the corresponding left eigenvector is $\bm{\ell}_0 = (1,1,...,1)^\mathsf{T}$.
From this result, we can show that the asymptotic value of $\bm{Q}_j(\tau)$
is $\lim_{\tau\to\infty} \bm{Q}_j(\tau) = \bigl[ \bm{\ell}_0 \cdot \bm{Q}_j(0) \bigr] \bm{r}_0 = \expval*{\hat{\sigma}_j^+ \hat{\sigma}_j^-}_{\mathrm{ss}} \bm{P}_{\mathrm{ss}}$,
where we used $\sum_n \hat{P}_n = 1$.
Therefore, we obtain $\lim_{\tau\to\infty} \sum_j \expval{\hat{\sigma}_j^+ \hat{n}(\tau) \hat{\sigma}_j^-}_{\mathrm{ss}} = \sum_j \sum_n n \expval*{\hat{\sigma}_j^+ \hat{\sigma}_j^-}_{\mathrm{ss}} P_{\mathrm{ss}}(n) = \expval*{\hat{n}}_{\mathrm{ss}}^2$, which leads to $\lim_{\tau\to\infty} g^{(2)}(\tau) = 1$.

Now we perturbatively estimate the eigenvalues by assuming $\tau_{\mathrm{rad}} \gamma \ll 1$.
For this purpose, we decompose $\Gamma$ as $\Gamma = \Gamma^0 + \gamma \Gamma^1$.
The unperturbed part $\Gamma^0$ is the matrix where $\gamma$ in Eq.~\eqref{ME_matrix} is replaced with zero.
The perturbation matrix $\Gamma_1$ has only two non-zero elements: $\Gamma^1_{0,0} = 1$ and $\Gamma^1_{N,0} = -1$.

Since $\Gamma^0$ is an upper triangular matrix,
its eigenvalues (the zeroth-order eigenvalues $\lambda^0_n$) are the diagonal elements of $\Gamma^0$, that is, $\lambda^0_n = n / \tau_{\mathrm{rad}}$ ($0 \le n \le N$).
The corresponding (zeroth-order) left and right eigenvectors, $\bm{\ell}^0_n$ and $\bm{r}^0_n$, are determined by
\begin{align}
  (\Gamma^0)^\mathsf{T} \bm{\ell}^0_n
   & = \lambda^0_n \bm{\ell}^0_n,
  \\
  \Gamma^0 \bm{r}^0_n
   & = \lambda^0_n \bm{r}^0_n,
\end{align}
with the normalization $\bm{\ell}^0_n \cdot \bm{r}^0_n = 1$.
After some algebraic calculation, we obtain
\begin{align}
  \bm{\ell}^0_n & = \Bigl( \underbrace{0, ..., 0}_{n}, 1, \tbinom{n+1}{1}, \tbinom{n+2}{2}, ..., \tbinom{N}{N-n} \Bigr)^\mathsf{T},
  \label{left_eigenv_0}
  \\
  \bm{r}^0_n    & = \Bigl( (-1)^n \tbinom{n}{n}, (-1)^{n-1} \tbinom{n}{n-1}, ..., (-1) \tbinom{n}{1}, 1, \underbrace{0, ..., 0}_{N-n} \Bigr)^\mathsf{T},
  \label{right_eigenv_0}
\end{align}
where $\binom{m}{k} = m! / [ k! (m-k)! ]$ is a binomial coefficient.

The perturbative analysis for a non-Hermitian matrix is almost the same as that for Hermitian cases in quantum mechanics \cite{Sakurai_2011}---the first order correction to the eigenvalue is
\begin{align}
  \lambda_n^1
   & = \gamma \bm{\ell}_n^0 \cdot \Gamma^1 \bm{r}_n^0
  \notag
  \\
   & =
  \begin{cases}
    0                                & (n=0)
    \\
    \gamma (-1)^{n+1} \binom{N}{N-n} & (1 \le n \le N).
  \end{cases}
\end{align}
Therefore, in the first order of $\tau_{\mathrm{rad}} \gamma$,
we obtain an approximate form of the eigenvalues of $\Gamma$ (except for the zero eigenvalue $\lambda_0 = 0$):
\begin{align}
  \lambda_n \simeq \frac{n}{\tau_{\mathrm{rad}}} \qty[ 1  + \frac{(-1)^{n+1} \tau_{\mathrm{rad}} \gamma}{n} \binom{N}{N-n} ]
  \quad (1 \le n \le N).
\end{align}
We thus estimate the decay rates $\{ \lambda_n \}_{n=1}^N$ of $\expval{\hat{\sigma}_j^+ \hat{n}(\tau) \hat{\sigma}_j^-}_{\mathrm{ss}}$.
In particular, when $N \tau_{\mathrm{rad}} \gamma \ll 1$,
the lowest rate is $\lambda_1 \simeq (1 / \tau_{\mathrm{rad}}) (1 + N \tau_{\mathrm{rad}} \gamma)$
and the second lowest is $\lambda_2 \simeq (2 / \tau_{\mathrm{rad}}) [1 - N (N-1) \tau_{\mathrm{rad}} \gamma / 4]$.

\bibliography{g2_CL_theory_abbrev.bib}

\begin{thebibliography}{61}%
\makeatletter
\providecommand \@ifxundefined [1]{%
 \@ifx{#1\undefined}
}%
\providecommand \@ifnum [1]{%
 \ifnum #1\expandafter \@firstoftwo
 \else \expandafter \@secondoftwo
 \fi
}%
\providecommand \@ifx [1]{%
 \ifx #1\expandafter \@firstoftwo
 \else \expandafter \@secondoftwo
 \fi
}%
\providecommand \natexlab [1]{#1}%
\providecommand \enquote  [1]{``#1''}%
\providecommand \bibnamefont  [1]{#1}%
\providecommand \bibfnamefont [1]{#1}%
\providecommand \citenamefont [1]{#1}%
\providecommand \href@noop [0]{\@secondoftwo}%
\providecommand \href [0]{\begingroup \@sanitize@url \@href}%
\providecommand \@href[1]{\@@startlink{#1}\@@href}%
\providecommand \@@href[1]{\endgroup#1\@@endlink}%
\providecommand \@sanitize@url [0]{\catcode `\\12\catcode `\$12\catcode
  `\&12\catcode `\#12\catcode `\^12\catcode `\_12\catcode `\%12\relax}%
\providecommand \@@startlink[1]{}%
\providecommand \@@endlink[0]{}%
\providecommand \url  [0]{\begingroup\@sanitize@url \@url }%
\providecommand \@url [1]{\endgroup\@href {#1}{\urlprefix }}%
\providecommand \urlprefix  [0]{URL }%
\providecommand \Eprint [0]{\href }%
\providecommand \doibase [0]{https://doi.org/}%
\providecommand \selectlanguage [0]{\@gobble}%
\providecommand \bibinfo  [0]{\@secondoftwo}%
\providecommand \bibfield  [0]{\@secondoftwo}%
\providecommand \translation [1]{[#1]}%
\providecommand \BibitemOpen [0]{}%
\providecommand \bibitemStop [0]{}%
\providecommand \bibitemNoStop [0]{.\EOS\space}%
\providecommand \EOS [0]{\spacefactor3000\relax}%
\providecommand \BibitemShut  [1]{\csname bibitem#1\endcsname}%
\let\auto@bib@innerbib\@empty
\bibitem [{\citenamefont {Stevens~Kalceff}\ and\ \citenamefont
  {Phillips}(1995)}]{Kalceff_etal_1995}%
  \BibitemOpen
  \bibfield  {author} {\bibinfo {author} {\bibfnamefont {M.~A.}\ \bibnamefont
  {Stevens~Kalceff}}\ and\ \bibinfo {author} {\bibfnamefont {M.~R.}\
  \bibnamefont {Phillips}},\ }\bibfield  {title} {\bibinfo {title}
  {Cathodoluminescence microcharacterization of the defect structure of
  quartz},\ }\href {https://doi.org/10.1103/PhysRevB.52.3122} {\bibfield
  {journal} {\bibinfo  {journal} {Phys. Rev. B}\ }\textbf {\bibinfo {volume}
  {52}},\ \bibinfo {pages} {3122} (\bibinfo {year} {1995})}\BibitemShut
  {NoStop}%
\bibitem [{\citenamefont {Mitsui}\ \emph {et~al.}(1996)\citenamefont {Mitsui},
  \citenamefont {Yamamoto}, \citenamefont {Tadokoro},\ and\ \citenamefont
  {Ohta}}]{Mitsui_etal_1996}%
  \BibitemOpen
  \bibfield  {author} {\bibinfo {author} {\bibfnamefont {T.}~\bibnamefont
  {Mitsui}}, \bibinfo {author} {\bibfnamefont {N.}~\bibnamefont {Yamamoto}},
  \bibinfo {author} {\bibfnamefont {T.}~\bibnamefont {Tadokoro}},\ and\
  \bibinfo {author} {\bibfnamefont {S.-i.}\ \bibnamefont {Ohta}},\ }\bibfield
  {title} {\bibinfo {title} {Cathodoluminescence image of defects and
  luminescence centers in {ZnS/GaAs(100)}},\ }\href
  {https://doi.org/10.1063/1.363770} {\bibfield  {journal} {\bibinfo  {journal}
  {J. Appl. Phys.}\ }\textbf {\bibinfo {volume} {80}},\ \bibinfo {pages} {6972}
  (\bibinfo {year} {1996})}\BibitemShut {NoStop}%
\bibitem [{\citenamefont {Tararan}\ \emph {et~al.}(2018)\citenamefont
  {Tararan}, \citenamefont {di~Sabatino}, \citenamefont {Gatti}, \citenamefont
  {Taniguchi}, \citenamefont {Watanabe}, \citenamefont {Reining}, \citenamefont
  {Tizei}, \citenamefont {Kociak},\ and\ \citenamefont
  {Zobelli}}]{Tararan_etal_2018}%
  \BibitemOpen
  \bibfield  {author} {\bibinfo {author} {\bibfnamefont {A.}~\bibnamefont
  {Tararan}}, \bibinfo {author} {\bibfnamefont {S.}~\bibnamefont
  {di~Sabatino}}, \bibinfo {author} {\bibfnamefont {M.}~\bibnamefont {Gatti}},
  \bibinfo {author} {\bibfnamefont {T.}~\bibnamefont {Taniguchi}}, \bibinfo
  {author} {\bibfnamefont {K.}~\bibnamefont {Watanabe}}, \bibinfo {author}
  {\bibfnamefont {L.}~\bibnamefont {Reining}}, \bibinfo {author} {\bibfnamefont
  {L.~H.~G.}\ \bibnamefont {Tizei}}, \bibinfo {author} {\bibfnamefont
  {M.}~\bibnamefont {Kociak}},\ and\ \bibinfo {author} {\bibfnamefont
  {A.}~\bibnamefont {Zobelli}},\ }\bibfield  {title} {\bibinfo {title} {Optical
  gap and optically active intragap defects in cubic {BN}},\ }\href
  {https://doi.org/10.1103/PhysRevB.98.094106} {\bibfield  {journal} {\bibinfo
  {journal} {Phys. Rev. B}\ }\textbf {\bibinfo {volume} {98}},\ \bibinfo
  {pages} {094106} (\bibinfo {year} {2018})}\BibitemShut {NoStop}%
\bibitem [{\citenamefont {Bidaud}\ \emph {et~al.}(2021)\citenamefont {Bidaud},
  \citenamefont {Moseley}, \citenamefont {Amarasinghe}, \citenamefont
  {Al-Jassim}, \citenamefont {Metzger},\ and\ \citenamefont
  {Collin}}]{Bidaud_etal_2021}%
  \BibitemOpen
  \bibfield  {author} {\bibinfo {author} {\bibfnamefont {T.}~\bibnamefont
  {Bidaud}}, \bibinfo {author} {\bibfnamefont {J.}~\bibnamefont {Moseley}},
  \bibinfo {author} {\bibfnamefont {M.}~\bibnamefont {Amarasinghe}}, \bibinfo
  {author} {\bibfnamefont {M.}~\bibnamefont {Al-Jassim}}, \bibinfo {author}
  {\bibfnamefont {W.~K.}\ \bibnamefont {Metzger}},\ and\ \bibinfo {author}
  {\bibfnamefont {S.}~\bibnamefont {Collin}},\ }\bibfield  {title} {\bibinfo
  {title} {Imaging {$\mathrm{CdCl}_2$} defect passivation and formation in
  polycrystalline {CdTe} films by cathodoluminescence},\ }\href
  {https://doi.org/10.1103/PhysRevMaterials.5.064601} {\bibfield  {journal}
  {\bibinfo  {journal} {Phys. Rev. Materials}\ }\textbf {\bibinfo {volume}
  {5}},\ \bibinfo {pages} {064601} (\bibinfo {year} {2021})}\BibitemShut
  {NoStop}%
\bibitem [{\citenamefont {Gustafsson}\ and\ \citenamefont
  {Samuelson}(1994)}]{Gustafsson_Samuelson_1994}%
  \BibitemOpen
  \bibfield  {author} {\bibinfo {author} {\bibfnamefont {A.}~\bibnamefont
  {Gustafsson}}\ and\ \bibinfo {author} {\bibfnamefont {L.}~\bibnamefont
  {Samuelson}},\ }\bibfield  {title} {\bibinfo {title} {Cathodoluminescence
  imaging of quantum wells: The influence of exciton transfer on the apparent
  island size},\ }\href {https://doi.org/10.1103/PhysRevB.50.11827} {\bibfield
  {journal} {\bibinfo  {journal} {Phys. Rev. B}\ }\textbf {\bibinfo {volume}
  {50}},\ \bibinfo {pages} {11827} (\bibinfo {year} {1994})}\BibitemShut
  {NoStop}%
\bibitem [{\citenamefont {Akiba}\ \emph {et~al.}(2004)\citenamefont {Akiba},
  \citenamefont {Yamamoto}, \citenamefont {Grillo}, \citenamefont {Genseki},\
  and\ \citenamefont {Watanabe}}]{Akiba_etal_2004}%
  \BibitemOpen
  \bibfield  {author} {\bibinfo {author} {\bibfnamefont {K.}~\bibnamefont
  {Akiba}}, \bibinfo {author} {\bibfnamefont {N.}~\bibnamefont {Yamamoto}},
  \bibinfo {author} {\bibfnamefont {V.}~\bibnamefont {Grillo}}, \bibinfo
  {author} {\bibfnamefont {A.}~\bibnamefont {Genseki}},\ and\ \bibinfo {author}
  {\bibfnamefont {Y.}~\bibnamefont {Watanabe}},\ }\bibfield  {title} {\bibinfo
  {title} {Anomalous temperature and excitation power dependence of
  cathodoluminescence from $\mathrm{InAs}$ quantum dots},\ }\href
  {https://doi.org/10.1103/PhysRevB.70.165322} {\bibfield  {journal} {\bibinfo
  {journal} {Phys. Rev. B}\ }\textbf {\bibinfo {volume} {70}},\ \bibinfo
  {pages} {165322} (\bibinfo {year} {2004})}\BibitemShut {NoStop}%
\bibitem [{\citenamefont {Merano}\ \emph {et~al.}(2005)\citenamefont {Merano},
  \citenamefont {Sonderegger}, \citenamefont {Crottini}, \citenamefont
  {Collin}, \citenamefont {Renucci}, \citenamefont {Pelucchi}, \citenamefont
  {Malko}, \citenamefont {Baier}, \citenamefont {Kapon}, \citenamefont
  {Deveaud},\ and\ \citenamefont {Gani{\`e}re}}]{Merano_etal_2005}%
  \BibitemOpen
  \bibfield  {author} {\bibinfo {author} {\bibfnamefont {M.}~\bibnamefont
  {Merano}}, \bibinfo {author} {\bibfnamefont {S.}~\bibnamefont {Sonderegger}},
  \bibinfo {author} {\bibfnamefont {A.}~\bibnamefont {Crottini}}, \bibinfo
  {author} {\bibfnamefont {S.}~\bibnamefont {Collin}}, \bibinfo {author}
  {\bibfnamefont {P.}~\bibnamefont {Renucci}}, \bibinfo {author} {\bibfnamefont
  {E.}~\bibnamefont {Pelucchi}}, \bibinfo {author} {\bibfnamefont
  {A.}~\bibnamefont {Malko}}, \bibinfo {author} {\bibfnamefont {M.~H.}\
  \bibnamefont {Baier}}, \bibinfo {author} {\bibfnamefont {E.}~\bibnamefont
  {Kapon}}, \bibinfo {author} {\bibfnamefont {B.}~\bibnamefont {Deveaud}},\
  and\ \bibinfo {author} {\bibfnamefont {J.-D.}\ \bibnamefont {Gani{\`e}re}},\
  }\bibfield  {title} {\bibinfo {title} {Probing carrier dynamics in
  nanostructures by picosecond cathodoluminescence},\ }\href
  {https://doi.org/10.1038/nature04298} {\bibfield  {journal} {\bibinfo
  {journal} {Nature}\ }\textbf {\bibinfo {volume} {438}},\ \bibinfo {pages}
  {479} (\bibinfo {year} {2005})}\BibitemShut {NoStop}%
\bibitem [{\citenamefont {Kuttge}\ \emph {et~al.}(2009)\citenamefont {Kuttge},
  \citenamefont {Vesseur}, \citenamefont {Koenderink}, \citenamefont {Lezec},
  \citenamefont {Atwater}, \citenamefont {Garc\'{\i}a~de Abajo},\ and\
  \citenamefont {Polman}}]{Kuttge_etal_2009}%
  \BibitemOpen
  \bibfield  {author} {\bibinfo {author} {\bibfnamefont {M.}~\bibnamefont
  {Kuttge}}, \bibinfo {author} {\bibfnamefont {E.~J.~R.}\ \bibnamefont
  {Vesseur}}, \bibinfo {author} {\bibfnamefont {A.~F.}\ \bibnamefont
  {Koenderink}}, \bibinfo {author} {\bibfnamefont {H.~J.}\ \bibnamefont
  {Lezec}}, \bibinfo {author} {\bibfnamefont {H.~A.}\ \bibnamefont {Atwater}},
  \bibinfo {author} {\bibfnamefont {F.~J.}\ \bibnamefont {Garc\'{\i}a~de
  Abajo}},\ and\ \bibinfo {author} {\bibfnamefont {A.}~\bibnamefont {Polman}},\
  }\bibfield  {title} {\bibinfo {title} {Local density of states, spectrum, and
  far-field interference of surface plasmon polaritons probed by
  cathodoluminescence},\ }\href {https://doi.org/10.1103/PhysRevB.79.113405}
  {\bibfield  {journal} {\bibinfo  {journal} {Phys. Rev. B}\ }\textbf {\bibinfo
  {volume} {79}},\ \bibinfo {pages} {113405} (\bibinfo {year}
  {2009})}\BibitemShut {NoStop}%
\bibitem [{\citenamefont {Yamamoto}\ \emph {et~al.}(2015)\citenamefont
  {Yamamoto}, \citenamefont {Javier Garc\'{\i}a~de Abajo},\ and\ \citenamefont
  {Myroshnychenko}}]{Yamamoto_etal_2015}%
  \BibitemOpen
  \bibfield  {author} {\bibinfo {author} {\bibfnamefont {N.}~\bibnamefont
  {Yamamoto}}, \bibinfo {author} {\bibfnamefont {F.}~\bibnamefont {Javier
  Garc\'{\i}a~de Abajo}},\ and\ \bibinfo {author} {\bibfnamefont
  {V.}~\bibnamefont {Myroshnychenko}},\ }\bibfield  {title} {\bibinfo {title}
  {Interference of surface plasmons and smith-purcell emission probed by
  angle-resolved cathodoluminescence spectroscopy},\ }\href
  {https://doi.org/10.1103/PhysRevB.91.125144} {\bibfield  {journal} {\bibinfo
  {journal} {Phys. Rev. B}\ }\textbf {\bibinfo {volume} {91}},\ \bibinfo
  {pages} {125144} (\bibinfo {year} {2015})}\BibitemShut {NoStop}%
\bibitem [{\citenamefont {Sannomiya}\ \emph {et~al.}(2020)\citenamefont
  {Sannomiya}, \citenamefont {Kone{\v{c}}n{\'a}}, \citenamefont {Matsukata},
  \citenamefont {Thollar}, \citenamefont {Okamoto}, \citenamefont
  {Garc{\'i}a~de Abajo},\ and\ \citenamefont {Yamamoto}}]{Sannomiya_etal_2020}%
  \BibitemOpen
  \bibfield  {author} {\bibinfo {author} {\bibfnamefont {T.}~\bibnamefont
  {Sannomiya}}, \bibinfo {author} {\bibfnamefont {A.}~\bibnamefont
  {Kone{\v{c}}n{\'a}}}, \bibinfo {author} {\bibfnamefont {T.}~\bibnamefont
  {Matsukata}}, \bibinfo {author} {\bibfnamefont {Z.}~\bibnamefont {Thollar}},
  \bibinfo {author} {\bibfnamefont {T.}~\bibnamefont {Okamoto}}, \bibinfo
  {author} {\bibfnamefont {F.~J.}\ \bibnamefont {Garc{\'i}a~de Abajo}},\ and\
  \bibinfo {author} {\bibfnamefont {N.}~\bibnamefont {Yamamoto}},\ }\bibfield
  {title} {\bibinfo {title} {Cathodoluminescence phase extraction of the
  coupling between nanoparticles and surface plasmon polaritons},\ }\href
  {https://doi.org/10.1021/acs.nanolett.9b04335} {\bibfield  {journal}
  {\bibinfo  {journal} {Nano Lett.}\ }\textbf {\bibinfo {volume} {20}},\
  \bibinfo {pages} {592} (\bibinfo {year} {2020})}\BibitemShut {NoStop}%
\bibitem [{\citenamefont {Fisher}\ \emph {et~al.}(2008)\citenamefont {Fisher},
  \citenamefont {Wessels}, \citenamefont {Dietz},\ and\ \citenamefont
  {Prendergast}}]{Fisher_etal_2008}%
  \BibitemOpen
  \bibfield  {author} {\bibinfo {author} {\bibfnamefont {P.~J.}\ \bibnamefont
  {Fisher}}, \bibinfo {author} {\bibfnamefont {W.~S.}\ \bibnamefont {Wessels}},
  \bibinfo {author} {\bibfnamefont {A.~B.}\ \bibnamefont {Dietz}},\ and\
  \bibinfo {author} {\bibfnamefont {F.~G.}\ \bibnamefont {Prendergast}},\
  }\bibfield  {title} {\bibinfo {title} {Enhanced biological
  cathodoluminescence},\ }\href {https://doi.org/10.1016/j.optcom.2007.04.069}
  {\bibfield  {journal} {\bibinfo  {journal} {Opt. Commun.}\ }\textbf {\bibinfo
  {volume} {281}},\ \bibinfo {pages} {1901} (\bibinfo {year} {2008})},\
  \bibinfo {note} {optics in Life Sciences}\BibitemShut {NoStop}%
\bibitem [{\citenamefont {Nagayama}\ \emph {et~al.}(2016)\citenamefont
  {Nagayama}, \citenamefont {Onuma}, \citenamefont {Ueno}, \citenamefont
  {Tamehiro},\ and\ \citenamefont {Minoda}}]{Nagayama_etal_2016}%
  \BibitemOpen
  \bibfield  {author} {\bibinfo {author} {\bibfnamefont {K.}~\bibnamefont
  {Nagayama}}, \bibinfo {author} {\bibfnamefont {T.}~\bibnamefont {Onuma}},
  \bibinfo {author} {\bibfnamefont {R.}~\bibnamefont {Ueno}}, \bibinfo {author}
  {\bibfnamefont {K.}~\bibnamefont {Tamehiro}},\ and\ \bibinfo {author}
  {\bibfnamefont {H.}~\bibnamefont {Minoda}},\ }\bibfield  {title} {\bibinfo
  {title} {Cathodoluminescence and electron-induced fluorescence enhancement of
  enhanced green fluorescent protein},\ }\href
  {https://doi.org/10.1021/acs.jpcb.5b08138} {\bibfield  {journal} {\bibinfo
  {journal} {The Journal of Physical Chemistry B}\ }\textbf {\bibinfo {volume}
  {120}},\ \bibinfo {pages} {1169} (\bibinfo {year} {2016})}\BibitemShut
  {NoStop}%
\bibitem [{\citenamefont {Akiba}\ \emph {et~al.}(2020)\citenamefont {Akiba},
  \citenamefont {Tamehiro}, \citenamefont {Matsui}, \citenamefont {Ikegami},\
  and\ \citenamefont {Minoda}}]{Akiba_etal_2020}%
  \BibitemOpen
  \bibfield  {author} {\bibinfo {author} {\bibfnamefont {K.}~\bibnamefont
  {Akiba}}, \bibinfo {author} {\bibfnamefont {K.}~\bibnamefont {Tamehiro}},
  \bibinfo {author} {\bibfnamefont {K.}~\bibnamefont {Matsui}}, \bibinfo
  {author} {\bibfnamefont {H.}~\bibnamefont {Ikegami}},\ and\ \bibinfo {author}
  {\bibfnamefont {H.}~\bibnamefont {Minoda}},\ }\bibfield  {title} {\bibinfo
  {title} {Cathodoluminescence of green fluorescent protein exhibits the
  redshifted spectrum and the robustness},\ }\href
  {https://doi.org/10.1038/s41598-020-74367-4} {\bibfield  {journal} {\bibinfo
  {journal} {Sci. Rep.}\ }\textbf {\bibinfo {volume} {10}},\ \bibinfo {pages}
  {17342} (\bibinfo {year} {2020})}\BibitemShut {NoStop}%
\bibitem [{\citenamefont {Tizei}\ and\ \citenamefont
  {Kociak}(2013)}]{Tizei_Kociak_2013}%
  \BibitemOpen
  \bibfield  {author} {\bibinfo {author} {\bibfnamefont {L.~H.~G.}\
  \bibnamefont {Tizei}}\ and\ \bibinfo {author} {\bibfnamefont
  {M.}~\bibnamefont {Kociak}},\ }\bibfield  {title} {\bibinfo {title}
  {Spatially resolved quantum nano-optics of single photons using an electron
  microscope},\ }\href {https://doi.org/10.1103/PhysRevLett.110.153604}
  {\bibfield  {journal} {\bibinfo  {journal} {Phys. Rev. Lett.}\ }\textbf
  {\bibinfo {volume} {110}},\ \bibinfo {pages} {153604} (\bibinfo {year}
  {2013})}\BibitemShut {NoStop}%
\bibitem [{\citenamefont {Meuret}\ \emph {et~al.}(2015)\citenamefont {Meuret},
  \citenamefont {Tizei}, \citenamefont {Cazimajou}, \citenamefont
  {Bourrellier}, \citenamefont {Chang}, \citenamefont {Treussart},\ and\
  \citenamefont {Kociak}}]{Meuret_etal_2015}%
  \BibitemOpen
  \bibfield  {author} {\bibinfo {author} {\bibfnamefont {S.}~\bibnamefont
  {Meuret}}, \bibinfo {author} {\bibfnamefont {L.~H.~G.}\ \bibnamefont
  {Tizei}}, \bibinfo {author} {\bibfnamefont {T.}~\bibnamefont {Cazimajou}},
  \bibinfo {author} {\bibfnamefont {R.}~\bibnamefont {Bourrellier}}, \bibinfo
  {author} {\bibfnamefont {H.~C.}\ \bibnamefont {Chang}}, \bibinfo {author}
  {\bibfnamefont {F.}~\bibnamefont {Treussart}},\ and\ \bibinfo {author}
  {\bibfnamefont {M.}~\bibnamefont {Kociak}},\ }\bibfield  {title} {\bibinfo
  {title} {Photon bunching in cathodoluminescence},\ }\href
  {https://doi.org/10.1103/PhysRevLett.114.197401} {\bibfield  {journal}
  {\bibinfo  {journal} {Phys. Rev. Lett.}\ }\textbf {\bibinfo {volume} {114}},\
  \bibinfo {pages} {197401} (\bibinfo {year} {2015})}\BibitemShut {NoStop}%
\bibitem [{\citenamefont {Solà-Garcia}\ \emph {et~al.}(2020)\citenamefont
  {Solà-Garcia}, \citenamefont {Meuret}, \citenamefont {Coenen},\ and\
  \citenamefont {Polman}}]{Garcia_etal_2020}%
  \BibitemOpen
  \bibfield  {author} {\bibinfo {author} {\bibfnamefont {M.}~\bibnamefont
  {Solà-Garcia}}, \bibinfo {author} {\bibfnamefont {S.}~\bibnamefont
  {Meuret}}, \bibinfo {author} {\bibfnamefont {T.}~\bibnamefont {Coenen}},\
  and\ \bibinfo {author} {\bibfnamefont {A.}~\bibnamefont {Polman}},\
  }\bibfield  {title} {\bibinfo {title} {Electron-induced state conversion in
  diamond {NV} centers measured with pump{--}probe cathodoluminescence
  spectroscopy},\ }\href {https://doi.org/10.1021/acsphotonics.9b01463}
  {\bibfield  {journal} {\bibinfo  {journal} {ACS Photonics}\ }\textbf
  {\bibinfo {volume} {7}},\ \bibinfo {pages} {232} (\bibinfo {year} {2020})},\
  \bibinfo {note} {pMID: 31976357}\BibitemShut {NoStop}%
\bibitem [{\citenamefont {Loudon}(2000)}]{Loudon_2000}%
  \BibitemOpen
  \bibfield  {author} {\bibinfo {author} {\bibfnamefont {R.}~\bibnamefont
  {Loudon}},\ }\href@noop {} {\emph {\bibinfo {title} {The Quantum Theory of
  Light}}},\ \bibinfo {edition} {3rd}\ ed.\ (\bibinfo  {publisher} {Oxford
  University Press, Oxford},\ \bibinfo {year} {2000})\BibitemShut {NoStop}%
\bibitem [{\citenamefont {Auff{\`{e}}ves}\ \emph {et~al.}(2011)\citenamefont
  {Auff{\`{e}}ves}, \citenamefont {Gerace}, \citenamefont {Portolan},
  \citenamefont {Drezet},\ and\ \citenamefont {Santos}}]{Auffeves_etal_2011}%
  \BibitemOpen
  \bibfield  {author} {\bibinfo {author} {\bibfnamefont {A.}~\bibnamefont
  {Auff{\`{e}}ves}}, \bibinfo {author} {\bibfnamefont {D.}~\bibnamefont
  {Gerace}}, \bibinfo {author} {\bibfnamefont {S.}~\bibnamefont {Portolan}},
  \bibinfo {author} {\bibfnamefont {A.}~\bibnamefont {Drezet}},\ and\ \bibinfo
  {author} {\bibfnamefont {M.~F.}\ \bibnamefont {Santos}},\ }\bibfield  {title}
  {\bibinfo {title} {Few emitters in a cavity: from cooperative emission to
  individualization},\ }\href {https://doi.org/10.1088/1367-2630/13/9/093020}
  {\bibfield  {journal} {\bibinfo  {journal} {New J. Phys.}\ }\textbf {\bibinfo
  {volume} {13}},\ \bibinfo {pages} {093020} (\bibinfo {year}
  {2011})}\BibitemShut {NoStop}%
\bibitem [{\citenamefont {Leymann}\ \emph {et~al.}(2015)\citenamefont
  {Leymann}, \citenamefont {Foerster}, \citenamefont {Jahnke}, \citenamefont
  {Wiersig},\ and\ \citenamefont {Gies}}]{Leymann_etal_2015}%
  \BibitemOpen
  \bibfield  {author} {\bibinfo {author} {\bibfnamefont {H.~A.~M.}\
  \bibnamefont {Leymann}}, \bibinfo {author} {\bibfnamefont {A.}~\bibnamefont
  {Foerster}}, \bibinfo {author} {\bibfnamefont {F.}~\bibnamefont {Jahnke}},
  \bibinfo {author} {\bibfnamefont {J.}~\bibnamefont {Wiersig}},\ and\ \bibinfo
  {author} {\bibfnamefont {C.}~\bibnamefont {Gies}},\ }\bibfield  {title}
  {\bibinfo {title} {Sub- and superradiance in nanolasers},\ }\href
  {https://doi.org/10.1103/PhysRevApplied.4.044018} {\bibfield  {journal}
  {\bibinfo  {journal} {Phys. Rev. Applied}\ }\textbf {\bibinfo {volume} {4}},\
  \bibinfo {pages} {044018} (\bibinfo {year} {2015})}\BibitemShut {NoStop}%
\bibitem [{\citenamefont {Jahnke}\ \emph {et~al.}(2016)\citenamefont {Jahnke},
  \citenamefont {Gies}, \citenamefont {A{\ss}mann}, \citenamefont {Bayer},
  \citenamefont {Leymann}, \citenamefont {Foerster}, \citenamefont {Wiersig},
  \citenamefont {Schneider}, \citenamefont {Kamp},\ and\ \citenamefont
  {H{\"o}fling}}]{Jahnke_etal_2016}%
  \BibitemOpen
  \bibfield  {author} {\bibinfo {author} {\bibfnamefont {F.}~\bibnamefont
  {Jahnke}}, \bibinfo {author} {\bibfnamefont {C.}~\bibnamefont {Gies}},
  \bibinfo {author} {\bibfnamefont {M.}~\bibnamefont {A{\ss}mann}}, \bibinfo
  {author} {\bibfnamefont {M.}~\bibnamefont {Bayer}}, \bibinfo {author}
  {\bibfnamefont {H.~A.~M.}\ \bibnamefont {Leymann}}, \bibinfo {author}
  {\bibfnamefont {A.}~\bibnamefont {Foerster}}, \bibinfo {author}
  {\bibfnamefont {J.}~\bibnamefont {Wiersig}}, \bibinfo {author} {\bibfnamefont
  {C.}~\bibnamefont {Schneider}}, \bibinfo {author} {\bibfnamefont
  {M.}~\bibnamefont {Kamp}},\ and\ \bibinfo {author} {\bibfnamefont
  {S.}~\bibnamefont {H{\"o}fling}},\ }\bibfield  {title} {\bibinfo {title}
  {Giant photon bunching, superradiant pulse emission and excitation trapping
  in quantum-dot nanolasers},\ }\href {https://doi.org/10.1038/ncomms11540}
  {\bibfield  {journal} {\bibinfo  {journal} {Nat. Commun.}\ }\textbf {\bibinfo
  {volume} {7}},\ \bibinfo {pages} {11540} (\bibinfo {year}
  {2016})}\BibitemShut {NoStop}%
\bibitem [{\citenamefont {Ridolfo}\ \emph {et~al.}(2010)\citenamefont
  {Ridolfo}, \citenamefont {Di~Stefano}, \citenamefont {Fina}, \citenamefont
  {Saija},\ and\ \citenamefont {Savasta}}]{Ridolfo_etal_2010}%
  \BibitemOpen
  \bibfield  {author} {\bibinfo {author} {\bibfnamefont {A.}~\bibnamefont
  {Ridolfo}}, \bibinfo {author} {\bibfnamefont {O.}~\bibnamefont {Di~Stefano}},
  \bibinfo {author} {\bibfnamefont {N.}~\bibnamefont {Fina}}, \bibinfo {author}
  {\bibfnamefont {R.}~\bibnamefont {Saija}},\ and\ \bibinfo {author}
  {\bibfnamefont {S.}~\bibnamefont {Savasta}},\ }\bibfield  {title} {\bibinfo
  {title} {Quantum plasmonics with quantum dot-metal nanoparticle molecules:
  Influence of the fano effect on photon statistics},\ }\href
  {https://doi.org/10.1103/PhysRevLett.105.263601} {\bibfield  {journal}
  {\bibinfo  {journal} {Phys. Rev. Lett.}\ }\textbf {\bibinfo {volume} {105}},\
  \bibinfo {pages} {263601} (\bibinfo {year} {2010})}\BibitemShut {NoStop}%
\bibitem [{\citenamefont {Zhao}\ \emph {et~al.}(2015)\citenamefont {Zhao},
  \citenamefont {Gu}, \citenamefont {Chen}, \citenamefont {Ren}, \citenamefont
  {Zhang},\ and\ \citenamefont {Gong}}]{Zhao_etal_2015}%
  \BibitemOpen
  \bibfield  {author} {\bibinfo {author} {\bibfnamefont {D.}~\bibnamefont
  {Zhao}}, \bibinfo {author} {\bibfnamefont {Y.}~\bibnamefont {Gu}}, \bibinfo
  {author} {\bibfnamefont {H.}~\bibnamefont {Chen}}, \bibinfo {author}
  {\bibfnamefont {J.}~\bibnamefont {Ren}}, \bibinfo {author} {\bibfnamefont
  {T.}~\bibnamefont {Zhang}},\ and\ \bibinfo {author} {\bibfnamefont
  {Q.}~\bibnamefont {Gong}},\ }\bibfield  {title} {\bibinfo {title} {Quantum
  statistics control with a plasmonic nanocavity: Multimode-enhanced
  interferences},\ }\href {https://doi.org/10.1103/PhysRevA.92.033836}
  {\bibfield  {journal} {\bibinfo  {journal} {Phys. Rev. A}\ }\textbf {\bibinfo
  {volume} {92}},\ \bibinfo {pages} {033836} (\bibinfo {year}
  {2015})}\BibitemShut {NoStop}%
\bibitem [{\citenamefont {Leymann}\ \emph {et~al.}(2013)\citenamefont
  {Leymann}, \citenamefont {Hopfmann}, \citenamefont {Albert}, \citenamefont
  {Foerster}, \citenamefont {Khanbekyan}, \citenamefont {Schneider},
  \citenamefont {H\"ofling}, \citenamefont {Forchel}, \citenamefont {Kamp},
  \citenamefont {Wiersig},\ and\ \citenamefont
  {Reitzenstein}}]{Leymann_etal_2013a}%
  \BibitemOpen
  \bibfield  {author} {\bibinfo {author} {\bibfnamefont {H.~A.~M.}\
  \bibnamefont {Leymann}}, \bibinfo {author} {\bibfnamefont {C.}~\bibnamefont
  {Hopfmann}}, \bibinfo {author} {\bibfnamefont {F.}~\bibnamefont {Albert}},
  \bibinfo {author} {\bibfnamefont {A.}~\bibnamefont {Foerster}}, \bibinfo
  {author} {\bibfnamefont {M.}~\bibnamefont {Khanbekyan}}, \bibinfo {author}
  {\bibfnamefont {C.}~\bibnamefont {Schneider}}, \bibinfo {author}
  {\bibfnamefont {S.}~\bibnamefont {H\"ofling}}, \bibinfo {author}
  {\bibfnamefont {A.}~\bibnamefont {Forchel}}, \bibinfo {author} {\bibfnamefont
  {M.}~\bibnamefont {Kamp}}, \bibinfo {author} {\bibfnamefont {J.}~\bibnamefont
  {Wiersig}},\ and\ \bibinfo {author} {\bibfnamefont {S.}~\bibnamefont
  {Reitzenstein}},\ }\bibfield  {title} {\bibinfo {title} {Intensity
  fluctuations in bimodal micropillar lasers enhanced by quantum-dot gain
  competition},\ }\href {https://doi.org/10.1103/PhysRevA.87.053819} {\bibfield
   {journal} {\bibinfo  {journal} {Phys. Rev. A}\ }\textbf {\bibinfo {volume}
  {87}},\ \bibinfo {pages} {053819} (\bibinfo {year} {2013})}\BibitemShut
  {NoStop}%
\bibitem [{\citenamefont {Redlich}\ \emph {et~al.}(2016)\citenamefont
  {Redlich}, \citenamefont {Lingnau}, \citenamefont {Holzinger}, \citenamefont
  {Schlottmann}, \citenamefont {Kreinberg}, \citenamefont {Schneider},
  \citenamefont {Kamp}, \citenamefont {Höfling}, \citenamefont {Wolters},
  \citenamefont {Reitzenstein},\ and\ \citenamefont
  {Lüdge}}]{Redlich_etal_2016}%
  \BibitemOpen
  \bibfield  {author} {\bibinfo {author} {\bibfnamefont {C.}~\bibnamefont
  {Redlich}}, \bibinfo {author} {\bibfnamefont {B.}~\bibnamefont {Lingnau}},
  \bibinfo {author} {\bibfnamefont {S.}~\bibnamefont {Holzinger}}, \bibinfo
  {author} {\bibfnamefont {E.}~\bibnamefont {Schlottmann}}, \bibinfo {author}
  {\bibfnamefont {S.}~\bibnamefont {Kreinberg}}, \bibinfo {author}
  {\bibfnamefont {C.}~\bibnamefont {Schneider}}, \bibinfo {author}
  {\bibfnamefont {M.}~\bibnamefont {Kamp}}, \bibinfo {author} {\bibfnamefont
  {S.}~\bibnamefont {Höfling}}, \bibinfo {author} {\bibfnamefont
  {J.}~\bibnamefont {Wolters}}, \bibinfo {author} {\bibfnamefont
  {S.}~\bibnamefont {Reitzenstein}},\ and\ \bibinfo {author} {\bibfnamefont
  {K.}~\bibnamefont {Lüdge}},\ }\bibfield  {title} {\bibinfo {title}
  {Mode-switching induced super-thermal bunching in quantum-dot microlasers},\
  }\href {https://doi.org/10.1088/1367-2630/18/6/063011} {\bibfield  {journal}
  {\bibinfo  {journal} {New J. Phys.}\ }\textbf {\bibinfo {volume} {18}},\
  \bibinfo {pages} {063011} (\bibinfo {year} {2016})}\BibitemShut {NoStop}%
\bibitem [{\citenamefont {Marconi}\ \emph {et~al.}(2018)\citenamefont
  {Marconi}, \citenamefont {Javaloyes}, \citenamefont {Hamel}, \citenamefont
  {Raineri}, \citenamefont {Levenson},\ and\ \citenamefont
  {Yacomotti}}]{Marconi_etal_2018}%
  \BibitemOpen
  \bibfield  {author} {\bibinfo {author} {\bibfnamefont {M.}~\bibnamefont
  {Marconi}}, \bibinfo {author} {\bibfnamefont {J.}~\bibnamefont {Javaloyes}},
  \bibinfo {author} {\bibfnamefont {P.}~\bibnamefont {Hamel}}, \bibinfo
  {author} {\bibfnamefont {F.}~\bibnamefont {Raineri}}, \bibinfo {author}
  {\bibfnamefont {A.}~\bibnamefont {Levenson}},\ and\ \bibinfo {author}
  {\bibfnamefont {A.~M.}\ \bibnamefont {Yacomotti}},\ }\bibfield  {title}
  {\bibinfo {title} {Far-from-equilibrium route to superthermal light in
  bimodal nanolasers},\ }\href {https://doi.org/10.1103/PhysRevX.8.011013}
  {\bibfield  {journal} {\bibinfo  {journal} {Phys. Rev. X}\ }\textbf {\bibinfo
  {volume} {8}},\ \bibinfo {pages} {011013} (\bibinfo {year}
  {2018})}\BibitemShut {NoStop}%
\bibitem [{\citenamefont {Leymann}\ \emph {et~al.}(2017)\citenamefont
  {Leymann}, \citenamefont {Vorberg}, \citenamefont {Lettau}, \citenamefont
  {Hopfmann}, \citenamefont {Schneider}, \citenamefont {Kamp}, \citenamefont
  {H\"ofling}, \citenamefont {Ketzmerick}, \citenamefont {Wiersig},
  \citenamefont {Reitzenstein},\ and\ \citenamefont
  {Eckardt}}]{Leymann_etal_2017}%
  \BibitemOpen
  \bibfield  {author} {\bibinfo {author} {\bibfnamefont {H.~A.~M.}\
  \bibnamefont {Leymann}}, \bibinfo {author} {\bibfnamefont {D.}~\bibnamefont
  {Vorberg}}, \bibinfo {author} {\bibfnamefont {T.}~\bibnamefont {Lettau}},
  \bibinfo {author} {\bibfnamefont {C.}~\bibnamefont {Hopfmann}}, \bibinfo
  {author} {\bibfnamefont {C.}~\bibnamefont {Schneider}}, \bibinfo {author}
  {\bibfnamefont {M.}~\bibnamefont {Kamp}}, \bibinfo {author} {\bibfnamefont
  {S.}~\bibnamefont {H\"ofling}}, \bibinfo {author} {\bibfnamefont
  {R.}~\bibnamefont {Ketzmerick}}, \bibinfo {author} {\bibfnamefont
  {J.}~\bibnamefont {Wiersig}}, \bibinfo {author} {\bibfnamefont
  {S.}~\bibnamefont {Reitzenstein}},\ and\ \bibinfo {author} {\bibfnamefont
  {A.}~\bibnamefont {Eckardt}},\ }\bibfield  {title} {\bibinfo {title}
  {Pump-power-driven mode switching in a microcavity device and its relation to
  bose-einstein condensation},\ }\href
  {https://doi.org/10.1103/PhysRevX.7.021045} {\bibfield  {journal} {\bibinfo
  {journal} {Phys. Rev. X}\ }\textbf {\bibinfo {volume} {7}},\ \bibinfo {pages}
  {021045} (\bibinfo {year} {2017})}\BibitemShut {NoStop}%
\bibitem [{\citenamefont {Schmidt}\ \emph {et~al.}(2021)\citenamefont
  {Schmidt}, \citenamefont {Grothe}, \citenamefont {Neumeier}, \citenamefont
  {Bremer}, \citenamefont {von Helversen}, \citenamefont {Zent}, \citenamefont
  {Melcher}, \citenamefont {Beyer}, \citenamefont {Schneider}, \citenamefont
  {H\"ofling}, \citenamefont {Wiersig},\ and\ \citenamefont
  {Reitzenstein}}]{Schmidt_etal_2021}%
  \BibitemOpen
  \bibfield  {author} {\bibinfo {author} {\bibfnamefont {M.}~\bibnamefont
  {Schmidt}}, \bibinfo {author} {\bibfnamefont {I.~H.}\ \bibnamefont {Grothe}},
  \bibinfo {author} {\bibfnamefont {S.}~\bibnamefont {Neumeier}}, \bibinfo
  {author} {\bibfnamefont {L.}~\bibnamefont {Bremer}}, \bibinfo {author}
  {\bibfnamefont {M.}~\bibnamefont {von Helversen}}, \bibinfo {author}
  {\bibfnamefont {W.}~\bibnamefont {Zent}}, \bibinfo {author} {\bibfnamefont
  {B.}~\bibnamefont {Melcher}}, \bibinfo {author} {\bibfnamefont
  {J.}~\bibnamefont {Beyer}}, \bibinfo {author} {\bibfnamefont
  {C.}~\bibnamefont {Schneider}}, \bibinfo {author} {\bibfnamefont
  {S.}~\bibnamefont {H\"ofling}}, \bibinfo {author} {\bibfnamefont
  {J.}~\bibnamefont {Wiersig}},\ and\ \bibinfo {author} {\bibfnamefont
  {S.}~\bibnamefont {Reitzenstein}},\ }\bibfield  {title} {\bibinfo {title}
  {Bimodal behavior of microlasers investigated with a two-channel
  photon-number-resolving transition-edge sensor system},\ }\href
  {https://doi.org/10.1103/PhysRevResearch.3.013263} {\bibfield  {journal}
  {\bibinfo  {journal} {Phys. Rev. Research}\ }\textbf {\bibinfo {volume}
  {3}},\ \bibinfo {pages} {013263} (\bibinfo {year} {2021})}\BibitemShut
  {NoStop}%
\bibitem [{\citenamefont {Meuret}\ \emph {et~al.}(2016)\citenamefont {Meuret},
  \citenamefont {Tizei}, \citenamefont {Auzelle}, \citenamefont {Songmuang},
  \citenamefont {Daudin}, \citenamefont {Gayral},\ and\ \citenamefont
  {Kociak}}]{Meuret_etal_2016}%
  \BibitemOpen
  \bibfield  {author} {\bibinfo {author} {\bibfnamefont {S.}~\bibnamefont
  {Meuret}}, \bibinfo {author} {\bibfnamefont {L.~H.~G.}\ \bibnamefont
  {Tizei}}, \bibinfo {author} {\bibfnamefont {T.}~\bibnamefont {Auzelle}},
  \bibinfo {author} {\bibfnamefont {R.}~\bibnamefont {Songmuang}}, \bibinfo
  {author} {\bibfnamefont {B.}~\bibnamefont {Daudin}}, \bibinfo {author}
  {\bibfnamefont {B.}~\bibnamefont {Gayral}},\ and\ \bibinfo {author}
  {\bibfnamefont {M.}~\bibnamefont {Kociak}},\ }\bibfield  {title} {\bibinfo
  {title} {Lifetime measurements well below the optical diffraction limit},\
  }\href {https://doi.org/10.1021/acsphotonics.6b00212} {\bibfield  {journal}
  {\bibinfo  {journal} {ACS Photonics}\ }\textbf {\bibinfo {volume} {3}},\
  \bibinfo {pages} {1157} (\bibinfo {year} {2016})}\BibitemShut {NoStop}%
\bibitem [{\citenamefont {Lourenço-Martins}\ \emph {et~al.}(2018)\citenamefont
  {Lourenço-Martins}, \citenamefont {Kociak}, \citenamefont {Meuret},
  \citenamefont {Treussart}, \citenamefont {Lee}, \citenamefont {Ling},
  \citenamefont {Chang},\ and\ \citenamefont
  {Galvão~Tizei}}]{Lourenco-Martins_etal_2018}%
  \BibitemOpen
  \bibfield  {author} {\bibinfo {author} {\bibfnamefont {H.}~\bibnamefont
  {Lourenço-Martins}}, \bibinfo {author} {\bibfnamefont {M.}~\bibnamefont
  {Kociak}}, \bibinfo {author} {\bibfnamefont {S.}~\bibnamefont {Meuret}},
  \bibinfo {author} {\bibfnamefont {F.}~\bibnamefont {Treussart}}, \bibinfo
  {author} {\bibfnamefont {Y.~H.}\ \bibnamefont {Lee}}, \bibinfo {author}
  {\bibfnamefont {X.~Y.}\ \bibnamefont {Ling}}, \bibinfo {author}
  {\bibfnamefont {H.-C.}\ \bibnamefont {Chang}},\ and\ \bibinfo {author}
  {\bibfnamefont {L.~H.}\ \bibnamefont {Galvão~Tizei}},\ }\bibfield  {title}
  {\bibinfo {title} {Probing plasmon-{NV$^0$} coupling at the nanometer scale
  with photons and fast electrons},\ }\href
  {https://doi.org/10.1021/acsphotonics.7b01093} {\bibfield  {journal}
  {\bibinfo  {journal} {ACS Photonics}\ }\textbf {\bibinfo {volume} {5}},\
  \bibinfo {pages} {324} (\bibinfo {year} {2018})}\BibitemShut {NoStop}%
\bibitem [{\citenamefont {Yanagimoto}\ \emph {et~al.}(2021)\citenamefont
  {Yanagimoto}, \citenamefont {Yamamoto}, \citenamefont {Sannomiya},\ and\
  \citenamefont {Akiba}}]{Yanagimoto_etal_2021}%
  \BibitemOpen
  \bibfield  {author} {\bibinfo {author} {\bibfnamefont {S.}~\bibnamefont
  {Yanagimoto}}, \bibinfo {author} {\bibfnamefont {N.}~\bibnamefont
  {Yamamoto}}, \bibinfo {author} {\bibfnamefont {T.}~\bibnamefont
  {Sannomiya}},\ and\ \bibinfo {author} {\bibfnamefont {K.}~\bibnamefont
  {Akiba}},\ }\bibfield  {title} {\bibinfo {title} {Purcell effect of
  nitrogen-vacancy centers in nanodiamond coupled to propagating and localized
  surface plasmons revealed by photon-correlation cathodoluminescence},\ }\href
  {https://doi.org/10.1103/PhysRevB.103.205418} {\bibfield  {journal} {\bibinfo
   {journal} {Phys. Rev. B}\ }\textbf {\bibinfo {volume} {103}},\ \bibinfo
  {pages} {205418} (\bibinfo {year} {2021})}\BibitemShut {NoStop}%
\bibitem [{\citenamefont {Meuret}\ \emph {et~al.}(2017)\citenamefont {Meuret},
  \citenamefont {Coenen}, \citenamefont {Zeijlemaker}, \citenamefont {Latzel},
  \citenamefont {Christiansen}, \citenamefont {Conesa-Boj},\ and\ \citenamefont
  {Polman}}]{Meuret_etal_2017}%
  \BibitemOpen
  \bibfield  {author} {\bibinfo {author} {\bibfnamefont {S.}~\bibnamefont
  {Meuret}}, \bibinfo {author} {\bibfnamefont {T.}~\bibnamefont {Coenen}},
  \bibinfo {author} {\bibfnamefont {H.}~\bibnamefont {Zeijlemaker}}, \bibinfo
  {author} {\bibfnamefont {M.}~\bibnamefont {Latzel}}, \bibinfo {author}
  {\bibfnamefont {S.}~\bibnamefont {Christiansen}}, \bibinfo {author}
  {\bibfnamefont {S.}~\bibnamefont {Conesa-Boj}},\ and\ \bibinfo {author}
  {\bibfnamefont {A.}~\bibnamefont {Polman}},\ }\bibfield  {title} {\bibinfo
  {title} {Photon bunching reveals single-electron cathodoluminescence
  excitation efficiency in {InGaN} quantum wells},\ }\href
  {https://doi.org/10.1103/PhysRevB.96.035308} {\bibfield  {journal} {\bibinfo
  {journal} {Phys. Rev. B}\ }\textbf {\bibinfo {volume} {96}},\ \bibinfo
  {pages} {035308} (\bibinfo {year} {2017})}\BibitemShut {NoStop}%
\bibitem [{\citenamefont {Meuret}\ \emph {et~al.}(2018)\citenamefont {Meuret},
  \citenamefont {Coenen}, \citenamefont {Woo}, \citenamefont {Ra},
  \citenamefont {Mi},\ and\ \citenamefont {Polman}}]{Meuret_etal_2018}%
  \BibitemOpen
  \bibfield  {author} {\bibinfo {author} {\bibfnamefont {S.}~\bibnamefont
  {Meuret}}, \bibinfo {author} {\bibfnamefont {T.}~\bibnamefont {Coenen}},
  \bibinfo {author} {\bibfnamefont {S.~Y.}\ \bibnamefont {Woo}}, \bibinfo
  {author} {\bibfnamefont {Y.-H.}\ \bibnamefont {Ra}}, \bibinfo {author}
  {\bibfnamefont {Z.}~\bibnamefont {Mi}},\ and\ \bibinfo {author}
  {\bibfnamefont {A.}~\bibnamefont {Polman}},\ }\bibfield  {title} {\bibinfo
  {title} {Nanoscale relative emission efficiency mapping using
  cathodoluminescence {$g^{(2)}$} imaging},\ }\href
  {https://doi.org/10.1021/acs.nanolett.7b04891} {\bibfield  {journal}
  {\bibinfo  {journal} {Nano Lett.}\ }\textbf {\bibinfo {volume} {18}},\
  \bibinfo {pages} {2288} (\bibinfo {year} {2018})}\BibitemShut {NoStop}%
\bibitem [{\citenamefont {{van
  Rijswijk}}(1976{\natexlab{a}})}]{van_Rijswijk_1976a}%
  \BibitemOpen
  \bibfield  {author} {\bibinfo {author} {\bibfnamefont {F.~C.}\ \bibnamefont
  {{van Rijswijk}}},\ }\bibfield  {title} {\bibinfo {title} {Photon statistics
  of characteristic cathodoluminescence radiation: I. theory},\ }\href
  {https://doi.org/https://doi.org/10.1016/0378-4363(76)90182-0} {\bibfield
  {journal} {\bibinfo  {journal} {Physica B+C}\ }\textbf {\bibinfo {volume}
  {82}},\ \bibinfo {pages} {193} (\bibinfo {year}
  {1976}{\natexlab{a}})}\BibitemShut {NoStop}%
\bibitem [{\citenamefont {{van
  Rijswijk}}(1976{\natexlab{b}})}]{van_Rijswijk_1976b}%
  \BibitemOpen
  \bibfield  {author} {\bibinfo {author} {\bibfnamefont {F.~C.}\ \bibnamefont
  {{van Rijswijk}}},\ }\bibfield  {title} {\bibinfo {title} {Photon statistics
  of characteristic cathodoluminescence radiation: Ii. experiment},\ }\href
  {https://doi.org/https://doi.org/10.1016/0378-4363(76)90183-2} {\bibfield
  {journal} {\bibinfo  {journal} {Physica B+C}\ }\textbf {\bibinfo {volume}
  {82}},\ \bibinfo {pages} {205} (\bibinfo {year}
  {1976}{\natexlab{b}})}\BibitemShut {NoStop}%
\bibitem [{\citenamefont {Solà-Garcia}\ \emph {et~al.}(2021)\citenamefont
  {Solà-Garcia}, \citenamefont {Mauser}, \citenamefont {Liebtrau},
  \citenamefont {Coenen}, \citenamefont {Christiansen}, \citenamefont
  {Meuret},\ and\ \citenamefont {Polman}}]{Garcia_etal_2021}%
  \BibitemOpen
  \bibfield  {author} {\bibinfo {author} {\bibfnamefont {M.}~\bibnamefont
  {Solà-Garcia}}, \bibinfo {author} {\bibfnamefont {K.~W.}\ \bibnamefont
  {Mauser}}, \bibinfo {author} {\bibfnamefont {M.}~\bibnamefont {Liebtrau}},
  \bibinfo {author} {\bibfnamefont {T.}~\bibnamefont {Coenen}}, \bibinfo
  {author} {\bibfnamefont {S.}~\bibnamefont {Christiansen}}, \bibinfo {author}
  {\bibfnamefont {S.}~\bibnamefont {Meuret}},\ and\ \bibinfo {author}
  {\bibfnamefont {A.}~\bibnamefont {Polman}},\ }\bibfield  {title} {\bibinfo
  {title} {Photon statistics of incoherent cathodoluminescence with continuous
  and pulsed electron beams},\ }\href
  {https://doi.org/10.1021/acsphotonics.0c01939} {\bibfield  {journal}
  {\bibinfo  {journal} {ACS Photonics}\ }\textbf {\bibinfo {volume} {8}},\
  \bibinfo {pages} {916} (\bibinfo {year} {2021})}\BibitemShut {NoStop}%
\bibitem [{\citenamefont {Feldman}\ \emph {et~al.}(2018)\citenamefont
  {Feldman}, \citenamefont {Dumitrescu}, \citenamefont {Bridges}, \citenamefont
  {Chisholm}, \citenamefont {Davidson}, \citenamefont {Evans}, \citenamefont
  {Hachtel}, \citenamefont {Hu}, \citenamefont {Pooser}, \citenamefont
  {Haglund},\ and\ \citenamefont {Lawrie}}]{Feldman_etal_2018}%
  \BibitemOpen
  \bibfield  {author} {\bibinfo {author} {\bibfnamefont {M.~A.}\ \bibnamefont
  {Feldman}}, \bibinfo {author} {\bibfnamefont {E.~F.}\ \bibnamefont
  {Dumitrescu}}, \bibinfo {author} {\bibfnamefont {D.}~\bibnamefont {Bridges}},
  \bibinfo {author} {\bibfnamefont {M.~F.}\ \bibnamefont {Chisholm}}, \bibinfo
  {author} {\bibfnamefont {R.~B.}\ \bibnamefont {Davidson}}, \bibinfo {author}
  {\bibfnamefont {P.~G.}\ \bibnamefont {Evans}}, \bibinfo {author}
  {\bibfnamefont {J.~A.}\ \bibnamefont {Hachtel}}, \bibinfo {author}
  {\bibfnamefont {A.}~\bibnamefont {Hu}}, \bibinfo {author} {\bibfnamefont
  {R.~C.}\ \bibnamefont {Pooser}}, \bibinfo {author} {\bibfnamefont {R.~F.}\
  \bibnamefont {Haglund}},\ and\ \bibinfo {author} {\bibfnamefont {B.~J.}\
  \bibnamefont {Lawrie}},\ }\bibfield  {title} {\bibinfo {title} {Colossal
  photon bunching in quasiparticle-mediated nanodiamond cathodoluminescence},\
  }\href {https://doi.org/10.1103/PhysRevB.97.081404} {\bibfield  {journal}
  {\bibinfo  {journal} {Phys. Rev. B}\ }\textbf {\bibinfo {volume} {97}},\
  \bibinfo {pages} {081404} (\bibinfo {year} {2018})}\BibitemShut {NoStop}%
\bibitem [{\citenamefont {Egerton}(2011)}]{Egerton_2011}%
  \BibitemOpen
  \bibfield  {author} {\bibinfo {author} {\bibfnamefont {R.~F.}\ \bibnamefont
  {Egerton}},\ }\href@noop {} {\emph {\bibinfo {title} {Electron Energy-Loss
  Spectroscopy in the Electron Microscope}}}\ (\bibinfo  {publisher} {Springer,
  New York},\ \bibinfo {year} {2011})\BibitemShut {NoStop}%
\bibitem [{\citenamefont {Yamamoto}(2010)}]{Yamamoto_2010}%
  \BibitemOpen
  \bibfield  {author} {\bibinfo {author} {\bibfnamefont {N.}~\bibnamefont
  {Yamamoto}},\ }\bibinfo {title} {Cathodoluminescence of nanomaterials},\ in\
  \href@noop {} {\emph {\bibinfo {booktitle} {Handbook of Nanophysics:
  Nanoelectronics and Nanophotonics}}},\ \bibinfo {editor} {edited by\ \bibinfo
  {editor} {\bibfnamefont {K.~D.}\ \bibnamefont {Sattler}}}\ (\bibinfo
  {publisher} {CRC Press, Boca Raton},\ \bibinfo {year} {2010})\ Chap.~\bibinfo
  {chapter} {21}\BibitemShut {NoStop}%
\bibitem [{\citenamefont {Rothwarf}(1973)}]{Rothwarf_1973}%
  \BibitemOpen
  \bibfield  {author} {\bibinfo {author} {\bibfnamefont {A.}~\bibnamefont
  {Rothwarf}},\ }\bibfield  {title} {\bibinfo {title} {Plasmon theory of
  electron--hole pair production: efficiency of cathode ray phosphors},\ }\href
  {https://doi.org/10.1063/1.1662257} {\bibfield  {journal} {\bibinfo
  {journal} {J. Appl. Phys.}\ }\textbf {\bibinfo {volume} {44}},\ \bibinfo
  {pages} {752} (\bibinfo {year} {1973})}\BibitemShut {NoStop}%
\bibitem [{\citenamefont {Yacobi}\ and\ \citenamefont
  {Holt}(1986)}]{Yacobi_Holt_1986}%
  \BibitemOpen
  \bibfield  {author} {\bibinfo {author} {\bibfnamefont {B.~G.}\ \bibnamefont
  {Yacobi}}\ and\ \bibinfo {author} {\bibfnamefont {D.~B.}\ \bibnamefont
  {Holt}},\ }\bibfield  {title} {\bibinfo {title} {Cathodoluminescence scanning
  electron microscopy of semiconductors},\ }\href
  {https://doi.org/10.1063/1.336491} {\bibfield  {journal} {\bibinfo  {journal}
  {J. Appl. Phys.}\ }\textbf {\bibinfo {volume} {59}},\ \bibinfo {pages} {R1}
  (\bibinfo {year} {1986})}\BibitemShut {NoStop}%
\bibitem [{\citenamefont {Varkentina}\ \emph {et~al.}(2022)\citenamefont
  {Varkentina}, \citenamefont {Auad}, \citenamefont {Woo}, \citenamefont
  {Zobelli}, \citenamefont {Bocher}, \citenamefont {Blazit}, \citenamefont
  {Li}, \citenamefont {Tencé}, \citenamefont {Watanabe}, \citenamefont
  {Taniguchi}, \citenamefont {Stéphan}, \citenamefont {Kociak},\ and\
  \citenamefont {Tizei}}]{Varkentina_etal_2022}%
  \BibitemOpen
  \bibfield  {author} {\bibinfo {author} {\bibfnamefont {N.}~\bibnamefont
  {Varkentina}}, \bibinfo {author} {\bibfnamefont {Y.}~\bibnamefont {Auad}},
  \bibinfo {author} {\bibfnamefont {S.~Y.}\ \bibnamefont {Woo}}, \bibinfo
  {author} {\bibfnamefont {A.}~\bibnamefont {Zobelli}}, \bibinfo {author}
  {\bibfnamefont {L.}~\bibnamefont {Bocher}}, \bibinfo {author} {\bibfnamefont
  {J.-D.}\ \bibnamefont {Blazit}}, \bibinfo {author} {\bibfnamefont
  {X.}~\bibnamefont {Li}}, \bibinfo {author} {\bibfnamefont {M.}~\bibnamefont
  {Tencé}}, \bibinfo {author} {\bibfnamefont {K.}~\bibnamefont {Watanabe}},
  \bibinfo {author} {\bibfnamefont {T.}~\bibnamefont {Taniguchi}}, \bibinfo
  {author} {\bibfnamefont {O.}~\bibnamefont {Stéphan}}, \bibinfo {author}
  {\bibfnamefont {M.}~\bibnamefont {Kociak}},\ and\ \bibinfo {author}
  {\bibfnamefont {L.~H.~G.}\ \bibnamefont {Tizei}},\ }\bibfield  {title}
  {\bibinfo {title} {Cathodoluminescence excitation spectroscopy: Nanoscale
  imaging of excitation pathways},\ }\href
  {https://doi.org/10.1126/sciadv.abq4947} {\bibfield  {journal} {\bibinfo
  {journal} {Sci. Adv.}\ }\textbf {\bibinfo {volume} {8}},\ \bibinfo {pages}
  {eabq4947} (\bibinfo {year} {2022})}\BibitemShut {NoStop}%
\bibitem [{\citenamefont {Gorini}\ \emph {et~al.}(1976)\citenamefont {Gorini},
  \citenamefont {Kossakowski},\ and\ \citenamefont
  {Sudarshan}}]{Gorini_etal_1976}%
  \BibitemOpen
  \bibfield  {author} {\bibinfo {author} {\bibfnamefont {V.}~\bibnamefont
  {Gorini}}, \bibinfo {author} {\bibfnamefont {A.}~\bibnamefont
  {Kossakowski}},\ and\ \bibinfo {author} {\bibfnamefont {E.~C.~G.}\
  \bibnamefont {Sudarshan}},\ }\bibfield  {title} {\bibinfo {title} {Completely
  positive dynamical semigroups of {N}-level systems},\ }\href
  {https://doi.org/10.1063/1.522979} {\bibfield  {journal} {\bibinfo  {journal}
  {J. Math. Phys.}\ }\textbf {\bibinfo {volume} {17}},\ \bibinfo {pages} {821}
  (\bibinfo {year} {1976})}\BibitemShut {NoStop}%
\bibitem [{\citenamefont {Lindblad}(1976)}]{Lindblad_1976}%
  \BibitemOpen
  \bibfield  {author} {\bibinfo {author} {\bibfnamefont {G.}~\bibnamefont
  {Lindblad}},\ }\bibfield  {title} {\bibinfo {title} {On the generators of
  quantum dynamical semigroups},\ }\href {https://doi.org/10.1007/BF01608499}
  {\bibfield  {journal} {\bibinfo  {journal} {Commun. Math. Phys.}\ }\textbf
  {\bibinfo {volume} {48}},\ \bibinfo {pages} {119} (\bibinfo {year}
  {1976})}\BibitemShut {NoStop}%
\bibitem [{\citenamefont {Breuer}\ and\ \citenamefont
  {Petruccione}(2002)}]{Breuer_Petruccione_2002}%
  \BibitemOpen
  \bibfield  {author} {\bibinfo {author} {\bibfnamefont {H.-P.}\ \bibnamefont
  {Breuer}}\ and\ \bibinfo {author} {\bibfnamefont {F.}~\bibnamefont
  {Petruccione}},\ }\href@noop {} {\emph {\bibinfo {title} {The Theory of Open
  Quantum Systems}}}\ (\bibinfo  {publisher} {Oxford University Press,
  Oxford},\ \bibinfo {year} {2002})\BibitemShut {NoStop}%
\bibitem [{\citenamefont {Carmichael}(1999)}]{Carmichael_1999}%
  \BibitemOpen
  \bibfield  {author} {\bibinfo {author} {\bibfnamefont {H.~J.}\ \bibnamefont
  {Carmichael}},\ }\href@noop {} {\emph {\bibinfo {title} {Statistical Methods
  in Quantum Optics 1: Master Equations and Fokker-Planck Equations}}}\
  (\bibinfo  {publisher} {Springer, Berlin},\ \bibinfo {year}
  {1999})\BibitemShut {NoStop}%
\bibitem [{\citenamefont {Scully}\ and\ \citenamefont
  {Zubairy}(1997)}]{Scully_Zubairy_1997}%
  \BibitemOpen
  \bibfield  {author} {\bibinfo {author} {\bibfnamefont {M.~O.}\ \bibnamefont
  {Scully}}\ and\ \bibinfo {author} {\bibfnamefont {M.~S.}\ \bibnamefont
  {Zubairy}},\ }\href@noop {} {\emph {\bibinfo {title} {Quantum Optics}}}\
  (\bibinfo  {publisher} {Cambridge University Press, Cambridge},\ \bibinfo
  {year} {1997})\BibitemShut {NoStop}%
\bibitem [{sup()}]{supplement}%
  \BibitemOpen
  \href@noop {} {}\bibinfo {note} {See supplemantal material.}\BibitemShut
  {Stop}%
\bibitem [{\citenamefont {Nakajima}(1958)}]{Nakajima_1958}%
  \BibitemOpen
  \bibfield  {author} {\bibinfo {author} {\bibfnamefont {S.}~\bibnamefont
  {Nakajima}},\ }\bibfield  {title} {\bibinfo {title} {On quantum theory of
  transport phenomena: Steady diffusion},\ }\href
  {https://doi.org/10.1143/PTP.20.948} {\bibfield  {journal} {\bibinfo
  {journal} {Prog. Theor. Phys.}\ }\textbf {\bibinfo {volume} {20}},\ \bibinfo
  {pages} {948} (\bibinfo {year} {1958})}\BibitemShut {NoStop}%
\bibitem [{\citenamefont {Zwanzig}(1960)}]{Zwanzig_1960}%
  \BibitemOpen
  \bibfield  {author} {\bibinfo {author} {\bibfnamefont {R.}~\bibnamefont
  {Zwanzig}},\ }\bibfield  {title} {\bibinfo {title} {Ensemble method in the
  theory of irreversibility},\ }\href {https://doi.org/10.1063/1.1731409}
  {\bibfield  {journal} {\bibinfo  {journal} {J. Chem. Phys.}\ }\textbf
  {\bibinfo {volume} {33}},\ \bibinfo {pages} {1338} (\bibinfo {year}
  {1960})}\BibitemShut {NoStop}%
\bibitem [{\citenamefont {Shibata}\ and\ \citenamefont
  {Arimitsu}(1980)}]{Shibata_Arimitsu_1980}%
  \BibitemOpen
  \bibfield  {author} {\bibinfo {author} {\bibfnamefont {F.}~\bibnamefont
  {Shibata}}\ and\ \bibinfo {author} {\bibfnamefont {T.}~\bibnamefont
  {Arimitsu}},\ }\bibfield  {title} {\bibinfo {title} {Expansion formulas in
  nonequilibrium statistical mechanics},\ }\href
  {https://doi.org/10.1143/JPSJ.49.891} {\bibfield  {journal} {\bibinfo
  {journal} {J. Phys. Soc. Jpn.}\ }\textbf {\bibinfo {volume} {49}},\ \bibinfo
  {pages} {891} (\bibinfo {year} {1980})}\BibitemShut {NoStop}%
\bibitem [{\citenamefont {Gardiner}(2009)}]{Gardiner_2009}%
  \BibitemOpen
  \bibfield  {author} {\bibinfo {author} {\bibfnamefont {C.}~\bibnamefont
  {Gardiner}},\ }\href@noop {} {\emph {\bibinfo {title} {Stochastic Methods: A
  Handbook for the Natural and Social Sciences}}}\ (\bibinfo  {publisher}
  {Springer, Berlin Heidelberg},\ \bibinfo {year} {2009})\BibitemShut {NoStop}%
\bibitem [{\citenamefont {Mandel}\ and\ \citenamefont
  {Wolf}(1995)}]{Mandel_Wolf_1995}%
  \BibitemOpen
  \bibfield  {author} {\bibinfo {author} {\bibfnamefont {L.}~\bibnamefont
  {Mandel}}\ and\ \bibinfo {author} {\bibfnamefont {E.}~\bibnamefont {Wolf}},\
  }\href@noop {} {\emph {\bibinfo {title} {Optical Coherence and Quantum
  Optics}}}\ (\bibinfo  {publisher} {Cambridge University Press, Cambridge},\
  \bibinfo {year} {1995})\BibitemShut {NoStop}%
\bibitem [{\citenamefont {Lettau}\ \emph {et~al.}(2018)\citenamefont {Lettau},
  \citenamefont {Leymann}, \citenamefont {Melcher},\ and\ \citenamefont
  {Wiersig}}]{Lettau_etal_2018}%
  \BibitemOpen
  \bibfield  {author} {\bibinfo {author} {\bibfnamefont {T.}~\bibnamefont
  {Lettau}}, \bibinfo {author} {\bibfnamefont {H.~A.~M.}\ \bibnamefont
  {Leymann}}, \bibinfo {author} {\bibfnamefont {B.}~\bibnamefont {Melcher}},\
  and\ \bibinfo {author} {\bibfnamefont {J.}~\bibnamefont {Wiersig}},\
  }\bibfield  {title} {\bibinfo {title} {Superthermal photon bunching in terms
  of simple probability distributions},\ }\href
  {https://doi.org/10.1103/PhysRevA.97.053835} {\bibfield  {journal} {\bibinfo
  {journal} {Phys. Rev. A}\ }\textbf {\bibinfo {volume} {97}},\ \bibinfo
  {pages} {053835} (\bibinfo {year} {2018})}\BibitemShut {NoStop}%
\bibitem [{\citenamefont {Grünwald}(2019)}]{Grunwald_2019}%
  \BibitemOpen
  \bibfield  {author} {\bibinfo {author} {\bibfnamefont {P.}~\bibnamefont
  {Grünwald}},\ }\bibfield  {title} {\bibinfo {title} {Effective second-order
  correlation function and single-photon detection},\ }\href
  {https://doi.org/10.1088/1367-2630/ab3ae0} {\bibfield  {journal} {\bibinfo
  {journal} {New J. Phys.}\ }\textbf {\bibinfo {volume} {21}},\ \bibinfo
  {pages} {093003} (\bibinfo {year} {2019})}\BibitemShut {NoStop}%
\bibitem [{\citenamefont {Loudon}(1983)}]{Loudon_1983}%
  \BibitemOpen
  \bibfield  {author} {\bibinfo {author} {\bibfnamefont {R.}~\bibnamefont
  {Loudon}},\ }\href@noop {} {\emph {\bibinfo {title} {The Quantum Theory of
  Light}}},\ \bibinfo {edition} {2nd}\ ed.\ (\bibinfo  {publisher} {Clarendon
  Press, Oxford},\ \bibinfo {year} {1983})\BibitemShut {NoStop}%
\bibitem [{\citenamefont {Eberly}\ and\ \citenamefont
  {W\'{o}dkiewicz}(1977)}]{Eberly_Wodkiewicz_1977}%
  \BibitemOpen
  \bibfield  {author} {\bibinfo {author} {\bibfnamefont {J.~H.}\ \bibnamefont
  {Eberly}}\ and\ \bibinfo {author} {\bibfnamefont {K.}~\bibnamefont
  {W\'{o}dkiewicz}},\ }\bibfield  {title} {\bibinfo {title} {The time-dependent
  physical spectrum of light$\ast$},\ }\href
  {https://doi.org/10.1364/JOSA.67.001252} {\bibfield  {journal} {\bibinfo
  {journal} {J. Opt. Soc. Am.}\ }\textbf {\bibinfo {volume} {67}},\ \bibinfo
  {pages} {1252} (\bibinfo {year} {1977})}\BibitemShut {NoStop}%
\bibitem [{\citenamefont {Yamaguchi}\ \emph {et~al.}(2021)\citenamefont
  {Yamaguchi}, \citenamefont {Lyasota},\ and\ \citenamefont
  {Yuge}}]{Yamaguchi_etal_2021}%
  \BibitemOpen
  \bibfield  {author} {\bibinfo {author} {\bibfnamefont {M.}~\bibnamefont
  {Yamaguchi}}, \bibinfo {author} {\bibfnamefont {A.}~\bibnamefont {Lyasota}},\
  and\ \bibinfo {author} {\bibfnamefont {T.}~\bibnamefont {Yuge}},\ }\bibfield
  {title} {\bibinfo {title} {Theory of fano effect in cavity quantum
  electrodynamics},\ }\href {https://doi.org/10.1103/PhysRevResearch.3.013037}
  {\bibfield  {journal} {\bibinfo  {journal} {Phys. Rev. Research}\ }\textbf
  {\bibinfo {volume} {3}},\ \bibinfo {pages} {013037} (\bibinfo {year}
  {2021})}\BibitemShut {NoStop}%
\bibitem [{\citenamefont {Temnov}\ and\ \citenamefont
  {Woggon}(2009)}]{Temnov_Woggon_2009}%
  \BibitemOpen
  \bibfield  {author} {\bibinfo {author} {\bibfnamefont {V.~V.}\ \bibnamefont
  {Temnov}}\ and\ \bibinfo {author} {\bibfnamefont {U.}~\bibnamefont
  {Woggon}},\ }\bibfield  {title} {\bibinfo {title} {Photon statistics in the
  cooperative spontaneous emission},\ }\href
  {https://doi.org/10.1364/OE.17.005774} {\bibfield  {journal} {\bibinfo
  {journal} {Opt. Express}\ }\textbf {\bibinfo {volume} {17}},\ \bibinfo
  {pages} {5774} (\bibinfo {year} {2009})}\BibitemShut {NoStop}%
\bibitem [{\citenamefont {Kamide}\ \emph {et~al.}(2014)\citenamefont {Kamide},
  \citenamefont {Iwamoto},\ and\ \citenamefont {Arakawa}}]{Kamide_etal_2014}%
  \BibitemOpen
  \bibfield  {author} {\bibinfo {author} {\bibfnamefont {K.}~\bibnamefont
  {Kamide}}, \bibinfo {author} {\bibfnamefont {S.}~\bibnamefont {Iwamoto}},\
  and\ \bibinfo {author} {\bibfnamefont {Y.}~\bibnamefont {Arakawa}},\
  }\bibfield  {title} {\bibinfo {title} {Impact of the dark path on quantum dot
  single photon emitters in small cavities},\ }\href
  {https://doi.org/10.1103/PhysRevLett.113.143604} {\bibfield  {journal}
  {\bibinfo  {journal} {Phys. Rev. Lett.}\ }\textbf {\bibinfo {volume} {113}},\
  \bibinfo {pages} {143604} (\bibinfo {year} {2014})}\BibitemShut {NoStop}%
\bibitem [{\citenamefont {Vorberg}\ \emph {et~al.}(2015)\citenamefont
  {Vorberg}, \citenamefont {Wustmann}, \citenamefont {Schomerus}, \citenamefont
  {Ketzmerick},\ and\ \citenamefont {Eckardt}}]{Vorberg_etal_2015}%
  \BibitemOpen
  \bibfield  {author} {\bibinfo {author} {\bibfnamefont {D.}~\bibnamefont
  {Vorberg}}, \bibinfo {author} {\bibfnamefont {W.}~\bibnamefont {Wustmann}},
  \bibinfo {author} {\bibfnamefont {H.}~\bibnamefont {Schomerus}}, \bibinfo
  {author} {\bibfnamefont {R.}~\bibnamefont {Ketzmerick}},\ and\ \bibinfo
  {author} {\bibfnamefont {A.}~\bibnamefont {Eckardt}},\ }\bibfield  {title}
  {\bibinfo {title} {Nonequilibrium steady states of ideal bosonic and
  fermionic quantum gases},\ }\href
  {https://doi.org/10.1103/PhysRevE.92.062119} {\bibfield  {journal} {\bibinfo
  {journal} {Phys. Rev. E}\ }\textbf {\bibinfo {volume} {92}},\ \bibinfo
  {pages} {062119} (\bibinfo {year} {2015})}\BibitemShut {NoStop}%
\bibitem [{\citenamefont {Sakurai}\ and\ \citenamefont
  {Napolitano}(2011)}]{Sakurai_2011}%
  \BibitemOpen
  \bibfield  {author} {\bibinfo {author} {\bibfnamefont {J.~J.}\ \bibnamefont
  {Sakurai}}\ and\ \bibinfo {author} {\bibfnamefont {J.}~\bibnamefont
  {Napolitano}},\ }\href@noop {} {\emph {\bibinfo {title} {Modern Quantum
  Mechanics}}},\ \bibinfo {edition} {2nd}\ ed.\ (\bibinfo  {publisher}
  {Addison-Wesley},\ \bibinfo {year} {2011})\BibitemShut {NoStop}%
\end{thebibliography}%

\clearpage
\widetext

\newcommand{\dbexpval}[1]{\bigl\langle\mspace{-3mu}\bigl\langle {#1} \bigr\rangle\mspace{-3mu}\bigr\rangle}

\setcounter{section}{0}
\setcounter{equation}{0}
\setcounter{figure}{0}
\renewcommand{\thesection}{S~\Roman{section}}
\renewcommand{\thesubsection}{S~\Roman{section}~\Alph{subsection}}
\renewcommand{\theequation}{S\arabic{equation}}
\renewcommand{\thefigure}{S\arabic{figure}}

\begin{center}
  \textbf{\large Supplemental Material for
    \smallskip\\
    ``Superbunching in cathodoluminescence: a master equation approach''}
  \bigskip\\
  Tatsuro Yuge
  \\
  \textit{\small Department of Physics, Shizuoka University, Shizuoka 422-8529, Japan}
  \medskip\\
  Naoki Yamamoto~~and~~Takumi Sannomiya
  \\
  \textit{\small Department of Materials Science and Engineering,
    School of Materials and Chemical Technology, Tokyo Institute of Technology,
    4259 Nagatsuta, Midoriku, Yokohama 226-8503, Japan}
  \medskip\\
  Keiichirou Akiba
  \\
  \textit{\small Takasaki Advanced Radiation Research Institute,
    National Institutes for Quantum Science and Technology,
    1233 Watanuki, Takasaki, Gunma 370-1292, Japan}
\end{center}
\vspace{8truemm}

In this supplemental material,
we consider two models (called Models 2 and 3) other than the model (called Model 1), which is proposed and mainly discussed in the main text.
We compare the numerical results among these models.
Model 2 justifies the feature of Model 1
where only the fixed number $N$ of emitters are excited for each excitation event.
Model 3 provides a justification of Model 1 from a semi-microscopic point of view,
considering the excitation process.

\section{Model 1}

In Sec.~II in the main text, we propose the QME to describe cathodoluminescence (CL):
\begin{align}
  \frac{d}{dt} \hat{\rho}(t)
   & = \mathcal{L} \hat{\rho}(t),
  \tag{\ref{QME}}
  \label{QME_SM}
  \\
  \mathcal{L} \hat{\rho}
   & = \frac{1}{i\hbar} [\hat{H}, \hat{\rho}]
  + \mathcal{D}_{\mathrm{rad}}\mspace{2mu}\hat{\rho}
  + \mathcal{D}_{\mathrm{ex}}\mspace{2mu}\hat{\rho},
  \tag{\ref{Liouvillian}}
  \label{Liouvillian_SM}
  \\
  \hat{H}
   & = \sum_{j=1}^N \frac{\hbar \omega_\mathrm{e}}{2} \hat{\sigma}_j^z,
  \tag{\ref{Hamiltonian}}
  \label{Hamiltonian_SM}
  \\
  \mathcal{D}_{\mathrm{rad}}\mspace{2mu}\hat{\rho}
   & = \frac{1}{\tau_\mathrm{rad}} \sum_{j=1}^N
  \qty( \hat{\sigma}_j^- \hat{\rho} \hat{\sigma}_j^+
  - \frac{1}{2} \qty{ \hat{\sigma}_j^+ \hat{\sigma}_j^-, \hat{\rho} } ),
  \tag{\ref{QME_2nd_term}}
  \label{QME_2nd_term_SM}
  \\
  \mathcal{D}_{\mathrm{ex}}\mspace{2mu}\hat{\rho}
   & = \gamma \qty( \hat{\Pi}^+ \hat{\rho} \hat{\Pi}^-
  - \frac{1}{2} \qty{ \hat{\Pi}^- \hat{\Pi}^+, \hat{\rho} } ),
  \tag{\ref{QME_3rd_term}}
  \label{QME_3rd_term_SM}
  \\
  \hat{\Pi}^{\pm}
   & = \bigotimes_{j=1}^{N} \hat{\sigma}_j^{\pm},
  \tag{\ref{Lindblad_op_3rd_term}}
  \label{Lindblad_op_3rd_term_SM}
\end{align}
where $N$ is the number of emitters that is excited by a single incident electron,
$\tau_{\mathrm{rad}}$ is the radiative lifetime of each emitter,
and $\gamma$ is the excitation rate.
In this supplemental material, we refer to this QME model as Model 1.

\section{Model 2: Generalization in excitation number}

In Sec.~IV in the main text, we also introduce Model 2, a generalization of Model 1, to incorporate the effect that emitters excited may differ for each incident electron.
The resulting excitation term reads
\begin{align}
  \mathcal{D}_{\mathrm{ex}}^{(2)} \mspace{2mu} \hat{\rho}
   & = \gamma_{2} \sum_{N_{\mathrm{ex}}=1}^{N_\mathrm{tot}}
  \eta^{N_{\mathrm{ex}}} (1 - \eta)^{N_{\mathrm{tot}}-N_{\mathrm{ex}}}
  \mspace{-6mu} \sum_{j_1 < j_2 < ... < j_{N_{\mathrm{ex}}}} \mspace{-3mu}
  \qty( \hat{\Pi}_{j_1,j_2,...,j_{N_{\mathrm{ex}}}}^+ ~ \hat{\rho} ~ \hat{\Pi}_{j_1,j_2,...,j_{N_{\mathrm{ex}}}}^-
  - \frac{1}{2} \qty{ \hat{\Pi}_{j_1,j_2,...,j_{N_{\mathrm{ex}}}}^- \hat{\Pi}_{j_1,j_2,...,j_{N_{\mathrm{ex}}}}^+ , \hat{\rho}} ),
  \tag{\ref{model2_dissipator}}
  \label{model2_dissipator_SM}
\end{align}
where $N_{\mathrm{tot}}$ is the total number of emitters located within the excited region,
and each index in the second sum on the right-hand side runs from $1$ to $N_\mathrm{tot}$
with satisfying the constraint of $j_1 < j_2 < ... < j_{N_{\mathrm{ex}}}$.
As mentioned in the main text, Model 2 is obtained
by replacing $N$ with $N_\mathrm{tot}$ and $\mathcal{D}_{\mathrm{ex}} \hat{\rho}$ with $\mathcal{D}_{\mathrm{ex}}^{(2)} \hat{\rho}$ in Model 1.

\subsection{Relation between Models 1 and 2}

As mentioned in the main text, Models 1 and 2 are connected by the following relations:
\begin{align}
  N ~      & =~ \overline{N_{\mathrm{ex}}} = \eta N_{\mathrm{tot}},
  \tag{\ref{relation_1_2_N}}
  \label{relation_1_2_N_SM}
  \\
  \gamma ~ & =~ \gamma_2.
  \tag{\ref{relation_1_2_gamma}}
  \label{relation_1_2_gamma_SM}
\end{align}

To illustrate the above relation between Models 1 and 2,
we here compare the results of the numerical simulations of the models.
In the simulation of Model~2, we set the parameter values as
$\eta = N / N_{\mathrm{tot}}$ [from Eq.~\eqref{relation_1_2_N_SM}]
and $\gamma_{2} / \gamma = 1$  [from Eq.~\eqref{relation_1_2_gamma_SM}].
We fix the total number of emitters to $N_{\mathrm{tot}} = 10$ and
set $N$ (the number of emitters in Model 1) to an integer value (ranging from $N=2$ to $N=N_{\mathrm{tot}} - 1$).

\begin{figure}[tb]
  \centering
  \includegraphics[width=\linewidth]{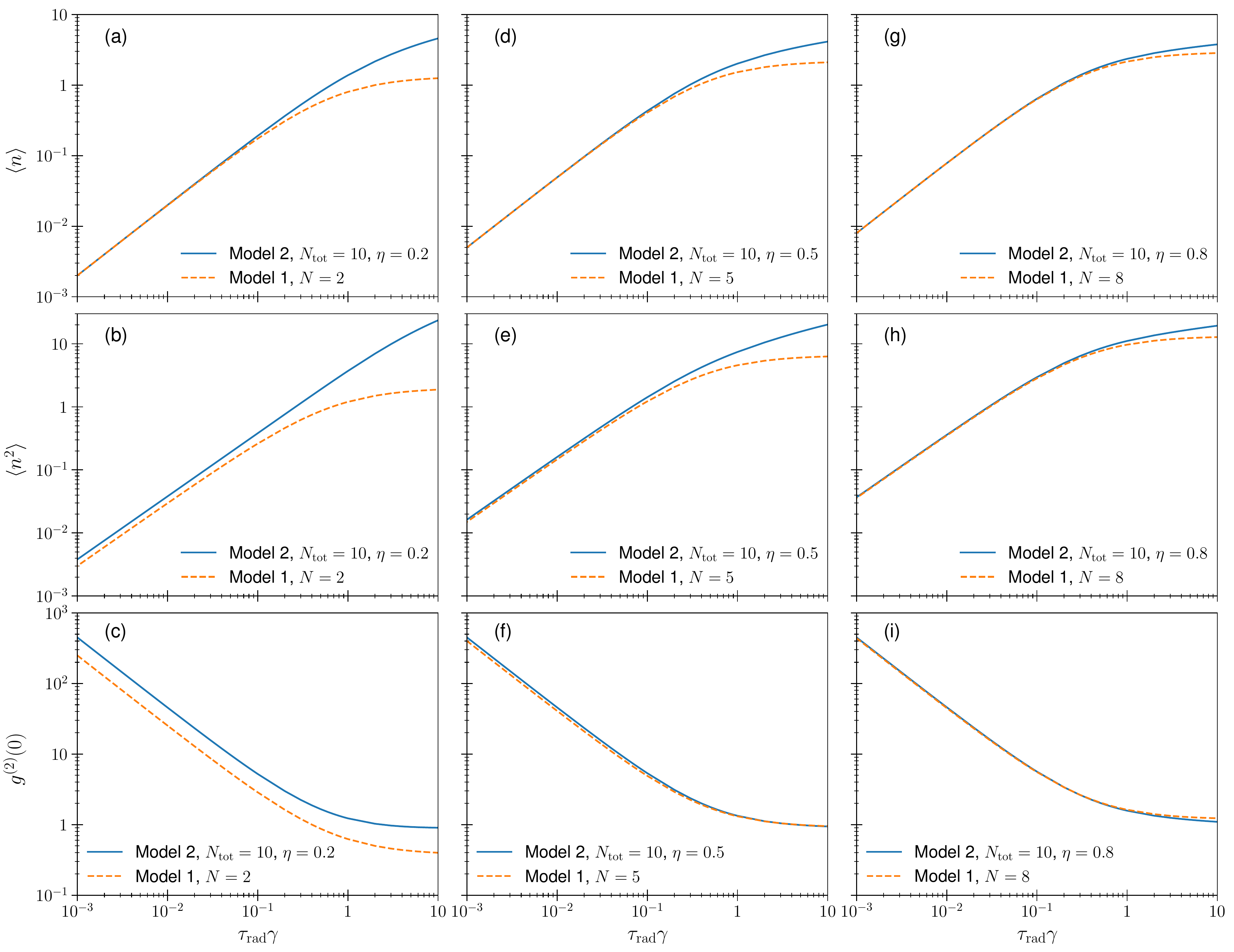}
  \caption{
    Comparison of the numerical results in Models 1 and 2.
    Left column [(a)--(c)]: $N=2$ in Model 1 and $N_{\mathrm{tot}} = 10$ in Model 2.
    Middle column [(d)--(f)]: $N=5$ in Model 1 and $N_{\mathrm{tot}} = 10$ in Model 2.
    Right column  [(g)--(i)]: $N=8$ in Model 1 and $N_{\mathrm{tot}} = 10$ in Model 2.
    The figures in the top [(a), (d), and (g)], middle [(b) (e), and (h)], and bottom [(c), (f), and (i)] are the $\gamma$-dependence of $\expval{\hat{n}}$, $\expval*{\hat{n}^2}$, and $g^{(2)}(0)$, respectively.
  }
  \label{fig:comparison_1_2}
\end{figure}

In Fig.~\ref{fig:comparison_1_2},
we plot the numerical results of $\expval{\hat{n}}$, $\expval*{\hat{n}^2}$, and $g^{(2)}(0)$ in Models 1 and 2 as functions of $\gamma$.
In the graphs in the left, middle, and right columns,
we show the results of Model 1 with $N=2$, $N=5$, and $N=8$, respectively,
and compare them with those of Model 2 with $N_{\mathrm{tot}} = 10$ and $\eta = 0.2$, $0.5$, and $0.8$, respectively.
In Fig.~\ref{fig:comparison_1_2} (a), (d), and (g),
we see that the result of $\expval{\hat{n}}$ in Model~2 is well approximated by Model~1
for $\tau_{\mathrm{rad}} \gamma \ll 1$.
As for $\expval*{\hat{n}^2}$ and $g^{(2)}(0)$ in Fig.~\ref{fig:comparison_1_2} (b), (c), (e), (f), (h), (i),
they show the similar $\gamma$-dependence for $\tau_{\mathrm{rad}} \gamma \ll 1$ in Models 1 and 2.
Moreover, although quantitative differences between the models are not small (in particular, the value of $g^{(2)}(0)$ for Model 2 is larger than that for Model 1) for $N=2$ ($\eta = 0.2$),
they get smaller as we set $N$ larger.

\begin{figure}[tb]
  \centering
  \includegraphics[width=.45\linewidth]{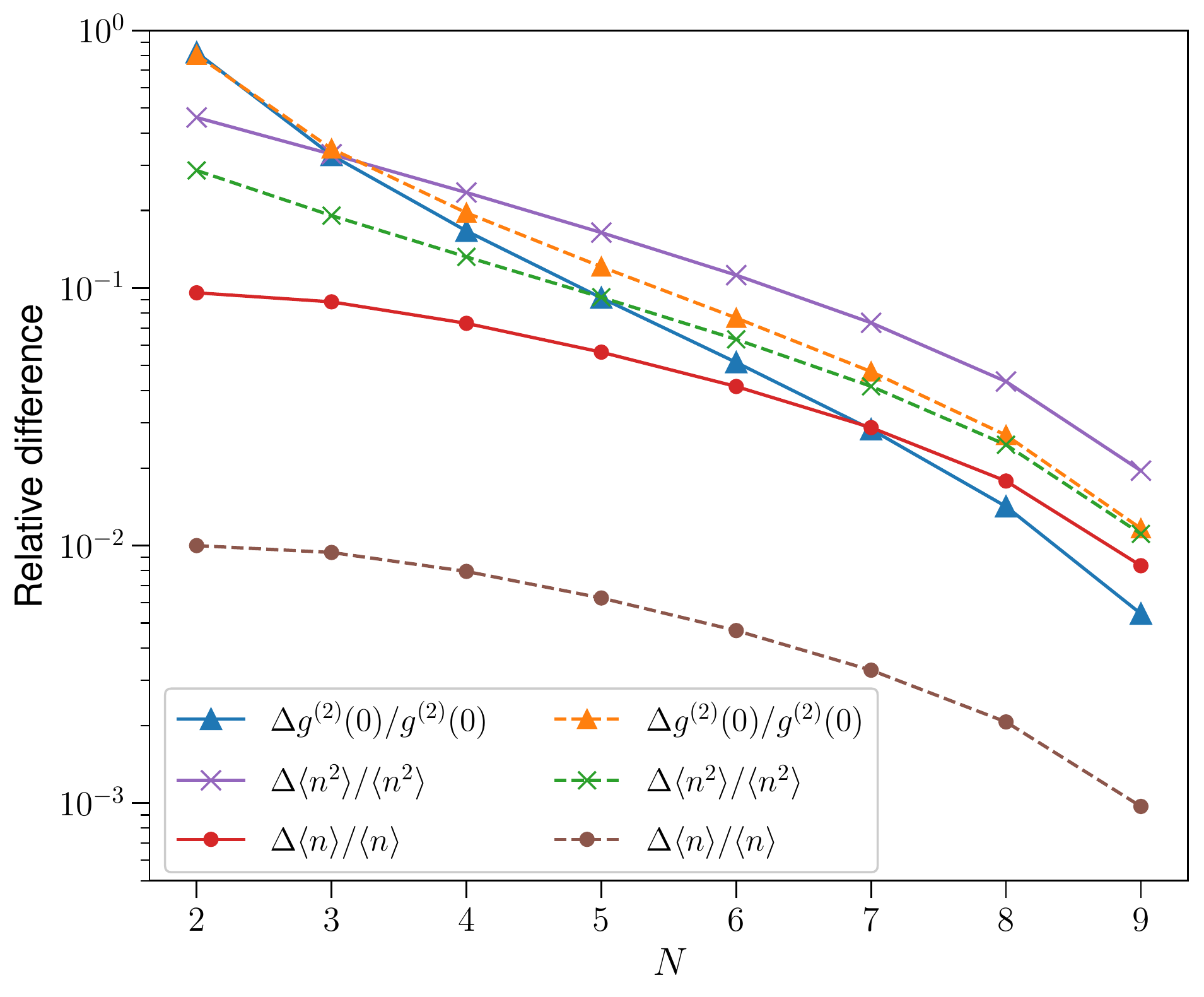}
  \caption{
    Relative differences between Model 1 ($N=2, 3, ..., 9$) and Model 2 ($N_\mathrm{tot} = 10$ and $\eta = N / N_\mathrm{tot}=0.2, 0.3,..., 0.9$) plotted against $N$.
    The circles, crosses, and triangles represent
    $\Delta \expval*{\hat{n}}_{\mathrm{ss}} / \expval*{\hat{n}}_{\mathrm{ss}}$,
    $\Delta \expval*{\hat{n}^2}_{\mathrm{ss}} / \expval*{\hat{n}^2}_{\mathrm{ss}}$,
    and $\Delta g^{(2)}(0)/ g^{(2)}_{\mathrm{model1}}(0)$, respectively.
    The symbols connected with the solid lines are the results for $\tau_{\mathrm{rad}} \gamma = 0.1$,
    and those with the dashed lines are for $\tau_{\mathrm{rad}} \gamma = 0.01$.
  }
  \label{fig:difference_1_2}
\end{figure}

In Fig.~\ref{fig:difference_1_2}, we can more clearly see this decreasing differences.
In this figure, we plot the $N$-dependence of the relative differences,
$\Delta \expval{\hat{n}}_{\mathrm{ss}} / \expval{\hat{n}}_{\mathrm{ss}}$,
$\Delta \expval*{\hat{n}^2}_{\mathrm{ss}} / \expval*{\hat{n}^2}_{\mathrm{ss}}$,
and $\Delta g^{(2)}(0)/ g^{(2)}_{\mathrm{model1}}(0)$.
Here, $\Delta \bullet$ is the difference of a quantity at the steady states of the two models;
e.g.,
$\Delta \expval{\hat{n}}_{\mathrm{ss}} = \big| \expval{\hat{n}}_{\mathrm{ss}} - \Tr \bigl[ \hat{\rho}_{\mathrm{ss}}^{[2]} \hat{n} \bigr] \big|$
where $\expval{\hat{n}}_{\mathrm{ss}}$ is the average number of excited emitters in the steady state in Model~1 [Eq.~(12) in the main text],
and $\Tr \bigl[ \hat{\rho}_{\mathrm{ss}}^{[2]} \hat{n} \bigr]$ is that in Model~2
($\hat{\rho}_{\mathrm{ss}}^{[2]}$ is the steady state of Model 2 and
$\hat{n} = \sum_{j=1}^{N_{\mathrm{tot}}} \hat{\sigma}_j^+ \hat{\sigma}_j^-$).
We divide the differences by the values of Model 1 to obtain the relative differences.
As seen in Fig.~\ref{fig:difference_1_2},
the differences between the models become smaller as we increase $N$.

To understand the origin of this behavior,
we note the relative fluctuation in the binomial distribution of the excitation number:
\begin{align}
  \frac{\sqrt{\overline{N_{\mathrm{ex}}^2} - \overline{N_{\mathrm{ex}}}^2}}{\overline{N_{\mathrm{ex}}}}
  = \sqrt{\frac{1 - \eta}{\eta N_{\mathrm{tot}}}}.
  \tag{\ref{relative_fluctuation_Nex}}
  \label{relative_fluctuation_Nex_SM}
\end{align}
This implies that the relative fluctuation of the number of emitters excited by an incident electron becomes smaller as $\eta = N / N_{\mathrm{tot}}$ increases.
Therefore, for larger $\eta$ (or $N$), we can assume that the excitation number is approximately the same single value, $\overline{N_{\mathrm{ex}}}$ ($=N$), for each electron.
That is, Model 2 is well approximated by Model 1 for larger $\eta$ (or $N$).

\begin{figure}[tb]
  \centering
  \includegraphics[width=.44\linewidth]{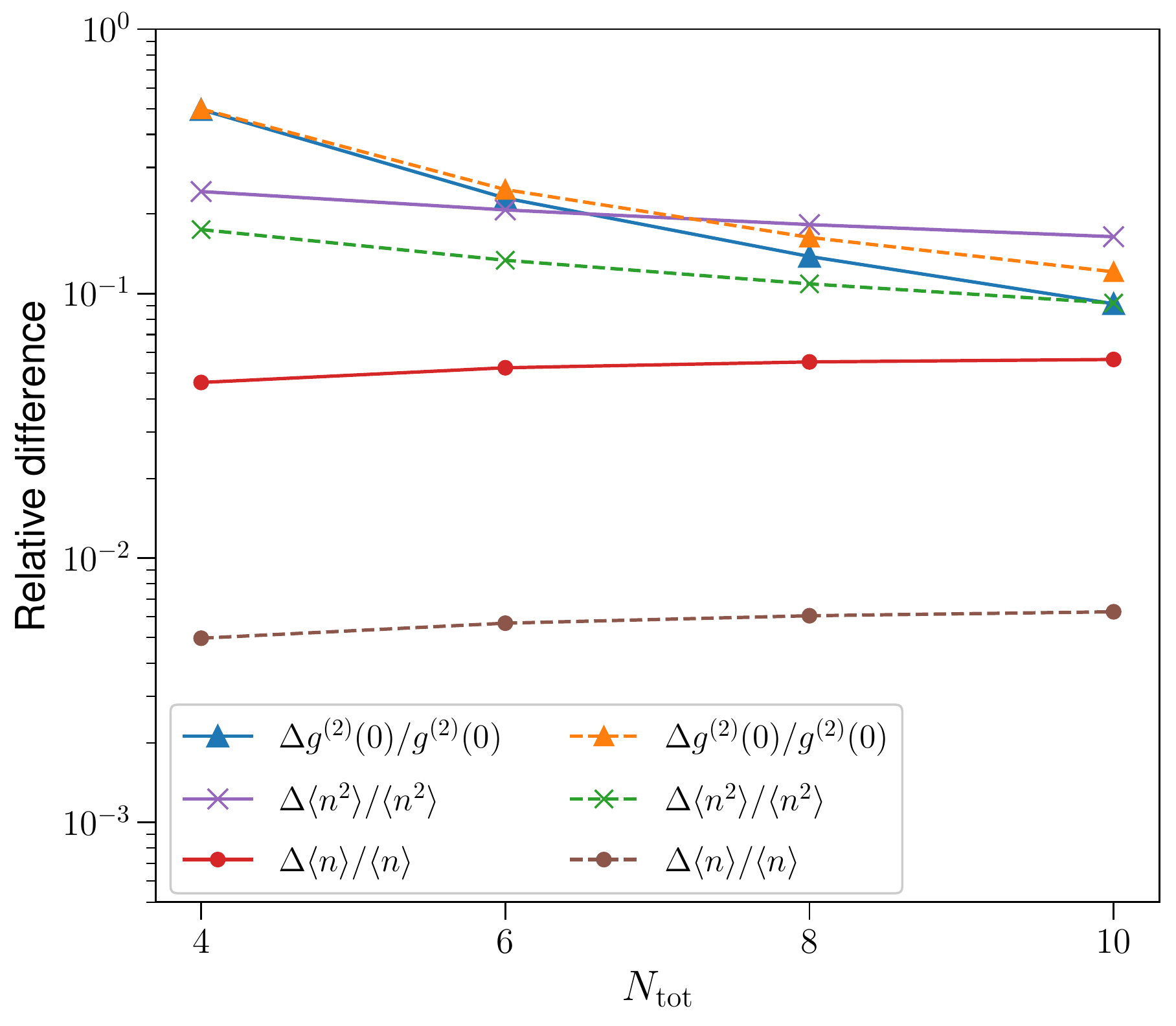}
  \quad
  \includegraphics[width=.44\linewidth]{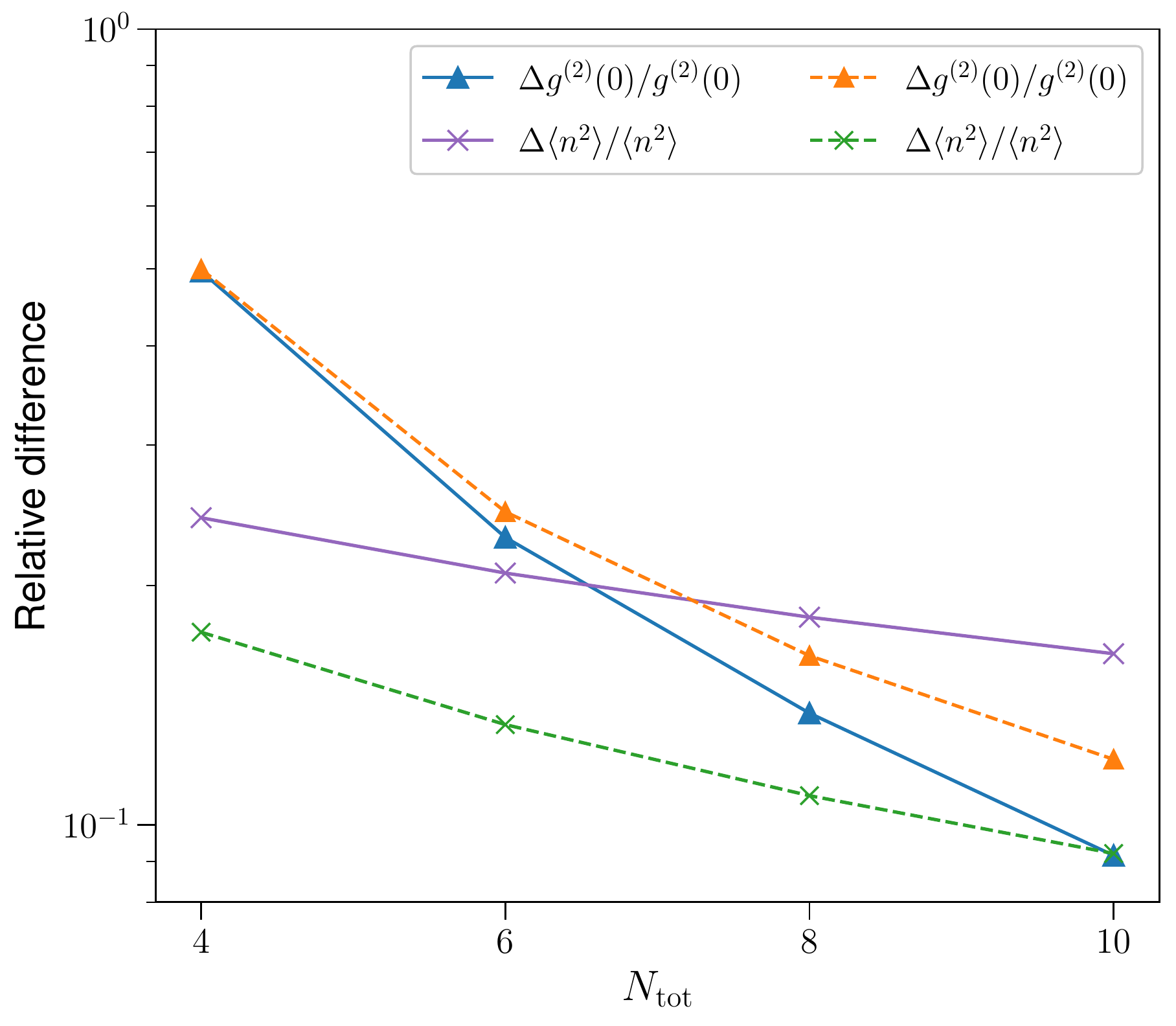}
  \caption{
    Relative differences between Model 1 ($N=2, 3, 4, 5$) and Model 2 ($\eta = 0.5$ and $N_\mathrm{tot} = N / \eta = 4, 6, 8, 10$) plotted against $N_\mathrm{tot}$.
    Left: the circles, crosses, and triangles represent
    $\Delta \expval*{\hat{n}}_{\mathrm{ss}} / \expval*{\hat{n}}_{\mathrm{ss}}$,
    $\Delta \expval*{\hat{n}^2}_{\mathrm{ss}} / \expval*{\hat{n}^2}_{\mathrm{ss}}$,
    and $\Delta g^{(2)}(0)/ g^{(2)}_{\mathrm{model1}}(0)$, respectively.
    The symbols connected with the solid lines are the results for $\tau_{\mathrm{rad}} \gamma = 0.1$,
    and those with the dashed lines are for $\tau_{\mathrm{rad}} \gamma = 0.01$.
    Right: a magnified plot for $\Delta \expval*{\hat{n}^2}_{\mathrm{ss}} / \expval*{\hat{n}^2}_{\mathrm{ss}}$
    and $\Delta g^{(2)}(0)/ g^{(2)}_{\mathrm{model1}}(0)$.
  }
  \label{fig:difference_1_2_2}
\end{figure}

Moreover, since the relative fluctuation in Eq.~\eqref{relative_fluctuation_Nex_SM} decreases as $N_{\mathrm{tot}}$ increases,
Model 2 should be well approximated by Model 1 also for larger $N_{\mathrm{tot}}$ with the same argument.
We numerically demonstrate this in Fig.~\ref{fig:difference_1_2_2},
where we plot the $N_{\mathrm{tot}}$-dependence of the relative differences for the case of $\eta$ fixed to $0.5$.
We see that $\Delta \expval*{\hat{n}^2}_{\mathrm{ss}} / \expval*{\hat{n}^2}_{\mathrm{ss}}$
and $\Delta g^{(2)}(0) / g^{(2)}_{\mathrm{model1}}(0)$ decreases as we increase $N_{\mathrm{tot}}$,
whereas $\Delta \expval*{\hat{n}^2}_{\mathrm{ss}} / \expval*{\hat{n}^2}_{\mathrm{ss}}$ is almost unchanged
(note that $\Delta \expval*{\hat{n}^2}_{\mathrm{ss}} / \expval*{\hat{n}^2}_{\mathrm{ss}}$ is small even for small $N_{\mathrm{tot}}$).

From these results, we conclude that Model 1 can well describe the features of Model 2 for lower excitation rate ($\tau_{\mathrm{rad}} \gamma \ll 1$) if $\eta$ or $N_{\mathrm{tot}}$ are large.

\subsection{Large excitation rate}

We also investigate the large $\tau_{\mathrm{rad}} \gamma$ region in Model 2.
As explained in the main text, Model 1 is valid only for $\tau_{\mathrm{rad}} \gamma \ll 1$
because the excitation term [Eq.~\eqref{QME_3rd_term_SM}] does not work well for large $\tau_{\mathrm{rad}} \gamma$.
In contrast, the excitation term [Eq.~\eqref{model2_dissipator_SM}] of Model 2 can work well also for the large $\tau_{\mathrm{rad}} \gamma$ region.

\begin{figure}[tb]
  \centering
  \includegraphics[width=.89\linewidth]{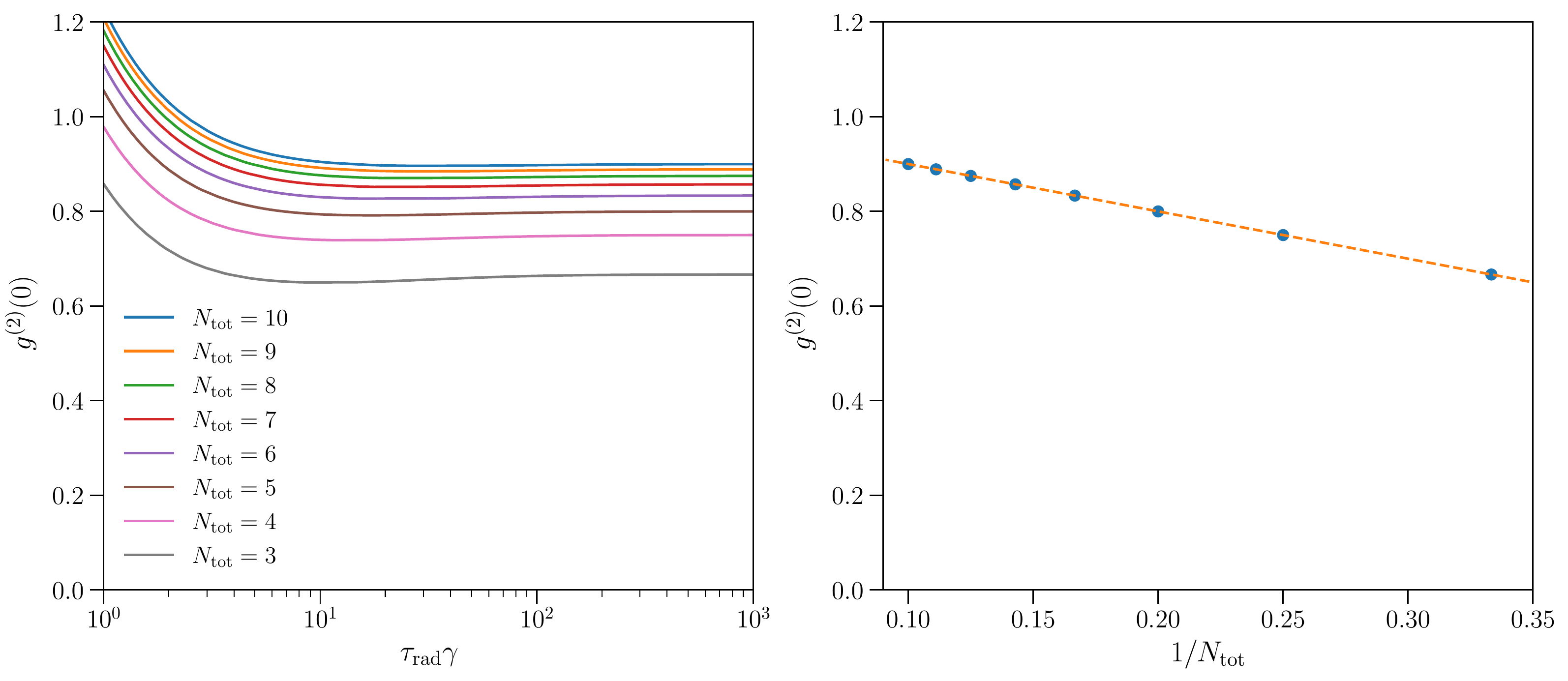}
  \caption{
  Left: $g^{(2)}(0)$ of Model 2 for large $\tau_{\mathrm{rad}} \gamma$.
  The curves from bottom to top show the results for $N_{\mathrm{tot}} = 3, 4, ..., 10$ (with $\eta = 2 / N_{\mathrm{tot}}$).
  Right: asymptotic value of $g^{(2)}(0)$ evaluated at $\tau_{\mathrm{rad}} \gamma = 10^3$, as a function of $1/N_{\mathrm{tot}}$.
  The dashed line is $1 - 1/N_{\mathrm{tot}}$.
  }
  \label{fig:model2_large_gamma}
\end{figure}

In Fig.~\ref{fig:model2_large_gamma}, we show $g^{(2)}(0)$ of Model 2 for large $\tau_{\mathrm{rad}} \gamma$.
As seen in the left panel, $g^{(2)}(0)$ for each $N_{\mathrm{tot}}$ approaches a value below one as $\tau_{\mathrm{rad}} \gamma$ becomes large.
And we find that
$\lim_{\tau_{\mathrm{rad}} \gamma \to \infty} g^{(2)}(0) = 1 - 1/N_{\mathrm{tot}}$,
as shown in the right panel.
This result of antibunching is consistent with that in the simulation in Ref.~\cite{Meuret_etal_2015} for a large beam current.
We can understand this antibunching behavior as follows:
since the emitters are excited more frequently than decaying in the large $\tau_{\mathrm{rad}} \gamma$ limit,
all the emitters are in the upper levels in most of the time, and the resulting photonic state is the Fock state of $N_{\mathrm{tot}}$ photons.

We note that the same antibunching behavior is also obtained in PL,
which is theoretically described by the QME obtained by replacing $\mathcal{D}_{\mathrm{ex}}^{(2)} \mspace{2mu} \hat{\rho}$ with
\begin{align}
  \mathcal{D}_{\mathrm{ex}}^{\mathrm{PL}} \mspace{2mu} \hat{\rho}
   & = \gamma_{\mathrm{PL}} \sum_{j=1}^{N_{\mathrm{tot}}}
  \qty( \hat{\sigma}_j^+ \hat{\rho} \hat{\sigma}_j^-
  - \frac{1}{2} \qty{ \hat{\sigma}_j^- \hat{\sigma}_j^+, \hat{\rho} } ),
  \label{CL_excitation}
\end{align}
where $\gamma_{\mathrm{PL}}$ is the excitation rate of PL.
This implies that the situation in large-current CL is almost the same as that in PL.
And the experimental result that $g^{(2)}(0)$ approaches one for large-current CL and that $g^{(2)}(0) \approx 1$ for PL \cite{Meuret_etal_2015} is understood from this (asymptotic) antibunching behavior with large number of emitters
(where $1 / N_{\mathrm{tot}}$ is smaller than the measurement noise level for $g^{(2)}$).

\section{Model 3: Higher-order quantum master equation}

In this section, we start from a semi-microscopic model
where an interaction term emulates the process that an electron, randomly incident on the sample, randomly excites the emitters while traveling through the sample.
Based on this model, we derive a QME of Model 3 that includes terms of multiple emitter excitation.

We numerically demonstrate that Model 3 is approximately described by Model 2.
Therefore, with the result in the previous section, Model 3 provides a semi-microscopic justification of Model 1.

\subsection{Setup}

As in Model 2, we consider $N_\mathrm{tot}$ emitters located within the excited region.
We model the effect of electron beam on the emitters by a random noise,
instead of treating detailed elementary processes via the secondary (quasi-)particles (which is why we call the starting model ``semi-microscopic'' model).
The Hamiltonian of this model of $N_\mathrm{tot}$ emitters reads
\begin{align}
  \hat{H}_{\mathrm{tot}}(t)
   & = \hat{H}_S + \hat{H}_\mathrm{ex}(t),
  \label{H_tot}
  \\
  \hat{H}_S
   & = \sum_{j=1}^{N_\mathrm{tot}} \frac{\hbar \omega_e}{2} \hat{\sigma}_j^z,
  \label{H_S}
  \\
  \hat{H}_\mathrm{ex}(t)
   & = \hbar \xi(t) \sum_{j=1}^{N_\mathrm{tot}} \bigl[ g_j(t) \hat{\sigma}_j^+ + g_j^*(t) \hat{\sigma}_j^- \bigr].
  \label{H_ex}
\end{align}
The random noise is described by the two types of stochastic processes, $\xi(t)$ and $g_j(t)$, as explained below.

\begin{figure}[b]
  \centering
  \includegraphics[width=0.6\textwidth]{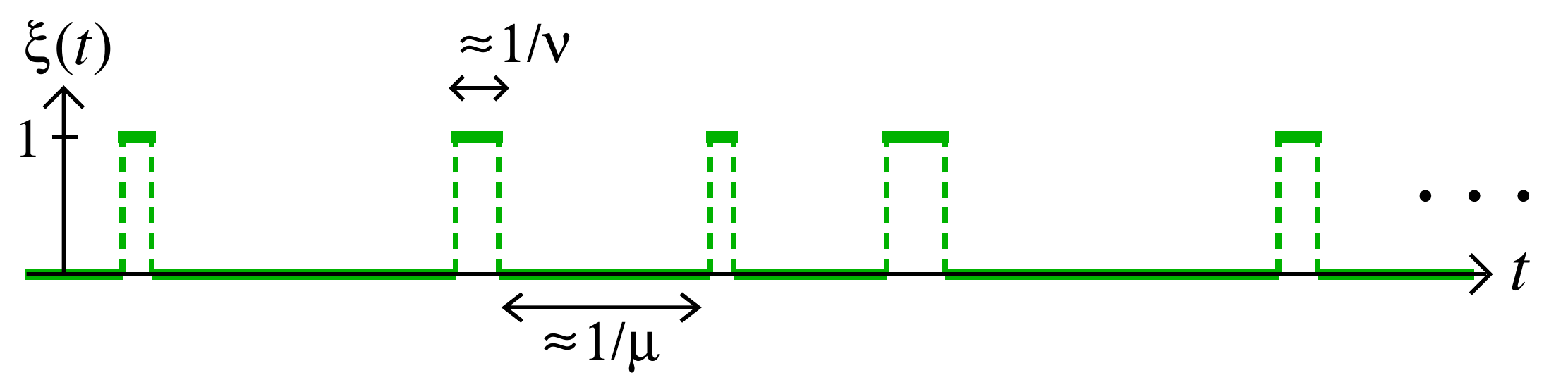}
  \caption{A schematic picture of random telegraph noise}
  \label{fig:RTN}
\end{figure}

$\xi(t)$ is a random telegraph noise (RTN).
$\xi(t)$ can have either of two dimensionless values 1 and 0,
which represent whether an incoming electron exists near the sample or not
($1=$``exist'' and $0=$``not exist'').
We write the unit-time switching probabilities as
\begin{align}
  p(0 \to 1) & = \mu,
  \\
  p(1 \to 0) & = \nu.
\end{align}
As seen in Fig.~\ref{fig:RTN},
this RTN represents a sequence of random pulses whose average duration time is $1/\nu$
and average inter-pulse time is $1/\mu$.
We note that $1/\nu$ corresponds to the time during which an incident electron passes through the sample and $\mu$ to the number of incident electrons per unit time, $I/e$.
Therefore, in the situation of CL, it is reasonable to assume
\begin{align}
  \mu \ll \nu.
\end{align}

On the other hand, we assume that $g_j(t)$, which corresponds to the noise strength on the $j$th emitter, is a stationary complex Gaussian noise whose average and correlations are
\begin{align}
   & \expval{g_j(t)}_g = 0,
  \label{noise_amplitude_avg_1}
  \\
   & \expval{g_{j_1}(t_1) g_{j_2}^*(t_2)}_g = \delta_{j_1,j_2} \mathfrak{g}^2 e^{-\nu |t_1 - t_2|},
  \label{noise_amplitude_avg_2}
  \\
   & \expval{g_{j_1}(t_1) g_{j_2}(t_2)}_g = 0,
  \label{noise_amplitude_avg_3}
\end{align}
where $\mathfrak{g}$ is a positive constant.
In Eq.~\eqref{noise_amplitude_avg_2}, we assumed that the correlation time $1/\nu$ is the same as the average duration time of a pulse in the RTN $\xi(t)$,
so that the amplitude of a pulse is almost surely uncorrelated with those of the other pulses.
As seen in the next subsection, the QME of Model 3 has a form of power series expansion in terms of $(\mathfrak{g}/\nu)^2$.
Therefore, we also assume
\begin{align}
  \mathfrak{g} \ll \nu.
\end{align}

In the following, we denote the averages of $\xi$ and $\{g_j\}$ by $\expval{\cdots}_\xi$ and $\expval{\cdots}_g$, respectively.
We also define $\dbexpval{\cdots} \equiv \expval*{\expval{\cdots}_\xi}_g$.

\subsection{QME of Model 3}

As we will explain in \S\S~III~D in detail,
using a higher-order perturbation expansion with respect to $\hat{H}_\mathrm{ex}$
together with a Markov approximation and a first-order approximation with respect to $\mu / \nu$,
we have a Markovian QME that describes the time evolution of the noise-averaged system state:
\begin{align}
  \frac{d}{dt} \hat{\rho}(t)
   & = \frac{1}{i\hbar} [\hat{H}_S, \hat{\rho}(t)]
  + \mathcal{D}_{\mathrm{rad}} \mspace{2mu} \hat{\rho}(t)
  + \mathcal{D}_{\mathrm{ex}}^{(3)} \mspace{2mu} \hat{\rho}(t),
\end{align}
where $\mathcal{D}_{\mathrm{rad}}$ is the same as that in Model 2
(i.e., Eq.~\eqref{QME_2nd_term_SM} with the replacement $N \to N_\mathrm{tot}$),
and $\mathcal{D}_{\mathrm{ex}}^{(3)}$ is given by
\begin{align}
  \mathcal{D}_{\mathrm{ex}}^{(3)}
   & \equiv \mu \sum_{N_{\mathrm{ex}}=1}^\infty
  \qty(-\frac{\mathfrak{g}^2}{\nu^2})^{\mspace{-3mu}N_{\mathrm{ex}}}
  \sum_{j_1,j_2,...,j_{N_{\mathrm{ex}}}=1}^{N_\mathrm{tot}}
  \mspace{4mu}
  \sum_{s_1,s_2,...,s_{N_{\mathrm{ex}}}=\pm}
  \mspace{4mu}
  \sum_{(b,\tilde{b})} \frac{1}{D(\alpha)}
  \mspace{3mu}
  \sideset{}{'}\prod_{m=1}^{2N_{\mathrm{ex}}}
  L_{j_{\beta(m)}}^{\alpha(m) s_{\beta(m)}}.
  \label{model3}
\end{align}
Here, $\alpha$ is a function of $m \in \{ 1, 2, ..., 2N_{\mathrm{ex}} \}$ that can have either of the two signs $+$ and $-$.
The superscript $\alpha(m) s_{\beta(m)}$ of $L_{j_{\beta(m)}}^{\alpha(m) s_{\beta(m)}}$
means the multiplication of the signs of $\alpha(m)$ and $s_{\beta(m)}$.
$D(\alpha)$ is a positive constant that depends on $\alpha$.
And $\beta$ is a function of $m \in \{ 1, 2, ..., 2N_{\mathrm{ex}} \}$ that can have one of $\{ 1, 2, ..., N_{\mathrm{ex}} \}$.
The superoperator $L_j^\pm$ is defined as:
\begin{align}
  L_j^\pm \hat{A} = [\hat{\sigma}_j^\pm, \hat{A}].
  \label{L_j_pm}
\end{align}
In the product $\prod_m'$,
the $2N$ superoperators are arranged in ascending order of $m$.
Detailed definitions of $\alpha$, $\beta$, $\sum_{(b,\tilde{b})}$, and $D(\alpha)$
are given in \S\S~III~D.

For $N_{\mathrm{ex}} \ge 2$, $\mathcal{D}_{\mathrm{ex}}^{(3)} \hat{\rho}$ contains
terms of the following form:
\begin{align}
  \mu \qty(\frac{\mathfrak{g}^2}{\nu^2})^{\mspace{-3mu}N_{\mathrm{ex}}}
  \mspace{3mu}
  C_{N_{\mathrm{ex}}}
  \sum_{j_1 < j_2 < ... < j_{N_{\mathrm{ex}}}}
  \hat{\Pi}_{j_1,j_2,...,j_{N_{\mathrm{ex}}}}^+ \mspace{3mu}
  \hat{\rho} \mspace{5mu} \hat{\Pi}_{j_1,j_2,...,j_{N_{\mathrm{ex}}}}^-,
  \label{multiple_excitation}
\end{align}
where $C_{N_{\mathrm{ex}}}$ is determined from $D(\alpha)$ and the number of terms of this form.
This expresses that an incident electron in Model 3 can excite multiple emitters.
Since $\mu$ represents the number of incoming electrons per unit time
and
$(\mathfrak{g}^2 / \nu^2)^{N_{\mathrm{ex}}} C_{N_{\mathrm{ex}}}$
corresponds to the probability of $N_{\mathrm{ex}}$-emitter excitation by a single incoming electron,
we can interpret the prefactor $\mu (\mathfrak{g}^2 / \nu^2)^{N_{\mathrm{ex}}} C_{N_{\mathrm{ex}}}$
as the excitation rate $\gamma_{N_{\mathrm{ex}}}$.

\subsection{Comparison of Models 2 and 3}

For the situation where bunching is observed in CL,
$\tau_{\mathrm{rad}} \gamma_{N_{\mathrm{ex}}} \ll 1$ (for $N_{\mathrm{ex}} \ge 2$) should be satisfied
similarly to $\tau_{\mathrm{rad}} \gamma \ll 1$ in Model 1.
In this case, almost all the emitters are in the lower levels for most of the time,
so that the term of the form in Eq.~\eqref{multiple_excitation} is expected to be dominant
for each $N_{\mathrm{ex}}$ in $\mathcal{D}_{\mathrm{ex}}^{(3)}$.
We thus expect that $\mathcal{D}_{\mathrm{ex}}^{(3)}$ (Model 3)
can be approximately described by $\mathcal{D}_{\mathrm{ex}}^{(2)}$ (Model 2).

This argument is rather heuristic.
Here, to illustrate this expected approximate relation between Models 2 and 3,
we compare the results of their numerical simulations.

In the simulation of Model 3,
we truncate the first sum in Eq.~\eqref{model3} to $N_{\mathrm{tot}}$
($\sum_{N_{\mathrm{ex}}=1}^\infty \to \sum_{N_{\mathrm{ex}}=1}^{N_{\mathrm{tot}}}$)
and fix the value of $(\mathfrak{g} / \nu)^2$ to $(\mathfrak{g} / \nu)^2 = 0.04$.
On the other hand, we adjust the parameters of Model 2
to see if Model 2 can reasonably approximate more realistic Model 3.
To do so, we determine the values of $\gamma_2 / \mu$ and $\eta$
by the optimization to minimize the total relative difference of $\expval{\hat{n}}$, $\expval*{\hat{n}^2}$, and $g^{(2)}(0)$:
\begin{align}
  \frac{\qty| \Tr \bigl[ \hat{\rho}_{\mathrm{ss}}^{[2]} \hat{n} \bigr] - \Tr \bigl[ \hat{\rho}_{\mathrm{ss}}^{[3]} \hat{n} \bigr] |}{\Tr \bigl[ \hat{\rho}_{\mathrm{ss}}^{[2]} \hat{n} \bigr]}
  + \frac{\qty| \Tr \bigl[ \hat{\rho}_{\mathrm{ss}}^{[2]} \hat{n}^2 \bigr] - \Tr \bigl[ \hat{\rho}_{\mathrm{ss}}^{[3]} \hat{n}^2 \bigr] |}{\Tr \bigl[ \hat{\rho}_{\mathrm{ss}}^{[2]} \hat{n}^2 \bigr]}
  + \frac{\qty| g^{(2)}_{\mathrm{Model 2}}(0) - g^{(2)}_{\mathrm{Model 3}}(0) |}{g^{(2)}_{\mathrm{Model 2}}(0)},
  \label{relative_diff}
\end{align}
where $\mu$ in Model 3 is fixed to a specific value in the optimization
(we use $\mu = 0.04 / \tau_{\mathrm{rad}}$).
We set the total number of emitters as $N_{\mathrm{tot}} = 5$.
In this case, we find that the optimized values are $\gamma_2 / \mu = 0.346$ and $\eta = 0.105$.

\begin{figure}[tb]
  \centering
  \includegraphics[width=.465\linewidth]{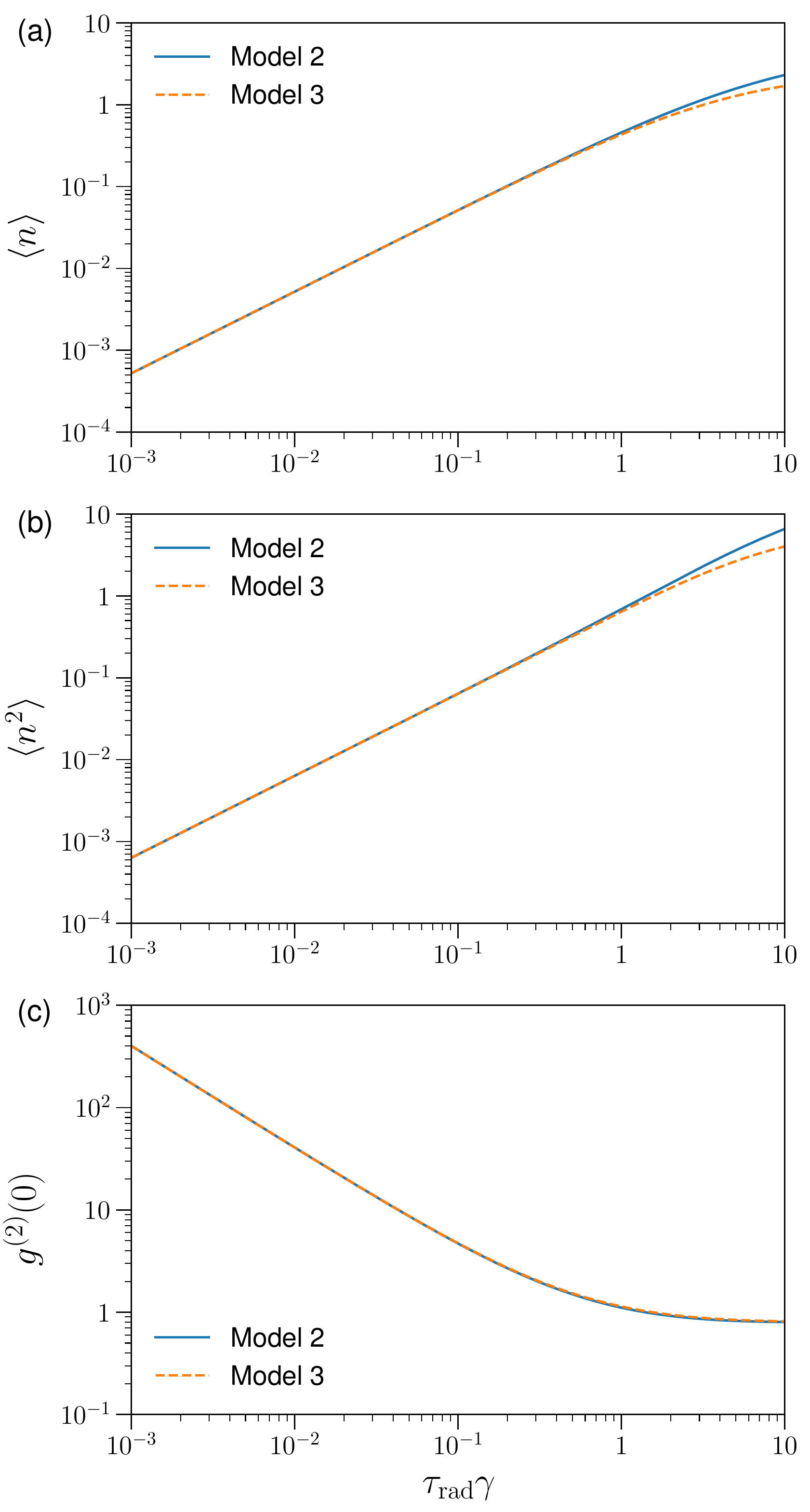}
  \hfill
  \includegraphics[width=.515\linewidth]{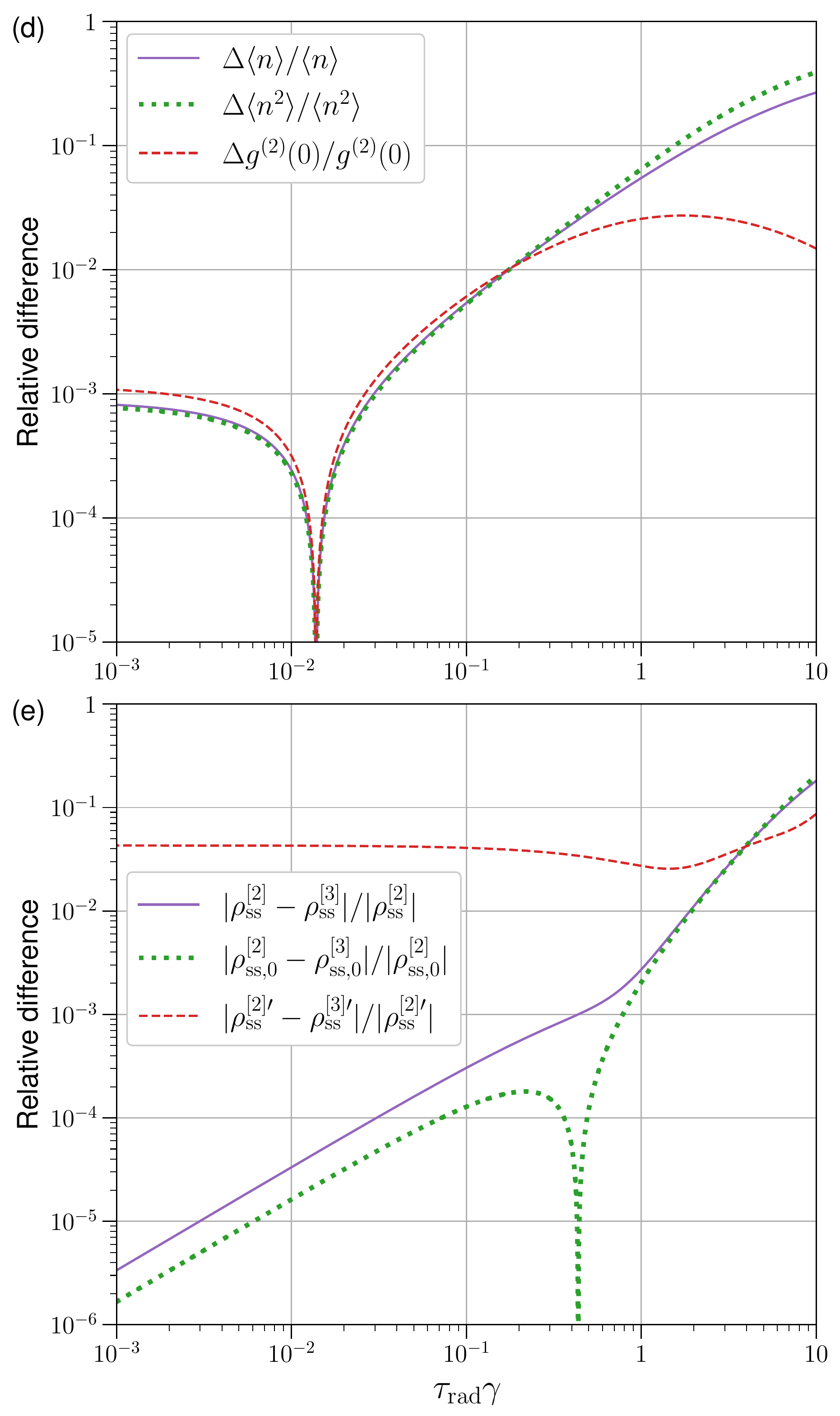}
  \caption{
  Comparison of Models 2 and 3.
  Left: (a), (b), and (c) are $\gamma$-dependence of the steady-state values of $\hat{n}$,  $\hat{n}^2$, and $g^{(2)}(0)$, respectively. The results of Models 2 and 3 are plotted by the solid and dashed curves, respectively.
  (d): the relative differences of the results in (a) (solid curve), (b) (dotted curve), and (c) (dashed curve).
  (e): the relative differences of density matrices at the steady state.
  The solid curve is the difference of the full density matrices $\hat{\rho}^{[\bullet]}_{\mathrm{ss}}$,
  the dotted curve is that of the populations $\rho^{[\bullet]}_{\mathrm{ss},0}$ of $\ket{0,0,...,0}$,
  and the dashed curve is that of $\hat{\rho}^{[\bullet]\prime}_{\mathrm{ss}} = \hat{\rho}^{[\bullet]}_{\mathrm{ss}} - \rho^{[\bullet]}_{\mathrm{ss},0} \ket{0,0,...,0} \bra{0,0,...,0}$.
  The total number of emitters is set to $N_{\mathrm{tot}} = 5$ in both Models 2 and 3.
  The other parameter values are:
  $\gamma_2 = \gamma$ and $\eta = 0.105$ for Model 2,
  and $\mu = (1/0.346) \gamma$ and $(\mathfrak{g} / \nu)^2 = 0.04$ for Model 3.
  }
  \label{fig:comparison_2_3}
\end{figure}

In the left of Fig.~\ref{fig:comparison_2_3} [(a)--(c)],
we plot the numerical results of $\expval{\hat{n}}$, $\expval*{\hat{n}^2}$, and $g^{(2)}(0)$ in Models 2 and 3 as functions of $\gamma~(=\gamma_2)$, where the optimized parameters are used.
We see that the results of Model 3 are well approximated by those of Model 2 for $\tau_{\mathrm{rad}} \gamma \lesssim 1$.

To see the closeness of the results more clearly,
we show the relative differences of these quantities in Fig.~\ref{fig:comparison_2_3} (d),
where the solid, dotted, and dashed curves corresponds to the individual terms in \eqref{relative_diff}, respectively.
At $\tau_{\mathrm{rad}} \gamma = 0.0138~(=0.04 \times 0.346)$, where $\gamma_2 / \mu$ and $\eta$ are optimized,
the differences take the minimum values that are less than $10^{-5}$.
Even in other regions of $\gamma$, if $\tau_{\mathrm{rad}} \gamma \ll 1$ is satisfied,
the relative differences remain small (less than $10^{-2}$ for $\tau_{\mathrm{rad}} \gamma < 0.1$).

Furthermore, we investigate the closeness of the steady states (density matrices) in Models 2 and 3.
The solid curve in Fig.~\ref{fig:comparison_2_3} (e) shows the relative difference of the steady states:
\begin{align}
  \frac{\qty| \hat{\rho}^{[2]}_{\mathrm{ss}} - \hat{\rho}^{[3]}_{\mathrm{ss}} |}{\qty| \hat{\rho}^{[2]}_{\mathrm{ss}} |},
  \label{rel_diff_states}
\end{align}
where $| \cdots |$ stands for the Frobenius norm
(i.e., $\bigl| \hat{A} \bigr| = \sqrt{\mathrm{Tr}[\hat{A}^\dag \hat{A}]}$ ),
and $\hat{\rho}^{[2]}_{\mathrm{ss}}$ and $ \hat{\rho}^{[3]}_{\mathrm{ss}}$ are the steady states for Models 2 and 3, respectively.
We see that the relative difference is less than $10^{-3}$ for $\tau_{\mathrm{rad}} \gamma < 0.1$.
In the steady states for $\tau_{\mathrm{rad}} \gamma \ll 1$,
almost all the emitters are in their lower levels:
the population $\rho^{[\bullet]}_{\mathrm{ss},0} = \bra{0,0,...,0} \hat{\rho}^{[\bullet]}_{\mathrm{ss}} \ket{0,0,...,0}$ should be near to one,
and the other matrix elements in $\hat{\rho}^{[2]}_{\mathrm{ss}}$ and $\hat{\rho}^{[3]}_{\mathrm{ss}}$ should be near to zero.
Therefore, one might think that the smallness of the relative difference in Eq.~\eqref{rel_diff_states} is determined only by the population $\rho^{[\bullet]}_{\mathrm{ss},0}$.
To investigate this effect, we also calculate the relative difference of the population $\rho^{[\bullet]}_{\mathrm{ss},0}$
and that of $\hat{\rho}^{[\bullet]\prime}_{\mathrm{ss}} \equiv \hat{\rho}^{[\bullet]}_{\mathrm{ss}} - \rho^{[\bullet]}_{\mathrm{ss},0} \ket{0,0,...,0} \bra{0,0,...,0}$, separately.
The dotted and dashed curves in Fig.~\ref{fig:comparison_2_3} (e) show the relative differences of $\rho^{[\bullet]}_{\mathrm{ss},0}$ and $\hat{\rho}^{[\bullet]\prime}_{\mathrm{ss}}$, respectively.
We see that indeed the extreme smallness (less than $10^{-3}$ for $\tau_{\mathrm{rad}} \gamma < 0.1$) of the relative difference is due to the population $\rho^{[\bullet]}_{\mathrm{ss},0}$,
but the difference of $\hat{\rho}^{[\bullet]\prime}_{\mathrm{ss}}$ is also small (less than $5 \times 10^{-2}$ for $\tau_{\mathrm{rad}} \gamma < 0.1$).

From these results, we conclude that we can use Model 2 as an approximation of Model 3,
not only for the quantities [$\expval{\hat{n}}$, $\expval*{\hat{n}^2}$, and $g^{(2)}(0)$]  but also for the steady state.

\subsection{Derivation of Model 3}

\subsubsection{Nakajima-Zwanzig equation}

We start from the von Neumann equation in the interaction picture:
\begin{align}
  \frac{d}{dt}\breve{\rho}(t)
   & = \breve{\mathcal{L}}_{\mathrm{ex}}(t) \breve{\rho}(t),
  \label{vonNeumannEq_int_pic}
  \\
  \breve{\mathcal{L}}_{\mathrm{ex}}(t) \hat{A}
   & \equiv \frac{1}{i\hbar} \bigl[ \breve{H}_{\mathrm{ex}}(t), \hat{A} \bigr],
\end{align}
where
$\breve{\rho}(t) = e^{-\hat{H}_S t / i\hbar} \hat{\rho}(t) e^{\hat{H}_S t / i\hbar}$
with $\hat{\rho}(t)$ being the density matrix of the system in the Schr\"odinger picture,
and
\begin{align}
  \breve{H}_{\mathrm{ex}}(t) = e^{-\hat{H}_S t / i\hbar} \hat{H}_{\mathrm{ex}}(t) e^{\hat{H}_S t / i\hbar}.
  \label{H_ex_int_pic}
\end{align}

To derive the QME of Model 3, which includes higher-order terms,
we first use the Nakajima-Zwanzig projection operator method \cite{Breuer_Petruccione_2002,Nakajima_1958,Zwanzig_1960}.
We define the projection superoperator $\mathcal{P}$ as
\begin{align}
  \mathcal{P} \hat{A} = \dbexpval{\hat{A}}
\end{align}
and also define the complementary projection superoperator $\mathcal{Q} = 1 - \mathcal{P}$.
Then, by applying the standard projection operator method to the von Neumann equation~\eqref{vonNeumannEq_int_pic},
we obtain the Nakajima-Zwanzig equation:
\begin{align}
  \frac{d}{dt} \mathcal{P}\breve{\rho}(t)
  = \int_{t_0}^t ds~\mathcal{K}(t,s) \mathcal{P}\breve{\rho}(s),
  \label{Nakajima_Zwanzig}
\end{align}
where
\begin{align}
  \mathcal{K}(t,s)
   & \equiv \mathcal{P} \breve{\mathcal{L}}_{\mathrm{ex}}(t) \mathcal{G}(t,s) \mathcal{Q} \breve{\mathcal{L}}_{\mathrm{ex}}(s) \mathcal{P},
  \label{kernel_NZ}
  \\
  \mathcal{G}(t,s)
   & \equiv \mathcal{T} \exp \qty[ \int_s^t ds' \mathcal{Q} \breve{\mathcal{L}}_{\mathrm{ex}}(s')],
  \label{propagator_QL}
\end{align}
and $\mathcal{T}$ is the chronological time ordering.
In deriving Eq.~\eqref{Nakajima_Zwanzig}, we used that the initial state is independent of the noise ($\mathcal{Q} \breve{\rho}(t_0) = 0$)
and that $\dbexpval{\hat{H}_{\mathrm{ex}}(t)}$ vanishes due to Eq.\eqref{noise_amplitude_avg_1}.

We next note that $\mathcal{G}(t,s)$ in Eq.~\eqref{propagator_QL} is explicitly expressed in a series of $\mathcal{Q} \breve{\mathcal{L}}_{\mathrm{ex}}$:
\begin{align}
  \mathcal{G}(t,s)
   & = 1 + \sum_{k=1}^\infty \mathcal{G}_k(t,s),
  \\
  \mathcal{G}_k(t,s)
   & \equiv \int_s^t \mspace{-2mu} dt_1 \int_s^{t_1} \mspace{-2mu} dt_2 \cdots \int_s^{t_{k-1}} \mspace{-4mu} dt_k
  \mspace{2mu}
  \mathcal{Q} \breve{\mathcal{L}}_{\mathrm{ex}}(t_1) \mathcal{Q} \breve{\mathcal{L}}_{\mathrm{ex}}(t_2) \cdots \mathcal{Q} \breve{\mathcal{L}}_{\mathrm{ex}}(t_k).
\end{align}
Then, we can rewrite the right-hand side of the Nakajima-Zwanzig equation \eqref{Nakajima_Zwanzig} as
\begin{align}
  \int_{t_0}^t ds~\mathcal{K}(t,s) \mathcal{P}\breve{\rho}(s)
   & = \sum_{k=2}^\infty \breve{\mathcal{R}}_k(t,s) \mathcal{P}\breve{\rho}(s),
  \\
  \breve{\mathcal{R}}_k(t,s) \mathcal{P}\breve{\rho}(s)
   & \equiv \int_{t_0}^t \mspace{-2mu} dt_1 \int_{t_0}^{t_1} \mspace{-2mu} dt_2 \cdots
  \int_{t_0}^{t_{k-3}} \mspace{-5mu} dt_{k-2} \int_{t_0}^{t_{k-2}} \mspace{-5mu} dt_{k-1}
  \mspace{2mu}
  \mathcal{P} \breve{\mathcal{L}}_{\mathrm{ex}}(t) \mathcal{Q} \breve{\mathcal{L}}_{\mathrm{ex}}(t_1) \mathcal{Q} \breve{\mathcal{L}}_{\mathrm{ex}}(t_2) \cdots \mathcal{Q} \breve{\mathcal{L}}_{\mathrm{ex}}(t_{k-1}) \mathcal{P} \breve{\rho}(t_{k-1}).
  \label{R_k_int_pic}
\end{align}

\subsubsection{Higher-order Markovian QME}

We now perform the Markov approximation
by replacing $\breve{\rho}(t_{k-1})$ with $\breve{\rho}(t)$
and taking the limit of $t_0 \to -\infty$ in Eq.~\eqref{R_k_int_pic}.
Going back to the Schr\"odinger picture,
we thus obtain a Markovian QME for the noise-averaged state $\overline{\rho}(t) \equiv \mathcal{P} \hat{\rho}(t)$ of the system:
\begin{align}
  \frac{d}{dt} \overline{\rho}(t)
   & = \frac{1}{i\hbar} [\hat{H}_S, \overline{\rho}(t)]
  + \sum_{k=2}^\infty \mathcal{R}_k \overline{\rho}(t),
  \label{Markov_QME}
  \\
  \mathcal{R}_k
   & \equiv \int_{-\infty}^t \mspace{-2mu} dt_1 \int_{-\infty}^{t_1} \mspace{-2mu} dt_2 \cdots
  \int_{-\infty}^{t_{k-3}} \mspace{-5mu} dt_{k-2} \int_{-\infty}^{t_{k-2}} \mspace{-5mu} dt_{k-1}
  \mspace{2mu}
  \mathcal{P} \mathcal{L}_{\mathrm{ex}}(t) \mathcal{Q} \mathcal{L}_{\mathrm{ex}}(t_1) \mathcal{Q} \mathcal{L}_{\mathrm{ex}}(t_2) \cdots \mathcal{Q} \mathcal{L}_{\mathrm{ex}}(t_{k-1}) \mathcal{P}.
  \label{R_k}
\end{align}
Here, $\mathcal{L}_{\mathrm{ex}}(t) \hat{A} = (1/i\hbar) [\hat{H}_{\mathrm{ex}}(t), \hat{A}]$,
which is rewritten as
\begin{align}
  \mathcal{L}_{\mathrm{ex}}(t)
  = -i \xi(t) \sum_{j=1}^{N_\mathrm{tot}} \bigl[ g_j(t) L_j^+ + g_j^*(t) L_j^- \bigr].
\end{align}
with Eq.~\eqref{L_j_pm}.

To rewrite the Markovian QME into a tractable form,
we note that the integrand in Eq.~\eqref{R_k} is expressed as
\begin{align}
  \mathcal{P} \mathcal{L}_{\mathrm{ex}}(t) \mathcal{Q} \mathcal{L}_{\mathrm{ex}}(t_1) \mathcal{Q} \mathcal{L}_{\mathrm{ex}}(t_2) \cdots \mathcal{Q} \mathcal{L}_{\mathrm{ex}}(t_{k-1}) \mathcal{P}
  =
  \Bigl\langle\mspace{-6mu}\Bigl\langle
  \mathcal{L}_{\mathrm{ex}}(t) \mathcal{L}_{\mathrm{ex}}(t_1) \mathcal{L}_{\mathrm{ex}}(t_2) \cdots \mathcal{L}_{\mathrm{ex}}(t_{k-1})
  \Bigr\rangle\mspace{-6mu}\Bigr\rangle_{\mspace{-3mu}\mathrm{pc}}
  \mathcal{P}.
  \label{integrand_Markov_QME}
\end{align}
Here, $\dbexpval{\cdots}_{\mathrm{pc}}$ is the ``partial cumulant'' \cite{Shibata_Arimitsu_1980}:
\begin{align}
   & \Bigl\langle\mspace{-6mu}\Bigl\langle
  \mathcal{L}_{\mathrm{ex}}(t) \mathcal{L}_{\mathrm{ex}}(t_1) \mathcal{L}_{\mathrm{ex}}(t_2) \cdots \mathcal{L}_{\mathrm{ex}}(t_{k-1})
  \Bigr\rangle\mspace{-6mu}\Bigr\rangle_{\mspace{-3mu}\mathrm{pc}}
  \notag
  \\
   & = \sideset{}{'}\sum (-1)^{q-1}~
  \dbexpval{\mathcal{L}_{\mathrm{ex}}(t) \cdots \mathcal{L}_{\mathrm{ex}}(t_{\ell_2-1})}
  \times
  \dbexpval{\mathcal{L}_{\mathrm{ex}}(t_{\ell_2}) \cdots \mathcal{L}_{\mathrm{ex}}(t_{\ell_3 -1})}
  \times \cdots \times
  \dbexpval{\mathcal{L}_{\mathrm{ex}}(t_{\ell_q})\cdots\mathcal{L}_{\mathrm{ex}}(t_{k-1})},
  \label{partial_cumulant}
\end{align}
where the sum is taken over all the possible divisions of the large bracket on the left-hand side into smaller brackets keeping the chronological order $t \ge t_1 \ge t_2 \ge \cdots \ge t_{k-1}$,
and $q$ is the number of averages (brackets) in the summand.
Thus, we next investigate the $n$-point correlation
$\dbexpval{\mathcal{L}_{\mathrm{ex}}(t_{\ell+1}) \mathcal{L}_{\mathrm{ex}}(t_{\ell+2})\cdots \mathcal{L}_{\mathrm{ex}}(t_{\ell + n})}$
with $t_{\ell+1} \ge t_{\ell+2} \ge \cdots \ge t_{\ell+n}$.
Since $\xi$ and $g_j$ are independent,
we can calculate their correlations separately:
\begin{align}
   & \dbexpval{\mathcal{L}_{\mathrm{ex}}(t_{\ell+1}) \mathcal{L}_{\mathrm{ex}}(t_{\ell+2})\cdots \mathcal{L}_{\mathrm{ex}}(t_{\ell + n})}
  \notag
  \\
   & = (-i)^n
  \bigl\langle \xi(t_{\ell+1}) \xi(t_{\ell+2}) \cdots \xi(t_{\ell+n}) \bigr\rangle_{\mspace{-1mu}\xi}
  \sum_{j_1=1}^{N_\mathrm{tot}} \sum_{j_2=1}^{N_\mathrm{tot}} \cdots \sum_{j_n=1}^{N_\mathrm{tot}}
  \mspace{2mu}
  \expval{\sideset{}{'}\prod_{m=1}^n \bigl[ g_{j_m}(t_{\ell+m}) L_{j_m}^+ + g_{j_m}^*(t_{\ell+m}) L_{j_m}^- \bigr]}_{\mspace{-6mu}g},
  \label{n_correlation}
\end{align}
where the terms in the product $\prod_{m}'$ are arranged in ascending order of $m$.

\subsubsection{Correlation functions of noises}

We first calculate the RTN part.
As is easily shown \cite{Gardiner_2009}, the stationary probability $P_{\mathrm{s}}$ of the RTN is given by
\begin{align}
  P_{\mathrm{s}}(\xi) =
  \begin{cases}
    \nu / (\nu + \mu) & (\xi = 0)
    \\[4pt]
    \mu / (\nu + \mu) & (\xi = 1),
  \end{cases}
\end{align}
and the conditional probability by
\begin{align}
  P(\xi', t' | \xi, t)
  = P_{\mathrm{s}}(\xi') + (2\xi' - 1) \qty(\xi - \frac{\mu}{\nu + \mu}) e^{-(\nu + \mu)(t' - t)}
  \quad (t' \ge t).
\end{align}
By using these probabilities, we obtain an expression of the correlation function as
\begin{align}
   & \bigl\langle \xi(t_{\ell+1}) \xi(t_{\ell+2}) \cdots \xi(t_{\ell+n}) \bigr\rangle_{\mspace{-1mu}\xi}
  \notag
  \\
   & = \sum_{\xi_1, \xi_2,..., \xi_n = 0,1} \xi_1 \xi_2 \cdots \xi_n
  \mspace{2mu}
  P(\xi_1, t_{\ell+1} | \xi_2, t_{\ell+2})
  P(\xi_2, t_{\ell+2} | \xi_3, t_{\ell+3})
  \cdots
  P(\xi_{n-1}, t_{\ell+n-1} | \xi_n, t_{\ell+n})
  P_{\mathrm{s}}(\xi_n)
  \notag
  \\
   & = \frac{\mu}{(\nu + \mu)^n}
  \prod_{m=1}^{n-1} \qty[ \mu + \nu e^{-(\nu + \mu)(t_{\ell+m} - t_{\ell+m+1})} ].
  \label{correlation_RTN}
\end{align}
Since $\mu \ll \nu$, the second term is dominant in each square bracket in the product if $|t_{\ell+m} - t_{\ell+m+1}|$ is not too large.
Therefore, we can approximate the correlation function as
\begin{align}
  \bigl\langle \xi(t_{\ell+1}) \xi(t_{\ell+2}) \cdots \xi(t_{\ell+n}) \bigr\rangle_{\mspace{-1mu}\xi}
  ~\simeq~ \frac{\mu \nu^{n-1}}{(\nu + \mu)^n} e^{-(\nu + \mu)(t_{\ell+1} - t_{\ell+n})}
  ~\simeq~ \frac{\mu}{\nu} e^{-\nu(t_{\ell+1} - t_{\ell+n})}.
  \label{correlation_RTN_approx}
\end{align}
From this result, we can estimate the magnitude of the terms composed of $q$ averages (brackets) in the partial cumulant [Eq.~\eqref{partial_cumulant}] as $O\bigl( (\mu/\nu)^q \bigr)$, so that the term with $q=1$ is the leading term when $\mu \ll \nu$.
Therefore, in the first order of $\mu/\nu$, we can replace the partial cumulant $\expval{\expval{\cdots}}_{\mathrm{pc}}$ in Eqs.~\eqref{integrand_Markov_QME} and \eqref{partial_cumulant} with the ordinary average $\expval{\expval{\cdots}}$.

We next calculate the Gaussian noise part.
Due to the properties of the Gaussian noise $\{g_j\}$ with Eqs.~\eqref{noise_amplitude_avg_1}--\eqref{noise_amplitude_avg_3},
$\expval{\cdots}_g$ in Eq.~\eqref{n_correlation} has non-zero contribution
only if all the following three conditions are satisfied:
\begin{itemize}
  \item $n$ is even,
  \item the $n$ index numbers ($1, 2,..., n$) for $m$ can be divided into $n/2$ pairs in each of which the index of emitter is the same (e.g., $j_m = j_{\widetilde{m}}$ if $\widetilde{m}$ is the pair to $m$), and
  \item for each pair (say, $m$ and $\widetilde{m}$) in the above division,
        if $g_{j_m}(t_{\ell+m}) L_{j_m}^+$ is chosen for $m$,
        then $g_{j_{\widetilde{m}}}^*(t_{\ell+\widetilde{m}}) L_{j_{\widetilde{m}}}^-$ is chosen for $\widetilde{m}$
        (and vice versa).
\end{itemize}
Therefore,
we can rewrite the latter part on the right-hand side of Eq.~\eqref{n_correlation} for $n=2N$ as
\begin{align}
   & \sum_{j_1=1}^{N_\mathrm{tot}} \sum_{j_2=1}^{N_\mathrm{tot}} \cdots \sum_{j_n=1}^{N_\mathrm{tot}}
  \expval{\sideset{}{'}\prod_{m=1}^n \bigl[ g_{j_m}(t_{\ell+m}) L_{j_m}^+ + g_{j_m}^*(t_{\ell+m}) L_{j_m}^- \bigr]}_{\mspace{-6mu}g}
  \notag
  \\[2pt]
   & = \mathfrak{g}^{2N} \sum_{(b,\tilde{b})}
  \exp \mspace{-3mu} \qty[ -\nu \sum_{i=1}^N \qty( t_{\ell + b(i)} - t_{\ell + \tilde{b}(i)} ) ]
  \mspace{2mu} \sum_{j_{b(1)}=1}^{N_\mathrm{tot}} \mspace{2mu} \sum_{j_{b(2)}=1}^{N_\mathrm{tot}} \cdots \mspace{-4mu} \sum_{j_{b(N)}=1}^{N_\mathrm{tot}}
  \mspace{3mu} \sum_{s_1=\pm} \mspace{2mu} \sum_{s_2=\pm} \cdots \mspace{-3mu} \sum_{s_N=\pm}
  \mathcal{O} \qty[ \prod_{i=1}^N L_{j_{b(i)}}^{s_i} L_{j_{\tilde{b}(i)}}^{-s_i} ].
  \label{Gaussian_noise_average}
\end{align}
Here, $(b,\tilde{b})$ represents a division of the set $\{ 1, 2, ..., 2N \}$ into $N$ pairs $\bigl\{ (b(i), \tilde{b}(i)) \bigr\}_{i=1}^N$
with the constraints of $b(i) < b(i+1)$ and $b(i) < \tilde{b}(i)$ for any $i$,
and $\sum_{(b,\tilde{b})}$ is the sum taken over all the possible pair divisions of this type
(there are $(2N)! / (2^N N!)$ divisions).
In the product $\mathcal{O} \bigl[ \prod_{i} L_{j_{b(i)}}^{s_i} L_{j_{\tilde{b}(i)}}^{-s_i}\bigr]$ in Eq.~\eqref{Gaussian_noise_average},
the $2N$ superoperators are ordered according to the ascending sort of $\{ b(i), \tilde{b}(i) \}_{i=1}^N$, and then $j_{\tilde{b}(i)}$ is replaced with $j_{b(i)}$.
We can furthermore rewrite this equation as
\begin{align}
  = \mathfrak{g}^{2N}
  \sum_{\hat{\jmath}_1=1}^{N_\mathrm{tot}} \mspace{2mu} \sum_{\hat{\jmath}_2=1}^{N_\mathrm{tot}} \cdots \mspace{-4mu} \sum_{\hat{\jmath}_N=1}^{N_\mathrm{tot}}
  \mspace{3mu} \sum_{s_1=\pm} \mspace{2mu} \sum_{s_2=\pm} \cdots \mspace{-3mu} \sum_{s_N=\pm}
  \mspace{3mu} \sum_{(b,\tilde{b})}
  \mspace{3mu} \sideset{}{'}\prod_{m=1}^{2N}
  e^{-\alpha(m) \nu t_{\ell+m}}
  L_{\hat{\jmath}_{\beta(m)}}^{\alpha(m) s_{\beta(m)}},
  \label{Gaussian_noise_average_re}
\end{align}
where $\prod_{m}'$ is ascending order of $m$.
Here, the maps $\alpha$ and $\beta$ of $m \in \{ 1,2,...,2N \}$ depends on the pair division $(b,\tilde{b})$,
though we did not explicitly write the dependence for notational simplicity.
These maps are defined as
\begin{align}
  \alpha(m) & =
  \begin{cases}
    + & (m \in B)
    \\[2pt]
    - & (m \in \tilde{B})
  \end{cases}
  \\[2pt]
  \beta(m)  & =
  \begin{cases}
    b^{-1}(m)         & (m \in B)
    \\[4pt]
    \tilde{b}^{-1}(m) & (m \in \tilde{B}),
  \end{cases}
\end{align}
where $B = \{ b(1), b(2), ..., b(N) \}$ and $\tilde{B} = \{ \tilde{b}(1), \tilde{b}(2), ..., \tilde{b}(N) \}$.
We note that $\alpha(1) = +$, $\alpha(2N)=-$,
$\sum_{m=1}^M \alpha(m)1 \ge 0$ (for $M < 2N$),~~and~~$\sum_{m=1}^{2N} \alpha(m)1 = 0$
are valid for all the pair divisions
due to the constraints in constructing the divisions.

\subsubsection{Higher-order Markovian QME of Model 3}

By combining the results of RTN and Gaussian noise parts,
the partial cumulant in Eq.~\eqref{partial_cumulant} vanish if $k$ is odd,
and if $k=2N_{\mathrm{ex}}$ ($N_{\mathrm{ex}} = 1, 2, 3,...$) it reduces to
\begin{align}
   & \Bigl\langle\mspace{-6mu}\Bigl\langle
  \mathcal{L}_{\mathrm{ex}}(t) \mathcal{L}_{\mathrm{ex}}(t_1) \mathcal{L}_{\mathrm{ex}}(t_2) \cdots \mathcal{L}_{\mathrm{ex}}(t_{2N_{\mathrm{ex}}-1})
  \Bigr\rangle\mspace{-6mu}\Bigr\rangle_{\mspace{-3mu}\mathrm{pc}}
  \notag
  \\
   & \simeq \Bigl\langle\mspace{-6mu}\Bigl\langle
  \mathcal{L}_{\mathrm{ex}}(t) \mathcal{L}_{\mathrm{ex}}(t_1) \mathcal{L}_{\mathrm{ex}}(t_2) \cdots \mathcal{L}_{\mathrm{ex}}(t_{2N_{\mathrm{ex}}-1})
  \Bigr\rangle\mspace{-6mu}\Bigr\rangle
  \notag
  \\
   & = (-1)^{N_{\mathrm{ex}}} \times
  \frac{\mu}{\nu} e^{ -\nu(t - t_{2N_{\mathrm{ex}}-1}) } \times
  \mathfrak{g}^{2N_{\mathrm{ex}}}
  \sum_{\hat{\jmath}_1=1}^{N_\mathrm{tot}} \mspace{2mu} \sum_{\hat{\jmath}_2=1}^{N_\mathrm{tot}} \cdots \mspace{-4mu} \sum_{\hat{\jmath}_{N_{\mathrm{ex}}}=1}^{N_\mathrm{tot}}
  \mspace{3mu} \sum_{s_1=\pm} \mspace{2mu} \sum_{s_2=\pm} \cdots \mspace{-3mu} \sum_{s_{N_{\mathrm{ex}}}=\pm}
  \mspace{3mu} \sum_{(b,\tilde{b})}
  \mspace{3mu} \sideset{}{'}\prod_{m=1}^{2N_{\mathrm{ex}}}
  e^{-\alpha(m) \nu t_{m-1}}
  L_{\hat{\jmath}_{\beta(m)}}^{\alpha(m) s_{\beta(m)}}
  \notag
  \\
   & = \qty(-\mathfrak{g}^2)^{N_{\mathrm{ex}}}
  \frac{\mu}{\nu}
  \mspace{2mu} \sum_{\hat{\jmath}_1=1}^{N_\mathrm{tot}} \mspace{2mu} \sum_{\hat{\jmath}_2=1}^{N_\mathrm{tot}} \cdots \mspace{-4mu} \sum_{\hat{\jmath}_{N_{\mathrm{ex}}}=1}^{N_\mathrm{tot}}
  \mspace{3mu} \sum_{s_1=\pm} \mspace{2mu} \sum_{s_2=\pm} \cdots \mspace{-3mu} \sum_{s_{N_{\mathrm{ex}}}=\pm}
  \mspace{3mu} \sum_{(b,\tilde{b})}
  \qty( e^{ -2\nu(t - t_{2N_{\mathrm{ex}}-1}) }
  \prod_{m=2}^{2N_{\mathrm{ex}}-1} e^{-\alpha(m) \nu t_{m-1}} )
  \sideset{}{'}\prod_{m=1}^{2N_{\mathrm{ex}}}
  L_{\hat{\jmath}_{\beta(m)}}^{\alpha(m) s_{\beta(m)}},
  \label{partial_cumulant_re}
\end{align}
where $t_0=t$ in the third line,
and $\alpha(1) = +$ and $\alpha(2N_{\mathrm{ex}})=-$ are used in the fourth line.
After substituting this result into Eq.~\eqref{R_k},
we calculate the time integral to obtain
\begin{align}
   & \int_{-\infty}^t \mspace{-2mu} dt_1 \int_{-\infty}^{t_1} \mspace{-2mu} dt_2 \cdots
  \int_{-\infty}^{t_{2N_{\mathrm{ex}}-3}} \mspace{-5mu} dt_{2N_{\mathrm{ex}}-2}
  \int_{-\infty}^{t_{2N_{\mathrm{ex}}-2}} \mspace{-5mu} dt_{2N_{\mathrm{ex}}-1}
  \mspace{2mu}
  \qty( e^{ -2\nu(t - t_{2N_{\mathrm{ex}}-1}) } \prod_{m=2}^{2N_{\mathrm{ex}}-1} e^{-\alpha(m) \nu t_{m-1}} )
  \notag
  \\
   & = e^{-2\nu t} \int_{-\infty}^t \mspace{-2mu} dt_1 \mspace{2mu} e^{-\alpha(2) \nu t_1}
  \int_{-\infty}^{t_1} \mspace{-2mu} dt_2 \mspace{3mu} e^{-\alpha(3) \nu t_2} \cdots
  \int_{-\infty}^{t_{2N_{\mathrm{ex}}-3}} \mspace{-5mu} dt_{2N_{\mathrm{ex}}-2} \mspace{3mu} e^{-\alpha(2N_{\mathrm{ex}}-1) \nu t_{2N_{\mathrm{ex}}-2}}
  \int_{-\infty}^{t_{2N_{\mathrm{ex}}-2}} \mspace{-5mu} dt_{2N_{\mathrm{ex}}-1} \mspace{3mu} e^{2\nu t_{2N_{\mathrm{ex}}-1}}
  \notag
  \\[2pt]
   & = \frac{1}{D(\alpha) \nu^{2N_{\mathrm{ex}}-1}},
\end{align}
where $D(\alpha)$ is a positive constant given by
\begin{align}
  D(\alpha)
  = 2 \prod_{m=2}^{2N_{\mathrm{ex}}-1} \qty(2 - \sum_{m'=m}^{2N_{\mathrm{ex}}-1} \alpha(m')1)
\end{align}
for $N_{\mathrm{ex}} \ge 2$, and $D(\alpha) = 2$ for $N_{\mathrm{ex}} = 1$.
Therefore, Eq.~\eqref{R_k} for $k = 2N_{\mathrm{ex}}$ becomes
\begin{align}
  R_{k=2N_{\mathrm{ex}}}
  = \mu \qty(-\frac{\mathfrak{g}^2}{\nu^2})^{\mspace{-3mu}N_{\mathrm{ex}}}
  \mspace{2mu} \sum_{\hat{\jmath}_1=1}^{N_\mathrm{tot}} \mspace{2mu} \sum_{\hat{\jmath}_2=1}^{N_\mathrm{tot}} \cdots \mspace{-4mu} \sum_{\hat{\jmath}_{N_{\mathrm{ex}}}=1}^{N_\mathrm{tot}}
  \mspace{3mu} \sum_{s_1=\pm} \mspace{2mu} \sum_{s_2=\pm} \cdots \mspace{-3mu} \sum_{s_{N_{\mathrm{ex}}}=\pm}
  \mspace{3mu} \sum_{(b,\tilde{b})} \frac{1}{D(\alpha)} \mspace{3mu}
  \sideset{}{'}\prod_{m=1}^{2N_{\mathrm{ex}}}
  L_{\hat{\jmath}_{\beta(m)}}^{\alpha(m) s_{\beta(m)}},
\end{align}
We thus obtain $\mathcal{D}_{\mathrm{ex}}^{(3)}$ in the QME of Model 3:
\begin{align}
  \mathcal{D}_{\mathrm{ex}}^{(3)}
  = \mu \sum_{N_{\mathrm{ex}}=1}^\infty
  \qty(-\frac{\mathfrak{g}^2}{\nu^2})^{\mspace{-3mu}N_{\mathrm{ex}}}
  \sum_{j_1,j_2,...,j_{N_{\mathrm{ex}}}=1}^{N_\mathrm{tot}}
  \mspace{4mu}
  \sum_{s_1,s_2,...,s_{N_{\mathrm{ex}}}=\pm}
  \mspace{4mu}
  \sum_{(b,\tilde{b})} \frac{1}{D(\alpha)}
  \mspace{3mu}
  \sideset{}{'}\prod_{m=1}^{2N_{\mathrm{ex}}}
  L_{j_{\beta(m)}}^{\alpha(m) s_{\beta(m)}}.
  \tag{\ref{model3}}
\end{align}

\end{document}